\documentclass[12pt]{article}
\usepackage{graphicx}
\usepackage{color}
\usepackage{amsmath, amssymb, amsthm, ascmac, caption, comment, enumerate, bm, newtxtext}
\usepackage{latexsym, mathabx, mathrsfs, mathtools, psfrag, setspace, subcaption}
\usepackage{booktabs}
\usepackage{natbib}
\usepackage{clipboard}
\usepackage{algorithm}
\usepackage{algorithmic}

\mathtoolsset{showonlyrefs=true}

\definecolor{mycolor}{cmyk}{0.82, 0.23, 0.26, 0}
\usepackage[colorlinks=true, linkcolor=mycolor, filecolor=mycolor, urlcolor=mycolor, citecolor=mycolor, driverfallback=dvipdfmx]{hyperref}
\usepackage[margin = 2.3cm]{geometry}

\usepackage{titlesec}
\titleformat*{\section}{\Large\bfseries\sffamily}
\titleformat*{\subsection}{\large\bfseries\sffamily}
\titleformat*{\subsubsection}{\large\bfseries\sffamily}

\DeclareMathOperator*{\argmin}{argmin}
\DeclareMathOperator*{\argmax}{argmax}

\DeclareMathOperator*{\Var}{\mathrm{Var}}

\renewcommand{\hat}{\widehat}
\renewcommand{\tilde}{\widetilde}

\renewcommand{\bar}{\overline}
\makeatletter
\@addtoreset{figure}{section}
\makeatother
\makeatletter
\@addtoreset{table}{section}
\makeatother
\numberwithin{equation}{section}
 
\allowdisplaybreaks[2]

\theoremstyle{definition}
\newtheorem{theorem}{Theorem}[section]
\newtheorem{assumption}{Assumption}[section]
\newtheorem{definition}{Definition}[section]
\newtheorem{corollary}{Corollary}[section]
\newtheorem{example}{Example}[section]
\newtheorem{lemma}{Lemma}[section]
\newtheorem{proposition}{Proposition}[section]
\newtheorem{remark}{Remark}[section]

\title{Quantile Vector Autoregression without Crossing}

\author{ 
    Tomohiro Ando\thanks{Melbourne Business School, the University of Melbourne. Email: \href{mailto:T.Ando@mbs.edu}{T.Ando@mbs.edu}},   
    Tadao Hoshino\thanks{
    Address Correspondence to: Tadao Hoshino,
    School of Political Science and Economics, Waseda University. Email: \href{mailto:thoshino@waseda.jp}{thoshino@waseda.jp}}, and
    Ruey Tsay\thanks{Booth School of Business, the University of Chicago. Email: \href{mailto:ruey.tsay@chicagobooth.edu}{ruey.tsay@chicagobooth.edu}} 
}

\date{}

\begin{document}

\maketitle
\begin{abstract}

    This paper considers estimation and model selection of quantile vector autoregression (QVAR).
    Conventional quantile regression often yields undesirable crossing quantile curves, violating the monotonicity of quantiles.
    To address this issue, we propose a simplex quantile vector autoregression (SQVAR) framework, which transforms the autoregressive (AR) structure of the original QVAR model into a simplex, ensuring that the estimated quantile curves remain monotonic across all quantile levels.
    In addition, we impose the smoothly clipped absolute deviation (SCAD) penalty on the SQVAR model to mitigate the explosive nature of the parameter space.
    We further develop a Bayesian information criterion (BIC)-based procedure for selecting the optimal penalty parameter and introduce new frameworks for impulse response analysis of QVAR models. 
    Finally, we establish asymptotic properties of the proposed method, including the convergence rate and asymptotic normality of the estimator, the consistency of AR order selection, and the validity of the BIC-based penalty selection.
    For illustration, we apply the proposed method to U.S. stock market data, highlighting the usefulness of our SQVAR method.

    \bigskip

\end{abstract}


\clearpage
\section{Introduction}

Vector autoregressive (VAR) models are key tools in macroeconometrics for analyzing multivariate time series.
Traditional VAR models focus on conditional means, which limits their ability to capture asymmetric and heterogeneous responses across the distribution of time-series outcomes.
Quantile VAR (QVAR) models (e.g., \citealp{ando2022quantile, ando2024scenario, chavleishvili2024forecasting}) overcome this limitation by extending the idea of quantile regression (\citealp{koenker1978regression}) to multivariate time series and provide a more flexible framework by modeling conditional quantiles rather than conditional means.

However, QVAR models encounter two major technical challenges.
The primary challenge is quantile crossing, which violates the fundamental requirement that quantile functions must be non-decreasing in the quantile level.
When quantile crossing occurs, the model suffers from poor interpretability, incoherent forecasts, and invalid econometric inference.
This problem becomes particularly relevant when an estimated QVAR model is used for simulation exercises such as impulse response analysis.
In these applications, one often needs to recover the quantile level associated with each outcome value at each time point; without monotonicity, this inversion need not be unique.
Thus, quantile crossing undermines both the practical usefulness and theoretical validity of the QVAR model.
The second challenge is the quadratic increase in the number of coefficient parameters with the number of time series, often resulting in a loss of estimation efficiency.
Although this issue also arises in standard VAR models, it is more severe in QVAR because the parameters to be estimated are functions of the quantile level rather than scalars.

In the literature, traditional remedies to the first challenge include post-processing rearrangement (\citealp{chernozhukov2010quantile}) and directly imposing monotonicity via isotonic regression (\citealp{bondell2010noncrossing}).
However, these approaches often involve ad hoc corrections, limiting the scalability and interpretability of the resulting models.
Recently, \cite{ando2025simplex} proposed an alternative approach by embedding the domain of the quantile function into a simplex, which they call simplex quantile regression (SQR).
The SQR method reformulates the quantile regression problem as an optimization over the simplex space, ensuring non-crossing, smoothness of the quantile function, and computational efficiency. 
In this paper, we incorporate the approach of \cite{ando2025simplex} into the QVAR framework and propose a simplex QVAR (SQVAR) approach, which retains the same advantages of the SQR method in the context of QVAR.

Our SQVAR model serves as a fundamental tool for capturing dynamic and heterogeneous interdependence among multiple time series.
However, similar to traditional VAR models, it may suffer from the "curse of dimensionality": as the number of time series and the order of AR lags increase, the variance of the estimator can increase rapidly.
This occurs because the number of parameters grows quadratically with the number of series under study, making statistical inference unreliable without dimension reduction.
Thus, as in conventional VAR models, a key challenge in the SQVAR framework is determining which lagged variables are truly relevant for the system's dynamics.

To address this dimensionality issue, we adopt the smoothly clipped absolute deviation (SCAD) regularization method (\citealp{fan2001variable}), given its ability to consistently select relevant lags while allowing us to estimate active parameters without asymptotic bias.
Note that, in order to recover the correct AR structure of the original QVAR model, the SCAD penalty must be properly translated to ensure an equivalent penalization structure in the context of SQVAR. Under a suitably constructed simplex embedding and the corresponding SCAD penalty, we show that our proposed estimator based on series approximation can consistently identify the correct AR structure, and enjoys consistency and asymptotic normality for the coefficients of active lags.
In addition, we propose a Bayesian information criterion (BIC)-based method for selecting the penalty parameter, and establish its theoretical validity.

Similar to standard VAR models, once a QVAR model is estimated, we can perform impulse response analysis to study the dynamic effects of macroeconomic or financial shocks on economic variables of interest over time.
In this paper, we propose two types of impulse response analysis.
The first type is the generalized impulse response analysis, which can be viewed as a QVAR version of \cite{pesaran1998generalized}.
In this approach, the response to a shock is measured in terms of expected outcomes.
As in conventional VAR models, this approach averages out the dependence across different quantile levels and may overlook important heterogeneous aspects of economic dynamics, especially in the presence of economic asymmetries, regime changes, or tail events such as financial crises.
Alternatively, \cite{ando2024scenario} introduced scenario-based forecasting error variance decomposition analysis.
As the second approach, we extend their idea to the QVAR framework.
This approach provides a more granular and informative depiction of dynamic relationships among the variables under particular economic conditions specified by the researcher. 

To illustrate an application of the proposed method, we analyze daily returns of U.S. exchange-traded funds (ETFs) covering major financial market segments.
The estimation results reveal heterogeneous dynamic interactions across these ETFs, providing a detailed view of how dependence patterns vary across different parts of the return distribution.
We then conduct a scenario-based impulse response analysis for two major events, the 2008 financial crisis and the COVID-19 pandemic, using a low-volatility market period as the baseline.

\paragraph{Paper organization.}
The rest of the paper is organized as follows.
Section \ref{sec:qvar} introduces our working model, the QVAR.
In this section, we discuss how to transform a QVAR model into an SQVAR form and provide the stationarity condition of the model.
In Section \ref{sec:estimation}, we describe our SCAD-penalized estimation procedure and a BIC-type model selection method.
Asymptotic properties of the proposed method, including the rate of convergence, asymptotic normality, and the consistency of model selection, are presented in Section \ref{sec:asymp}.
In Section \ref{sec:irf}, we introduce two new impulse response analysis approaches.
Section \ref{sec:MC} presents numerical studies, including a set of Monte Carlo simulations and the empirical application to daily returns of U.S. ETFs.
Section \ref{sec:conclude} concludes.
All technical proofs are relegated to the appendix.

\paragraph{Notation.} 
For an integer $a$, we write $[a] \coloneqq \{1, 2, \ldots, a\}$.
$\sigma(X)$ denotes the $\sigma$-field generated by $X$.
For a $p \times q$ matrix $M \coloneqq (m_{ij})$, define the norms $||M|| \coloneqq \left(\sum_{i=1}^p\sum_{j=1}^q m_{ij}^2\right)^{1/2}$, $||M||_2 \coloneqq \left(\bar{\text{eig}}(M^\top M)\right)^{1/2}$, and $||M||_\infty \coloneqq \max_{1 \leq i \leq p}\sum_{j=1}^q |m_{ij}|$, where $M^\top$ is the transpose of $M$, and, for a nonnegative definite matrix $A$, $\bar{\text{eig}}(A)$ and $\underbar{\text{eig}}(A)$ denote its maximum and minimum eigenvalues, respectively.
For a function $f$ on $[0,1]$, we denote $||f||_2 \coloneqq \left(\int_0^1 |f(u)|^2 \mathrm{d}u \right)^{1/2}$, and $||f||_\infty \coloneqq \sup_{u \in [0,1]}|f(u)|$.
For a random variable $X$, define $||X||_q \coloneqq \left( \mathbb{E}|X|^q \right)^{1/q}$ for $q \ge 1$.
Finally, $c$, $\bar c$, and $\underline{c}$ possibly with subscripts denote generic positive constants.

\section{Quantile VAR without Crossing}\label{sec:qvar}

\subsection{The model}

Consider an $n$-dimensional vector time series $Y_t = (y_{1t}, \ldots ,y_{nt})^\top$ 
with $T$ equally-spaced observations $\{Y_1, \ldots , Y_T\}$.
For these $n$ time series, we assume that they follow a quantile vector autoregressive (QVAR) model of order $p$, in which their conditional quantile functions are given by
\begin{align}\label{eq:model}
    \begin{array}{c}
    Q_{y_{1t}}(\tau \mid \mathcal F_{t-1}) =  \theta_{01}(\tau) + \sum_{j=1}^p \left[ \theta_{11}^{(j)}(\tau) y_{1, t - j}  + \cdots + \theta_{n1}^{(j)}(\tau) y_{n, t - j} \right]\\
    \quad \vdots \\
    Q_{y_{nt}}(\tau \mid \mathcal F_{t-1}) =  \theta_{0n}(\tau) + \sum_{j=1}^p \left[ \theta_{1n}^{(j)}(\tau) y_{1, t - j}  + \cdots + \theta_{nn}^{(j)}(\tau) y_{n, t - j} 
    \right],\end{array}
\end{align}
for $\tau \in (0,1)$, where $\mathcal F_t = \sigma\{Y_s : s \le t\}$ is the information set available at time $t$, and $\theta_{0i}$ and $\theta_{li}^{(j)}$ denote unknown functions to be estimated, for $l,i \in [n]$ and $j \in [p]$.

Similar to \cite{koenker2006quantile}, we can characterize the data generating process (DGP) of $Y_t$ at each $t$ via a random-coefficient specification by introducing an $n$-dimensional uniform rank variable $\bm{U}_t = (U_{1t}, \ldots, U_{nt})$ in place of $\tau$ in \eqref{eq:model}.
That is, the DGP for the $i$-th time series $y_{it}$ can be expressed as
\begin{align}\label{eq:DGP}
    y_{it} = \mu_i + \sum_{j=1}^p \left[ 
    \theta_{1i}^{(j)}(U_{it}) y_{1, t - j} + \cdots + \theta_{ni}^{(j)}(U_{it}) y_{n, t - j} \right] + \varepsilon_{it},
\end{align}
where $\mu_i \coloneqq \mathbb E [\theta_{0i}(\text{Uniform}[0,1])]$ and $\varepsilon_{it} \coloneqq \theta_{0i}(U_{it}) - \mu_i$.
Here, $\varepsilon_{it}$ is interpreted as the innovation error term. 
The randomness of each $y_{it}$ is due to the randomness from all of $\bm U_t$, $\bm U_{t-1}$, $\ldots$.
For the stochastic process $\bm U_t$, we impose the following assumption.

\begin{assumption}\label{as:U} 
The process $\bm U_t$ is independent over $t$.
For each $t$, $\bm U_t$ has a time-invariant joint distribution with uniform marginal distributions $\text{Uniform}[0,1]$.
\end{assumption}

Assumption \ref{as:U} allows for arbitrary cross-sectional dependence across the elements of $\bm U_t$.
Thus, as in the traditional VAR model, correlation across the innovation terms $(\varepsilon_{1t}, \ldots, \varepsilon_{nt})$ is permitted through the correlation of $U_{it}$'s.

As another important feature of our model, we allow the model \eqref{eq:model} to be potentially over-parameterized in that the following two sets may be nonempty:
\begin{align}
    \mathcal{S}_{0i}
    \coloneqq \left\{(l,j) \in \mathcal S: \left\| \theta_{li}^{(j)}\right\|_2 = 0\right\}, \qquad
    \mathcal{S}_{1i}
    \coloneqq \left\{(l,j) \in \mathcal S: \left\| \theta_{li}^{(j)}\right\|_2 \neq 0\right\},
\end{align}
where $\mathcal S \coloneqq [n] \times [p]$.
This sparse structure is important in practice because researchers usually have no prior information about the lag order $p$, and hence may wish to adopt a relatively large $p$ to avoid misspecification.
However, as $p$ increases, the number of coefficients to be estimated grows rapidly, causing a severe finite-sample efficiency issue.
Note that this issue is more serious in QVAR than in the standard VAR, in which the parameters to be estimated are scalars, while they are functional parameters in QVAR.
To mitigate this problem, we advocate the use of a SCAD-penalized estimator that can automatically detect $\mathcal{S}_{0i}$ and $\mathcal{S}_{1i}$.

\subsection{Simplex QVAR representation}

The quantile functions in \eqref{eq:model} are non-decreasing by definition.
To effectively incorporate the monotonicity into the estimation procedure, we introduce a notion similar to the maximum effective region (MER) proposed in \cite{ando2025simplex}: the set of past outcome values where the model satisfies the quantile monotonicity. 

Define $N \coloneqq np$, $\Theta_i^{(j)}(\tau) = (\theta_{1i}^{(j)}(\tau), \ldots, \theta_{ni}^{(j)}(\tau))^\top$, 
$\underbracket{\bm{\Theta}_i(\tau)}_{(N + 1) \times 1} = (\theta_{0i}(\tau), \Theta_i^{(1)}(\tau)^\top, \ldots, \Theta_i^{(p)}(\tau)^\top)^\top$, 
and $\underbracket{W_t}_{(N + 1) \times 1} = (1, Y_{t - 1}^\top, \ldots, Y_{t - p}^\top)^\top$.
Then, for each $i \in [n]$, we can succinctly write \eqref{eq:model} as
\begin{align}
    Q_{y_{it}}(\tau \mid \mathcal F_{t-1}) = W_t^\top \bm{\Theta}_i(\tau).
\end{align}
Next, we define the strict MER (SMER) as follows.
\begin{definition}[SMER]
For each $i \in [n]$, 
\begin{align}
    \text{SMER}_i \coloneqq \left\{ w \in \mathbb{R}^{N + 1} : \frac{\text{d} }{\text{d} u} \left( w^\top \bm{\Theta}_i(u) \right) > 0 \;\; \text{for all} \;\; u \in (0,1)\right\}.
\end{align}
\end{definition}

The SMER is a stronger version of the MER in \cite{ando2025simplex} by requiring that the conditional quantile function of $y_{it}$ given $W_t = w$ is strictly increasing in $\tau$ for any $w \in \text{SMER}_i$.
When $\text{SMER}_i$ is required to coincide with the entire support of $W_t$, this holds if the conditional cumulative distribution function (CDF) of $y_{it}$ given $W_t = w$ is strictly increasing for any $w$ in the support of $W_t$, which is a common requirement in the quantile regression literature.
However, as shown below, our estimation procedure requires further that certain points potentially outside the support of $W_t$ also be included in $\text{SMER}_i$.
Thus, to guarantee this property, we impose some structural assumptions directly on the form of the true coefficient functions (see Assumption \ref{as:SMER}(ii) below). 
This assumption is a cost of our SQVAR method, incurred in exchange for its nice properties, and is usually not required in other monotone quantile regression approaches.

\bigskip

Following \cite{ando2025simplex}, we transform the QVAR model into a simplex QVAR (SQVAR) model with monotonic coefficient functions.  
Let $\bm v_0, \bm v_1, \ldots, \bm v_N \in \mathbb{R}^{N+1}$ be affine independent reference vertices such that, for any $w$ in the support of $W_t$, there exist nonnegative weights $\{c_j(w)\}_{j=0}^N$ with $\sum_{j=0}^N c_j(w)=1$ satisfying
\begin{align}\label{eq:conv}
    w = \sum_{j=0}^N c_j(w)\, \bm v_j .
\end{align}
In other words, $\{c_0(w), c_1(w), \ldots, c_N(w)\}$ represent the barycentric coordinates of $w$ with respect to the vertices $\bm v_0, \bm v_1, \ldots, \bm v_N$.  
Once such a coordinate system is specified (an example will be given later), the model can be rewritten as
\begin{align}
    y_{it} = W_t^\top \bm \Theta_i(U_{it}) 
      = \sum_{j=0}^N c_j(W_t) \bm v_j^\top \bm \Theta_i(U_{it}) 
      = C_t^\top \bm \Phi_i(U_{it}),
\end{align}
where $C_t = (c_0(W_t), \ldots, c_N(W_t))^\top$,  
$\bm \Phi_i(U_{it}) = (\phi_{0i}(U_{it}), \ldots, \phi_{Ni}(U_{it}))^\top$, and  
$\phi_{ji}(U_{it}) \coloneqq \bm v_j^\top \bm \Theta_i(U_{it})$.  
By construction, if $\bm v_j \in \text{SMER}_i$, then the SQVAR coefficient function $u \mapsto \phi_{ji}(u)$ is strictly increasing.
Therefore, if this property holds for all $j$, the nonnegativity of $C_t$ ensures that estimating the $\phi_{ji}$'s under monotonicity constraints automatically yields a monotone conditional quantile curve estimate, as desired.
Note also that, for each $w$ to be embedded in the barycentric coordinate system, the support of $W_t$ must be bounded.
We formally state these requirements in the following assumption.

\begin{assumption}\label{as:SMER}
    (i) For all $i \in [n]$ and $t \in [T]$, there exist $-\infty < \text{lb}_i, \text{ub}_i < \infty$ such that $\text{lb}_i \le y_{it} \le \text{ub}_i$.  
    (ii) There exist $\bm v_0, \bm v_1, \ldots, \bm v_N \in \text{SMER}_i$ satisfying \eqref{eq:conv} for all $w$ in the support of $W_t$.  
\end{assumption}

Assumption \ref{as:SMER}(i) requires that the observations of $y_{it}$ are bounded within a fixed range.
As discussed in \cite{bondell2010noncrossing}, in the context of linear quantile regression, estimating non-crossing quantile curves on an unbounded domain automatically requires that the conditional quantile function has constant slopes (i.e., a location-shift model).
Thus, the boundedness condition in Assumption \ref{as:SMER}(i) allows the coefficient functions to vary with $\tau$.
The bound $[\text{lb}_i, \text{ub}_i]$ does not need to be the tightest possible range for $y_{it}$, but it is assumed to be known.
In practice, one can simply use the empirical minimum and maximum of $y_{it}$ for $\text{lb}_i$ and $\text{ub}_i$, respectively.
Assumption \ref{as:SMER}(ii) requires $\text{SMER}_i$ to be sufficiently larger than the support of $W_t$, because the simplex generated by the reference vertices $\bm v_0, \bm v_1, \ldots, \bm v_N$ must contain all observed $W_t$.
Therefore, in the presence of a few extreme observations, the resulting simplex may become unnecessarily large for most observations and impose a strong global shape restriction.
To avoid this issue, one may instead construct the simplex using only a subset of observations of interest, such as the $5\%$--$95\%$ region of $W_t$.
This gives a partial shape restriction and does not guarantee global monotonicity, but it may be useful in practice.
Nevertheless, for simplicity of presentation, we focus only on the case of full enclosure.

Under Assumption \ref{as:SMER}, we obtain the following result, which parallels Theorem 3.1 of \cite{ando2025simplex}.

\begin{lemma}[SQVAR]\label{lem:nc-qvar}
    Suppose that Assumption \ref{as:SMER} holds.
    Then, we have
\begin{align}\label{SQVAR}
        y_{it} = C_t^\top \bm{\Phi}_i(U_{it}),
\end{align}
    where  the elements of $\bm{\Phi}_i$ are all strictly increasing over $(0,1)$. 
    In particular, the SQVAR coefficients $\bm{\Phi}_i(U_{it})$ and the original QVAR coefficients $\bm{\Theta}_i(U_{it})$ are related in the following manner: $\bm{\Theta}_i(U_{it}) = \bm V^{-1}\bm \Phi_i(U_{it})$, where $\bm V = (\bm v_0, \bm v_1, \ldots, \bm v_N)^\top$.
\end{lemma}

The last part of Lemma \ref{lem:nc-qvar} implies that 
\begin{align}
     \left\|  \tilde V_l^{(j)} \bm \Phi_i(\cdot) \right\|_2 = 0 \;\; \text{for all $(l,j) \in \mathcal S_{0i}$}
\end{align}
must hold, where $\tilde V_l^{(j)}$ denotes the corresponding $(l,j)$-th row of $\bm V^{-1}$.
Note also that since $W_t ^\top \bm \Theta_i(U_{it}) = W_t ^\top \bm V^{-1}\bm \Phi_i(U_{it}) = C_t^\top \bm \Phi_i(U_{it})$, this implies $(\bm V^{-1})^{\top} W_t = C_t$; thus, each element of $C_t$ is represented as a linear combination of $(1, y_{1,t-1}, \ldots, y_{n,t-p})$.

Hereinafter, to be consistent with the indexing in the original QVAR model and to improve readability, we shall also use the notations $\bm{\Phi}_i = (\phi_{0i},\phi_{1i}^{(1)}, \ldots, \phi_{ni}^{(1)}, \ldots, \phi_{1i}^{(p)}, \ldots, \phi_{ni}^{(p)})^\top$, $\bm V = (\bm v_0, \bm v_1^{(1)}, \ldots, \bm v_n^{(1)}, \ldots, \bm v_1^{(p)}, \ldots, \bm v_n^{(p)})^\top$, and $C_t = (c_{0t}, c_{1,t-1}, \ldots, c_{n,t-1}, \ldots, c_{1,t-p}, \ldots, c_{n,t-p})^\top$.

\begin{example}\label{ex:max-min}
    A particularly convenient and practical coordinate system is obtained by setting
    \begin{align}
        \bm v_0 
        & = (1, \text{lb}_1, \ldots, \text{lb}_n, \ldots, \text{lb}_1, \ldots, \text{lb}_n) \\
        \bm v_l^{(j)}
        & = (1, \underbracket{\text{lb}_1, \ldots, \text{lb}_n, \ldots, \text{lb}_{l-1}}_{n(j-1) + l - 1\: \text{elements}}, \text{lb}_l + N \Delta_l, \text{lb}_{l + 1}, \ldots, \text{lb}_1, \ldots, \text{lb}_n),
    \end{align}
    where $\Delta_l \coloneqq \text{ub}_l - \text{lb}_l$.
    In this case, we have 
    \begin{align}\label{eq:max-min}
    & c_{it} \coloneqq \frac{y_{it} - \text{lb}_i}{N \Delta_i}, \qquad
    c_{0t} \coloneqq 1 - \sum_{(i,j) \in \mathcal S} c_{i,t - j}.
    \end{align} 
    Furthermore, the inverse matrix $\bm V^{-1}$ has a closed-form expression, and we have
    \begin{align}\label{eq:transform}
        \begin{split}
        \phi_{0i}(\tau) 
        = \theta_{0i}(\tau) + \sum_{(l,j) \in \mathcal{S}_{1i}} \text{lb}_l \theta^{(j)}_{li}(\tau), \qquad
        \phi_{li}^{(j)}(\tau)
        =  N \Delta_l \theta^{(j)}_{li}(\tau) + \phi_{0i}(\tau).
        \end{split}
    \end{align}
    From the second expression, we can find that restricting $\theta^{(j)}_{li}$ to zero means that $ \phi_{li}^{(j)}$ and $\phi_{0i}$ are the same function.
    The derivation of the above result is provided in Appendix \ref{app:max-min}.
\end{example}

\subsection{Stationarity}

Stationarity is a fundamental concept in the analysis of VAR models. 
Whether the time-series process is stationary or not determines how we estimate and interpret the QVAR model. 
In the following, we derive the condition for our QVAR model to generate stationary processes.

Let $A_j(\bm U_t)$ be the $n\times n$ matrix whose $(i,l)$-th element is $\theta^{(j)}_{li}(U_{it})$ and $E(\bm{U}_t) \coloneqq (\theta_{01}(U_{1t}), \ldots, \theta_{0n}(U_{nt}))^\top$. 
The QVAR model becomes 
\begin{align}
    Y_t 
    & = A_1(\bm{U}_t) Y_{t - 1} + \cdots + A_p(\bm{U}_t) Y_{t - p} + E(\bm{U}_t).
\end{align}
This can be further written as an extended QVAR model of order one: 
\begin{align}
    \bm{Y}_t = \mathcal{A}(\bm{U}_t) \bm{Y}_{t-1} + \bm{E}(\bm{U}_t),
\end{align}
where
\begin{align}
    \underbracket{\bm{Y}_t}_{N \times 1} 
    = \left(\begin{array}{c}
    Y_t \\
    Y_{t-1} \\
    \vdots \\
    Y_{t - p + 1}
    \end{array}\right), \quad
    \underbracket{\mathcal{A}(\bm{U}_t)}_{N \times N}
    \coloneqq \left(\begin{array}{cccc}
    A_1(\bm{U}_t) & A_2(\bm{U}_t) & \cdots & A_p(\bm{U}_t) \\
    I_n & \bm{0}_{n \times n} & \cdots & \bm{0}_{n \times n} \\
    \vdots & \ddots & \ddots & \vdots \\
    \bm{0}_{n \times n} & \cdots & I_n & \bm{0}_{n \times n}
    \end{array}\right), \quad 
    \underbracket{\bm{E}(\bm{U}_t)}_{N \times 1} 
    = \left(\begin{array}{c}
    E(\bm{U}_t) \\
    \bm{0}_n \\
    \vdots \\
    \bm{0}_n
    \end{array}\right).
\end{align}
Moreover, let $\Gamma_{t,0} \coloneqq I_N$ and, for $k \geq 1$,
define 
\begin{align}
    \Gamma_{t,k} = \Gamma_{t,k}(\bm{U}_{t}, \ldots, \bm{U}_{t-k+1}) \coloneqq \prod_{l = 1}^k \mathcal{A}(\bm{U}_{t - l + 1}).
\end{align} 
Then, we obtain the moving-average expression 
\begin{align}
    \bm{Y}_t 
    & = \mathcal{A}(\bm{U}_t) \bm{Y}_{t-1} + \bm{E}(\bm{U}_t) \\
    & = \mathcal{A}(\bm{U}_t) \mathcal{A}(\bm{U}_{t - 1}) \bm{Y}_{t-2} + \bm{E}(\bm{U}_t) + \mathcal{A}(\bm{U}_t)\bm{E}(\bm{U}_{t-1})
    = \cdots = \sum_{k = 0}^\infty \Gamma_{t,k} \bm{E}(\bm{U}_{t-k}),
\end{align}
if the following assumption is satisfied.
\begin{assumption}\label{as:MA}
     There exists $|\rho| < 1$ such that $\|\Gamma_{t,k}\|_2 \lesssim \rho^k$ for all $k \ge 0$.
\end{assumption}
Assumption \ref{as:MA} leads to $\left\|\Gamma_{t,k}\right\|_2 \to 0$ as $k \to \infty$, which gives the above expression under Assumption \ref{as:SMER}(i).
In addition, under Assumption \ref{as:U}, we have
    \begin{align}
    \mathbb{E}[\bm{Y}_t] 
    = \sum_{k = 0}^\infty \mathbb{E}[\Gamma_{t,k}] \bm \mu, \; \text{where} \;
    \bm \mu
    = \left(\begin{array}{c}
    (\mu_1, \ldots, \mu_n)^\top \\
    \bm{0}_n \\
    \vdots \\
    \bm{0}_n
    \end{array}\right).
\end{align}
Note that $\mathbb{E}[\Gamma_{t,k}]$ is independent of $t$ under Assumption \ref{as:U}, implying the mean stationarity.

Assumption \ref{as:MA} restricts the relative magnitudes of autoregressive parameters.
However, its implications for individual coefficients are difficult to comprehend.
The next lemma provides an alternative, easier-to-interpret sufficient condition.

\begin{lemma}\label{lem:MA}
    Let $\varrho \coloneqq \sum_{j \in [p]} ( \sup_{\bm u \in [0,1]^n}\left\| A_j(\bm{u}) \right\|_2$), and suppose that $\varrho \in [0,1/p)$.
    Then, Assumption \ref{as:MA} holds.
\end{lemma}

The condition $\varrho \in [0,1/p)$ implies that the dependence on past outcomes needs to be sufficiently weak to achieve stationarity.

\section{Estimation}\label{sec:estimation}

\subsection{SCAD-penalized SQVAR estimator}

As shown in Lemma \ref{lem:nc-qvar}, the SQVAR coefficient functions are monotonically increasing by construction.
To preserve this monotonicity, we estimate the coefficient functions globally using a series approximation, rather than performing pointwise quantile regression repeatedly.

Let $\{b_h: h = 1,2, \ldots\}$ be a sequence of monotone-spline basis functions, such as I-splines (\citealp{ramsay1988monotone}) or monotone B-splines (\citealp{he1998monotone}), and denote $\bm{b}_H(\tau) = (b_1(\tau), \ldots, b_H(\tau))^\top$, where $H$ is the order of the basis functions chosen by the researcher. 
The first element of $\bm{b}_H(\tau)$ is typically a constant.
Suppose that there exist $H \times 1$ vectors $\gamma_{0i}$ and $\gamma^{(j)}_{li}$, for $j=1,\ldots,p$, such that
\begin{align}
    \phi_{0i}(\tau) \approx \bm{b}_H(\tau)^\top \gamma_{0i}, \quad
    \phi_{li}^{(j)}(\tau) \approx \bm{b}_H(\tau)^\top \gamma^{(j)}_{li},
\end{align}
where the approximation errors vanish as $H \to \infty$.
Define $\underbracket{\bm B_H(\tau)}_{(N + 1) \times H(N + 1)} = \text{diag}(\bm{b}_H(\tau)^\top, \ldots, \bm{b}_H(\tau)^\top)$, and $\underbracket{\bm{\gamma}_i}_{H(N + 1) \times 1} = (\gamma_{0i}^\top, \gamma_{1i}^{(1)\top}, \ldots, \gamma_{ni}^{(1)\top}, \ldots, \gamma_{1i}^{(p)\top}, \ldots, \gamma_{ni}^{(p)\top})^\top$.
To be consistent with the sparsity structure considered here, $\bm{\gamma}_i$ satisfies
\begin{align}\label{eq:restrict}
     \left\|  \tilde V_l^{(j)} \bm B_H(\cdot) \bm{\gamma}_i \right\|_2 = 0 
     \;\; \text{for all $(l,j) \in \mathcal S_{0i}$}.
\end{align}
Then, the SQVAR model is approximated as follows:
\begin{align}
    y_{it} 
     = C_t^\top \bm{\Phi}_i(U_{it}) \approx C_t^\top \bm B_H(U_{it}) \bm{\gamma}_i 
     = \xi_t(U_{it})^\top \bm{\gamma}_i,
\end{align}
where $\xi_t(U_{it}) \coloneqq C_t \otimes  \bm{b}_H(U_{it})$.

To estimate $\bm{\gamma}_i$, we perform a penalized monotone quantile regression of $y_{it}$ on $\xi_t(\cdot)$.  
As for the choice of the penalty function, we adopt the SCAD penalty (\citealp{fan2001variable}) due to its ability to automatically select relevant variables while producing asymptotically unbiased estimates.  
Based on \eqref{eq:restrict}, we consider the following penalty structure:
\begin{align}
    P_\lambda(\bm{\gamma}) = \sum_{(l, j) \in \mathcal S} s_\lambda\left( \left\| \tilde V_l^{(j)} \bm B_H(\cdot) \bm{\gamma} \right\|_2\right),
\end{align}
where $s_\lambda$ is the SCAD penalty function defined as
\begin{align}
    s_\lambda(x) = \begin{cases}
    \lambda x & \text{if } 0 \le x \le \lambda \\
    -\frac{x^2 - 2a\lambda x + \lambda^2}{2(a - 1)} & \text{if } \lambda < x \le a \lambda \\
    \frac{(a + 1)\lambda^2}{2} & \text{if } x > a \lambda.
    \end{cases}
\end{align}
Here, $\lambda$ is a penalty parameter satisfying $\lambda \to 0$ as $T \to \infty$, and $a$ is a pre-specified constant, which is typically set as  $a = 3.7$.

When one uses the coordinate system \eqref{eq:max-min} in Example \ref{ex:max-min}, the restriction \eqref{eq:restrict} is greatly simplified as $\gamma^{(j)}_{li} = \gamma_{0i}$ for all $(l, j) \in \mathcal S_{0i}$.
Accordingly, the penalty term takes the following form in this case:
\begin{align}
    P_\lambda(\bm{\gamma}) = \sum_{(l, j) \in \mathcal S}   s_\lambda \left( \left\| \bm{b}_H(\cdot)^\top\{\gamma_l^{(j)} - \gamma_0 \} \right\|_2 \right),
\end{align} 
which is conceptually similar to the group-SCAD penalty in \cite{wang2007group, wang2008variable}.

The choice of $\lambda$ governs both the sparsity pattern and the estimation quality.  
If $\lambda$ is too small, irrelevant lag effects may remain in the model, leading to higher variance and potential overfitting.  
If $\lambda$ is too large, relevant lag effects may be excluded, resulting in larger bias and reduced explanatory power.  
Thus, $\lambda$ must be carefully chosen in a data-adaptive manner to deal with the bias-variance tradeoff.  
Below, we propose a BIC-type method for selecting the optimal value of $\lambda$.

Let $\{\tau_1, \ldots, \tau_L\}$ be a set of equally spaced grid points in $(0,1)$: $\tau_\ell = \ell/(1 + L)$, where $L \to \infty$ as $T \to \infty$.
Then, we define our SCAD-penalized SQVAR estimator as
\begin{align}
\label{PenQloss}
    \hat{\bm{\gamma}}_i(\lambda) = \argmin_{\bm{\gamma} \in \mathcal{G}_H} \frac{1}{LT} \sum_{\ell \in [L]} \sum_{t \in [T]} \rho_{\tau_\ell}\left( y_{it} - \xi_t(\tau_\ell)^\top \bm{\gamma}\right) + P_\lambda(\bm{\gamma}),
\end{align}
where $\mathcal{G}_H = \prod_{j = 1}^{N + 1}\mathcal{G}_{j,H}$ with $\mathcal{G}_{j,H}$ denoting the $j$-th parameter space.
The choice of parameter space varies with the basis functions employed.  
For any basis functions, we assume $|\bm{b}_H(u)^\top \gamma| < \infty$ uniformly in $u \in [0,1]$ for any $\gamma \in \mathcal{G}_{j,H}$, uniformly in $H$.  
Additionally, for I-splines, each $\mathcal{G}_{j,H}$ must be a subset of $\{\gamma = (\gamma_1, \ldots, \gamma_H) \in \mathbb{R}^H : \gamma_h \ge 0, \; h \in \{2,3, \ldots, H\}\}$, where the first component of $\gamma$ corresponds to the "location" of the function.
Similarly, for monotone B-splines, each $\mathcal{G}_{j,H}$ is a subset of $\{\gamma = (\gamma_1, \ldots, \gamma_H) \in \mathbb{R}^H : \gamma_2 \le \cdots \le \gamma_H\}$.
The dependence of $\hat{\bm{\gamma}}_i(\lambda)$ on $\lambda$ is omitted when there is no confusion, and we simply write $\hat{\bm{\gamma}}_i$.

Once $\hat{\bm \gamma}_i$ is obtained, we can estimate $\bm \Phi_i(\tau)$ by $\hat{\bm \Phi}_i(\tau) \coloneqq \bm B_H(\tau) \hat{\bm \gamma}_i$.  
Then, the original QVAR coefficients can be recovered by  
$\hat{\bm \Theta}_i(\tau) \coloneqq \bm V^{-1} \hat{\bm \Phi}_i(\tau)$. 
Moreover, the set of inactive coefficients $\mathcal{S}_{0i}$ is estimated as 
\begin{align}
    \hat{\mathcal{S}}_{0i} \coloneqq \left\{ (l, j) \in \mathcal S : \left\| \hat \theta_{li}^{(j)} \right\|_2 = 0 \right\}.
\end{align}
In particular, in the case of coordinate system \eqref{eq:max-min}, the estimator of $\mathcal{S}_{0i}$ is given by
\begin{align}
\hat{\mathcal{S}}_{0i} \coloneqq \left\{ (l, j) \in \mathcal S : \left\| \hat \phi_{li}^{(j)} - \hat \phi_{0i} \right\|_2 = 0 \right\},
\end{align}
and the original QVAR coefficients are estimated by
\begin{align}
    \hat \theta_{li}^{(j)}(\tau) 
    \coloneqq \frac{\hat \phi_{li}^{(j)}(\tau) - \hat \phi_{0i}(\tau)}{N \Delta_l}, \qquad
    \hat \theta_{0i}(\tau)
    \coloneqq \hat \phi_{0i}(\tau) - \sum_{(l,j) \in \hat{\mathcal{S}}_{1i}} \text{lb}_l \hat \theta_{li}^{(j)}(\tau),
\end{align}
where $\hat{\mathcal{S}}_{1i} \coloneqq \left\{ (l, j) \in \mathcal S : \left\| \hat \theta_{li}^{(j)} \right\|_2 \neq 0 \right\}$.

\subsection{Regularization parameter selection}

In the analysis of VAR models, determining the lagged variables to include in the system is always a central concern.  
In our SCAD penalization procedure, the choice of lags is controlled by $\lambda$.  
Considering the consistency of BIC in selecting the true model in the context of quantile regression (e.g., \citealp{lian2012note, lee2014model}), we employ the following BIC-type criterion:
\begin{align}\label{eq:BIC}
    {\rm BIC}(\lambda) \coloneqq \ln\left( \frac{1}{LT} \sum_{\ell \in [L]} \sum_{t \in [T]} \rho_{\tau_\ell}\left( y_{it} - \xi_t(\tau_\ell)^\top \hat{ \bm \gamma}_i(\lambda) \right) \right) + \frac{\hat s_{1i}(\lambda) H \ln T}{2 T},
\end{align}
where $\hat s_{1i}(\lambda)$ denotes the cardinality of $\hat{\mathcal S}_{1i}$ under $\lambda$. 
Then, the optimal $\lambda$ can be obtained by  minimizing ${\rm BIC}(\lambda)$.

\bigskip

We summarize our SQVAR estimation procedure as follows:

\begin{algorithm}[H]
\caption{SCAD-penalized SQVAR estimation}
\begin{algorithmic}[1]
\item[] \textbf{Input:} Observed data $\{Y_t\}_{t=1}^T$; tuning parameters $H$, $L$, $(a = 3.7)$.
\STATE Find $\lambda$ that minimizes $\rm BIC(\lambda)$.
\STATE Obtain the SCAD-penalized coefficient estimator:
\begin{align}
    \hat{\bm \gamma}_i = \argmin_{\bm{\gamma} \in \mathcal{G}_H} \frac{1}{LT}\sum_{\ell \in [L]} \sum_{t \in [T]} \rho_{\tau_\ell} \left( y_{it} - \xi_t(\tau_\ell)^\top \bm{\gamma} \right) + P_\lambda(\bm{\gamma}).
\end{align}
\STATE Compute the SQVAR coefficients: 
$\hat{\bm \Phi}_i(\tau) \coloneqq \bm B_H(\tau)\hat{\bm \gamma}_i$.
\STATE Compute the QVAR coefficients: 
$\hat{\bm \Theta}_i(\tau) \coloneqq \bm V^{-1} \hat{\bm \Phi}_i(\tau)$.
\STATE Estimate the set of inactive coefficients: 
$\hat{\mathcal{S}}_{0i} \coloneqq \left\{ (l,j) \in \mathcal S : \left\|\hat \theta_{li}^{(j)}\right\|_2 = 0 \right\}$.
\item[] \textbf{Output:} Estimated QVAR coefficient functions 
$\{\hat \theta_{0i}(\tau), \hat \theta_{li}^{(j)}(\tau)\}$; estimated inactive set $\hat{\mathcal{S}}_{0i}$.
\end{algorithmic}
\end{algorithm}

\begin{remark}[Preliminary variable screening]
    An important limitation of the simplex embedding is that, as the dimension of the time series $n$ becomes large, the barycentric coordinates collapse in the limit.
    Thus, it is desirable to roughly screen relevant variables prior to applying the SQVAR method.
    In the literature, \citet{FanLv2008} proposed the sure independence screening (SIS) methodology for linear regression, which effectively reduces dimensionality by screening variables based on their marginal utilities.
    Building on this idea, \citet{HeWangHong2013} developed a quantile-adaptive framework that extends SIS to nonlinear variable screening in high-dimensional heterogeneous data settings.
    We can further extend their screening procedure to accommodate high-dimensional time series data so that it can be used for our purposes.
    For details about this method, see Appendix \ref{app:screen}.
\end{remark}

\begin{remark}[Estimation of the joint innovation distribution]\label{rem:copula}

When conducting prediction or impulse response analysis based on the estimated QVAR model, it is necessary to recover the joint distribution of $\bm U_t$.  
If the dimension $n$ is small, a nonparametric approach may be feasible.  
However, for a large or moderate $n$, any nonparametric method suffers from the curse of dimensionality.  
Thus, we suggest a parametric approach and  assume that the joint distribution is characterized by a known copula function $\Pi_\kappa$ with correlation parameter(s) $\kappa \in \mathcal{K}$. 
The parameter space $\mathcal{K}$ depends on the choice of copula.  
To estimate the copula parameter, we first need to recover the rank of each observation, $U_{it}$, by inverting the estimated quantile function:
\begin{align}
      \hat{U}_{it} \coloneqq \hat Q_{y_{it}}^{-1}(y_{it} \mid \mathcal{F}_{t-1}),
\end{align}
where $\hat Q_{y_{it}}(\tau \mid \mathcal{F}_{t-1}) \coloneqq \xi_t(\tau)^\top \hat{\bm{\gamma}}_i$. 
Note that the inverse function is well-defined because of the strict monotonicity of $\hat Q_{y_{it}}(\tau \mid \mathcal{F}_{t-1})$ under the proposed approach.
Once $\{(\hat{U}_{1t}, \ldots, \hat{U}_{nt})\}_{t=1}^T$ are obtained, the copula parameter $\kappa$ is estimated by the maximum likelihood:
\begin{align}
  \hat \kappa = \argmax_{\kappa \in \mathcal{K}} \sum_{t=1}^{T} \ln f_\kappa(\hat U_{1t}, \ldots, \hat U_{nt}),
\end{align}
where $f_\kappa$ denotes the copula density.
Studying the asymptotic properties of $\hat \kappa$ is beyond the scope of this paper.
\end{remark}

\section{Asymptotic Theory}\label{sec:asymp}

\subsection{Asymptotic properties of the penalized SQVAR estimator} 

In this section, we first study the convergence rate and the limiting distribution of our SQVAR estimator.  
For expositional simplicity, the basic model assumptions introduced in Section \ref{sec:qvar} (Assumptions \ref{as:U}-\ref{as:MA}) are maintained implicitly throughout the following discussion.  
In addition, we introduce the following assumptions.

\begin{assumption}[QVAR coefficients]\label{as:coef}
    (i) $\theta_{0i}$ and $\theta_{li}^{(j)}$ are continuous on $[0,1]$ for all $(l,j) \in \mathcal S$.
    (ii) $\bm V$ is nonsingular and $||\bm V^{-1} ||_\infty \le c_{\bm V}$, where 
    $\bm V$ is as given in Lemma \ref{lem:nc-qvar}. 
\end{assumption}

\begin{assumption}[Conditional distribution]\label{as:dist}
    (i) The conditional CDF of $y_{it}$ given $\mathcal{F}_{t - 1}$, $F_{it}(\cdot \mid \mathcal{F}_{t - 1})$, has a bounded Lipschitz continuous density $f_{it}(\cdot \mid \mathcal{F}_{t - 1})$.
    (ii) $f_{it}(y \mid \mathcal{F}_{t-1}) \ge \underbar{c}_f > 0$ uniformly in $y$ in the neighborhood of $C_t^\top \bm{\Phi}_i(\tau)$ for all $\tau \in (0,1)$.
\end{assumption}

\begin{assumption}[Basis function]\label{as:basis}
    (i) For all $h \ge 1$, $b_h(u)$ is continuous in $u \in [0,1]$.
    (ii) For all $H$, there exists $\bm \gamma_i$ and $\pi > 0$ such that $\left\|\phi_{0i} - \bm{b}_H(\cdot)^\top \gamma_{0i}\right\|_\infty = O(H^{-\pi})$ and $\left\|\phi_{li}^{(j)} - \bm{b}_H(\cdot)^\top \gamma^{(j)}_{li}\right\|_\infty = O(H^{-\pi})$ for all $(l,j) \in \mathcal S$, where $\bm \gamma_i$ satisfies \eqref{eq:restrict} and lies away from the boundary of $\mathcal G_H$.
    (iii) $\bar{\text{eig}}\left( \int_0^1 \bm{b}_H(u) \bm{b}_H(u)^\top \text{d}u \right) \le \bar c_b$, and $\underbar{\text{eig}}\left( \int_0^1 \bm{b}_H(u) \bm{b}_H(u)^\top \text{d}u \right) \ge \underbar{c}_b > 0$.
\end{assumption}

\begin{assumption}\label{as:matrix}
For all sufficiently large $T$,
\small\begin{align}
    \text{(i)} \;\;
    & \bar{\text{eig}}\Bigl((L T)^{-1} \sum_{\ell \in [L]} \sum_{t \in [T]} \mathbb{E}\left[ \xi_t(\tau_\ell) \xi_{t}(\tau_{\ell})^\top  \right] \Bigr) \le \bar c_1, \; \underbar{\text{eig}} \Bigl((L T)^{-1} \sum_{\ell \in [L]} \sum_{t \in [T]} \mathbb{E}\left[ \xi_t(\tau_\ell) \xi_{t}(\tau_{\ell})^\top  \right] \Bigr) \ge \underbar{c}_1 > 0. \\
    \text{(ii)} \;\; 
    & \bar{\text{eig}}\Bigl((L^2 T)^{-1} \sum_{\ell, \ell' \in [L]} \sum_{t \in [T]} \mathbb{E}\left[ \xi_t(\tau_\ell) \xi_{t}(\tau_{\ell'})^\top  \right] \min\{\tau_\ell, \tau_{\ell'}\}(1 - \max\{\tau_\ell, \tau_{\ell'}\})\Bigr) \le \bar c_2, \\ 
    & \underbar{\text{eig}}\Bigl( (L^2 T)^{-1} \sum_{\ell, \ell' \in [L]} \sum_{t \in [T]} \mathbb{E}\left[ \xi_t(\tau_\ell) \xi_{t}(\tau_{\ell'})^\top  \right] \min\{\tau_\ell, \tau_{\ell'}\}(1 - \max\{\tau_\ell, \tau_{\ell'}\})\Bigr)\ge \underbar{c}_2 > 0. \\
    \text{(iii)} \;\;
    & \bar{\text{eig}} \Bigl( (L T)^{-1} \sum_{\ell \in [L]} \sum_{t \in [T]} \mathbb{E}\left[ f_{it}(C_t^\top \bm{\Phi}_i(\tau_\ell)\mid \mathcal{F}_{t-1})   \xi_t(\tau_\ell) \xi_t(\tau_\ell)^\top \right] \Bigr) \le \bar c_3, \\
    & \underbar{\text{eig}} \Bigl( (L T)^{-1} \sum_{\ell \in [L]} \sum_{t \in [T]} \mathbb{E}\left[ f_{it}(C_t^\top \bm{\Phi}_i(\tau_\ell)\mid \mathcal{F}_{t-1}) \xi_t(\tau_\ell) \xi_t(\tau_\ell)^\top \right] \Bigr) \ge \underbar{c}_3 > 0.
\end{align}\normalsize
\end{assumption}

\begin{assumption}[Tuning parameters]\label{as:tune}
As $T \to \infty$, 
(i) $\lambda \to 0$ such that $\frac{\lambda}{T^{-1/2} + H^{1/2 - \pi}} \to \infty$, and
(ii) $H \to \infty$ such that $T^{1/2} H^{1/2 - \pi} = O(1)$ and $H/\sqrt{T} \to 0$, where $\pi$ is as given in Assumption \ref{as:basis}(ii). 
\end{assumption}

Assumption \ref{as:coef}(i) ensures the boundedness of the coefficient functions, and hence the conditional quantile function as well over $[0,1]$ by Assumption \ref{as:SMER}(i).  
In view of Assumption \ref{as:basis}(ii), we eventually require a specific smoothness condition, not mere continuity (e.g., a H\"{o}lder class of appropriate order).  
However, for simplicity and generality, we do not impose such a condition here.  
Assumption \ref{as:coef}(ii) is a weak technical requirement.

Assumption \ref{as:dist} should be standard in the literature on quantile regression.  
Assumption \ref{as:basis}(i) implies the boundedness of the basis functions, which yields $\sup_{u \in [0,1]} ||\bm{b}_H(u)|| = O(\sqrt{H})$. 
The constant $\pi$ in condition \ref{as:basis}(ii) represents the smoothness of the functions $\theta_{0i}$ and $\theta_{li}^{(j)}$.  
Since there is a rich discussion and guidance on this topic in the literature, we omit the details here (see, for example, \cite{CHEN20075549}).
Condition \ref{as:basis}(iii) is standard.

Assumption \ref{as:matrix} is a collection of conditions requiring the nonsingularity of some matrices.  
Note that the dimensions of these matrices increase with $H$.  
Hence, Assumption \ref{as:matrix} implicitly requires $L$ to increase to infinity along with $H$. 
Assumption \ref{as:tune} controls the rates of the tuning parameters $\lambda$ and $H$.  

Under the above assumptions, we obtain the following convergence results.

\begin{theorem}[Rate of convergence]\label{thm:convrate}
Suppose that Assumptions \ref{as:coef} - \ref{as:tune} hold.
Then, we have
\begin{align}
    \text{(i)} & \quad \left\| \hat \theta_{0i} - \theta_{0i} \right\|_2 = O_P(1/\sqrt{T}), 
    \quad \text{(ii)} \quad \left\| \hat \theta_{li}^{(j)} - \theta^{(j)}_{li} \right\|_2 = O_P(1/\sqrt{T}), \\
    \text{(iii)} & \quad \left\| \hat \theta_{0i} - \theta_{0i} \right\|_\infty = O_P(\sqrt{H/T}),
    \quad \text{(iv)} \quad \left\| \hat \theta_{li}^{(j)} - \theta^{(j)}_{li} \right\|_\infty = O_P(\sqrt{H/T}).
\end{align}
\end{theorem}

A notable implication of Theorem \ref{thm:convrate} is that the coefficient functions can be estimated at the parametric rate in the $L_2$-norm.  
This result is intuitively understandable when viewing our estimator as essentially connecting pointwise QR estimates (which themselves converge at the parametric rate) across different quantile levels using splines.  
A similar finding is reported in \cite{ando2025functional}.

\bigskip

Next, we study the limiting distribution of $(\hat \theta_{0i}(\tau), \hat \theta_{li}^{(j)}(\tau))$.  
Since deriving the limiting distribution of the SQVAR estimator under a general coordinate system complicates the analysis and is less convenient for applications, in the next two theorems, we restrict our attention to the coordinate system given in \eqref{eq:max-min}.
The next theorem states that, if the penalty parameter is chosen appropriately as in Assumption \ref{as:tune}(i), the inactive set of coefficient functions can be consistently estimated.

\begin{theorem}[Consistent model selection]\label{thm:lag}
    Suppose that Assumptions \ref{as:coef} - \ref{as:tune} hold.
    Then, we have $\Pr\left( \left\| \hat \theta_{li}^{(j)} \right\|_2 = 0 \right) \to 1$ as $T \to \infty$ for any $(l,j) \in \mathcal S_{0i}$.
\end{theorem}

Furthermore, in view of Theorem \ref{thm:convrate}(ii), the result in Theorem \ref{thm:lag} also implies the consistency of $\hat{\mathcal S}_{1i}$.
Given these results, we can derive the asymptotic distribution of the active coefficient functions by restricting our attention to the event $\{\hat{\mathcal S}_{1i} = \mathcal S_{1i}\}$, since this event occurs with probability approaching one.

\begin{theorem}[Asymptotic normality of the active coefficients]\label{thm:normality}
Suppose that Assumptions \ref{as:coef} - \ref{as:tune} hold.
If  $(H^{3/2}\ln T)/\sqrt{T} \to 0$ and $T^{1/2} H^{-\pi}/ \left\| \bm b_H(\tau) \right\| \to 0$ are additionally satisfied, for a given $\tau \in (0,1)$, we have
\begin{align}
    & \text{(i)} \quad
    \frac{\sqrt{T}(\hat \theta_{0i}(\tau) - \theta_{0i}(\tau))}{\nu_{0i}(\tau)} \overset{d}{\to} N(0, 1), \\
    &\text{(ii)} \quad
    \frac{N \Delta_l\sqrt{T}(\hat \theta_{li}^{(j)}(\tau) - \theta_{li}^{(j)}(\tau))}{\nu_{li}^{(j)}(\tau)} \overset{d}{\to} N(0, 1) \;\; \text{for $(l,j) \in \mathcal S_{1i}$},
\end{align}
where the definitions of $\nu_{0i}(\tau)$ and $\nu_{li}^{(j)}(\tau)$ are presented in Appendix \ref{app:prep}.
\end{theorem}

\begin{remark}[Choice of $H$]
    Suppose that $H$ is of order $T^d$ for some constant $d > 0$.
    Then, Theorem \ref{thm:normality} implies that $d$ must satisfy $1/(2\pi - 1) \le d < 1/3$.
    This condition automatically requires that the coefficient functions are sufficiently smooth, with the smoothness parameter $\pi$ strictly greater than 2.
    In the literature, $\pi$ typically represents the smoothness order of a H\"{o}lder class (e.g., \citealp{CHEN20075549}), and in this case, the above condition can be met if the coefficient functions are smoother than twice continuously differentiable.
    In practical terms, when the sample size is moderate, setting $H = 6$ or $7$ would be a reasonable default. 
\end{remark}

\subsection{Consistency of the BIC criterion}

In this subsection, we establish the consistency of the BIC criterion \eqref{eq:BIC}.  
We continue to assume that the coordinate system \eqref{eq:max-min} is employed.  
Define the pseudo true coefficient parameter under a potentially misspecified active coefficient set $\mathcal S_1 \subseteq \mathcal S$:
\begin{align}
    \bm{\gamma}_i(\mathcal S_1)
    \coloneqq \argmin_{\bm{\gamma} \in \mathcal{G}_H} 
    \frac{1}{LT} \sum_{\ell \in [L]} \sum_{t \in [T]} 
    \mathbb{E}\left[ \rho_{\tau_\ell}\left( y_{it} - \xi_t(\tau_\ell)^\top \bm{\gamma}\right)\right] \quad \text{subject to } \gamma_l^{(j)} = \gamma_0 \text{ for } (l,j) \notin \mathcal S_1,
\end{align}
and $\epsilon_{it}(\tau, \mathcal S_1) \coloneqq y_{it} - \xi_t(\tau)^\top \bm{\gamma}_i(\mathcal S_1)$.  
Furthermore, let 
\begin{align}\label{eq:BIClambda}
    \hat \lambda \coloneqq \argmin_{\lambda \ge 0} {\rm BIC}(\lambda),
\end{align}
and denote the estimator of the true active coefficient set $\mathcal S_{1i}$ under $\hat \lambda$ by $\hat{\mathcal S}_{1i}(\hat \lambda)$.

Note that overfitting only results in a loss of estimation efficiency and is therefore less problematic than underfitting.
To study the behavior of the estimator under underfitted misspecified models, we introduce the following additional assumptions.

\begin{assumption}[Underfitted models]\label{as:BIC}
    For any $\mathcal S_{1i} \nsubseteq \mathcal S_1$, (i) for all $\tau \in (0,1)$, the conditional CDF of $\epsilon_{it}(\tau, \mathcal S_1)$ given $\mathcal{F}_{t - 1}$, $G_{\mathcal S_1, it}(\cdot \mid \mathcal{F}_{t - 1})$, has a bounded continuous density $g_{\mathcal S_1, it}(\cdot \mid \mathcal{F}_{t - 1})$, and
    (ii) $\underbar{\text{eig}} \Bigl( (L T)^{-1} \sum_{\ell \in [L]} \sum_{t \in [T]} \mathbb{E}\left[ g_{\mathcal S_1, it}(0 \mid \mathcal{F}_{t-1}) \xi_t(\tau_\ell, \mathcal S_1) \xi_t(\tau_\ell, \mathcal S_1)^\top \right] \Bigr) \ge \underbar{c}_4 > 0$, where $\xi_t(\tau, \mathcal S_1) = (c_{0t} + \sum_{(l,j) \notin \mathcal S_1} c_{l,t-j}, \{c_{l,t-j}\}_{(l,j) \in \mathcal S_1} )^\top \otimes \bm b_H(\tau)$.
    (iii) There exists a constant $c_\gamma > 0$ such that, for all $(l,j) \in \mathcal S_{1i}$, $||\gamma_{li}^{(j)} - \gamma_{0i}|| \ge c_\gamma$.
\end{assumption}

Assumptions \ref{as:BIC}(i) and (ii) are parallel with Assumptions \ref{as:dist}(i) and \ref{as:matrix}(iii), respectively. 
Assumption \ref{as:BIC}(iii) implies that, if $\mathcal S_{1i}\nsubseteq\mathcal S_1$, then there exists a constant $c > 0$ such that $||\bm \gamma_i(\mathcal S_1) - \bm \gamma_i|| \ge c$ uniformly over $\mathcal S_1$.
This is essentially the same as the "beta-min" condition in the model selection literature.
They are used to derive the convergence rate of the underfitted QR estimator to its pseudo-true value.

\begin{theorem}[Consistency of the BIC criterion]\label{thm:BIC}
    Suppose that Assumptions \ref{as:coef} - \ref{as:matrix}, \ref{as:tune}(ii), and \ref{as:BIC} hold.
    Then, we have $\Pr\left( \hat{\mathcal S}_{1i}(\hat \lambda) = \mathcal S_{1i}\right) \to 1$ as $T \to \infty$.
\end{theorem}

Theorem \ref{thm:BIC} shows that the SCAD-penalized estimator consistently selects the active lag set when the penalty parameter is chosen according to \eqref{eq:BIClambda}.
Note that directly solving the minimization problem in \eqref{eq:BIClambda} is computationally intensive.  
Thus, in practice, we suggest using a grid search.  

\section{Impulse Response Analysis}\label{sec:irf}

In this section, we study impulse response analysis based on the QVAR model.  
Impulse responses measure how the system of time series reacts to exogenous shocks to a particular series.
In the context of QVAR, we consider the shock as a shift in the rank variable $U$. 

We propose two types of impulse response analysis.  
The first type is the generalized impulse response analysis, which can be viewed as a QVAR counterpart to \cite{pesaran1998generalized}.  
In this approach, the response to a shock is measured in terms of expected outcomes.  
While this type of analysis is standard in conventional VAR models, it averages out the dependence at different quantile levels.
On the other hand, there are situations in which a particular quantile is of interest. 
For instance, when assessing Value-at-Risk (VaR) dynamics, it is essential to focus on 
the tail behavior.  
To address this need, we propose a second type of impulse response analysis: a scenario-based quantile impulse response approach.


\subsection{Generalized impulse response analysis}

We define the generalized impulse response function as follows:
\begin{align}
    \text{IRF}_{ij}(h,\tau^\star)
    \coloneqq \mathbb{E}[y_{i,t + h} \mid U_{jt} = \tau^\star, \mathcal{F}_{t-1}] - \mathbb{E}[y_{i,t + h} \mid \mathcal{F}_{t-1}].
\end{align}
That is, $\text{IRF}_{ij}(h,\tau^\star)$ represents the change in the $h$-step ahead expected outcome $y_{i,t+h}$ when an exogenous shock that shifts the rank variable $U_{jt}$ to a specific quantile level $\tau^\star$ occurs, holding all other information at time $t-1$ fixed.  
This definition generalizes the conventional impulse response function by characterizing the magnitude of the shock through a shift in the rank variable, rather than through the size of the innovation term.

The computation of $\text{IRF}_{ij}(h,\tau^\star)$ involves Monte Carlo simulation.
Specifically, assume that the joint distribution of $\bm U_t$ is represented by a copula function $\Pi_{\kappa}$, as in Remark \ref{rem:copula}.
Then, $\mathbb{E}[y_{i,t + h} \mid U_{jt} = \tau^\star, \mathcal{F}_{t-1}]$ can be estimated by simulating future paths of the rank variables $\bm U_t, \bm U_{t+1}, \ldots$ drawn from the estimated conditional copula $\Pi_{\hat \kappa \mid U_{jt} = \tau^\star}$, and averaging the resulting simulated future outcomes.
Similarly, $\mathbb{E}[y_{i,t + h} \mid \mathcal{F}_{t-1}]$ can be estimated by simulating future paths of $\bm U_t, \bm U_{t+1}, \ldots$ drawn from the unconditional copula $\Pi_{\hat \kappa}$.

\subsection{Scenario-based analysis}

\cite{ando2024scenario} recently proposed a scenario-based quantile network connectedness framework to study how liquidity shocks propagate across U.S. large bank holding companies, particularly under stress scenarios like the COVID-19 pandemic. 
Based on their idea, we can conduct a scenario-based impulse response analysis in our context. 

Consider a rank trajectory $\bm \tau_{ih} \coloneqq \{\tau_{i0}, \tau_{i1}, \ldots, \tau_{ih}\}$ for time series $i$ of length $h + 1$.
We refer to this $\bm \tau_{ih}$ as the ``scenario" of interest for the $i$-th series.
The scenario-based impulse response analysis then investigates how future forecasts change when the rank variable evolves according to the specified scenario.

The choice of scenario can be arbitrary.
For example, if we are interested in particular historical events that occurred at time $t$ for time series $i$, we first estimate the actual quantile levels $\{U_{it}, U_{i,t+1}, \ldots, U_{i,t+h}\}$ from our estimated model.
Then, we can use the estimated sequence $\{\hat U_{it}, \hat U_{i,t+1}, \ldots, \hat U_{i,t+h}\}$ as $\bm \tau_{ih}$.
As an alternative to focusing on real historical scenarios, one may consider a hypothetical scenario $\bm \tau_{ih}$ based on empirical interest.
In this case, it is possible to directly incorporate the assumptions or beliefs of researchers into the scenario.

The scenario-based impulse response analysis can be implemented in the following procedure:
\begin{description}
    \item[Step 1:] Choose a scenario $\bm \tau_{ih}$ for each time series $i \in [n]$.
    \item[Step 2:] Compute $y_{it}^*(\bm \tau_{ih}) = \xi_t(\tau_{i0})^\top \hat{\bm{\gamma}}_i$ for all $i \in [n]$.
    \item[Step 3:] Compute $y_{i,t+1}^*(\bm \tau_{ih}) = \xi^*_{t+1}(\tau_{i1})^\top \hat{\bm{\gamma}}_i$ for all $i \in [n]$, where $\xi^*_{t+1}$ is constructed by replacing $\{y_{it}\}_{i \in [n]}$ in $\xi_{t+1}$ with $\{y_{it}^*(\bm \tau_{ih})\}_{i \in [n]}$.
    \item[Step 4:] Repeat Step 3 until $\{y_{i,t+h}^*(\bm \tau_{ih})\}_{i \in[n]}$ are obtained.
\end{description}
When there is a reference scenario $\bar{\bm \tau}_{ih}$ to measure the impact of a scenario of interest $\bm \tau_{ih}$, we define the impulse response function in this context as
\begin{align}
    \text{IRF}_i(h,\bm \tau_{ih}, \bar{\bm \tau}_{ih})
    \coloneqq y_{i,t+h}^*(\bm \tau_{ih}) - y_{i,t+h}^*(\bar{\bm \tau}_{ih}).
\end{align}
That is, $\text{IRF}_i(h,\bm \tau_{ih}, \bar{\bm \tau}_{ih})$ captures the deviation of the forecasted outcome under the scenario of interest from the reference scenario.  
Note that, in this scenario-based analysis, it is not necessary to recover the joint distribution of $\bm U_t$.

\section{Numerical Results}\label{sec:MC}

\subsection{Simulation study 1: Comparison with the standard QR method}

In this subsection, we numerically compare the performance of the proposed penalized SQVAR method and the standard QR method.
Since asymptotically the standard QR also produces quantile curves without crossing, both estimators should perform similarly when the sample size is sufficiently large.
However, in small samples, the monotonicity constraint explicitly imposed on the SQVAR method may help reduce the estimation variance under weak identification.

\begin{figure}[ht]
  \begin{subfigure}[b]{0.48\textwidth}
    \centering
    \includegraphics[width=\textwidth]{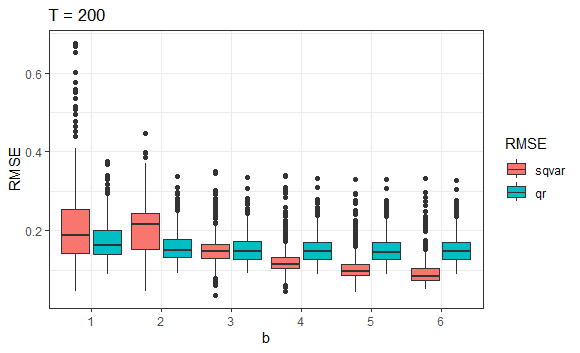}
    \caption{Red: SQVAR, Blue: QR ($T=200$)}
  \end{subfigure}
  \hfill
  \begin{subfigure}[b]{0.48\textwidth}
    \centering
    \includegraphics[width=\textwidth]{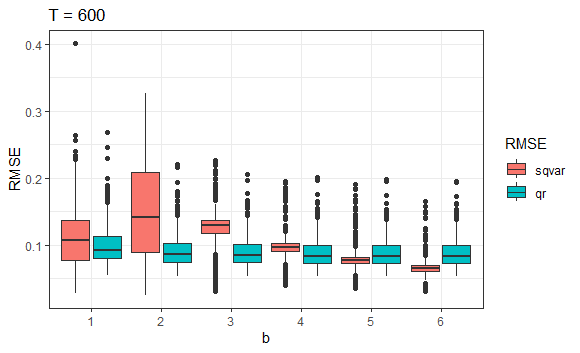}
    \caption{Red: SQVAR, Blue: QR ($T=600$)}
  \end{subfigure}
  \caption{Boxplots of RMSEs for SQVAR and standard QR methods}
  \label{fig:box}
\end{figure}

For the DGP, we consider a trivariate QVAR model with lag order $p=2$, where the first-order lags are all active and the second-order lags are all inactive.
The active and inactive lags are selected using the BIC criterion with grid search.
The strength of model identification is controlled by scaling the QVAR coefficients $\theta_{li}^{(1)}$ by $1/b$, where $b$ is chosen from $b \in \{1,2, \ldots, 6\}$; the larger $b$ is, the weaker the identification.
In Figure \ref{fig:box}, panel (a) reports the root mean squared error (RMSE) averaged over the coefficients for the two estimators when $T = 200$, and panel (b) reports those for $T = 600$, where the number of Monte Carlo replications is set to 500.
For more detailed information about the DGP and estimation procedure, see Appendix \ref{app:MC}.

From Figure \ref{fig:box}, we find that the standard QR estimator tends to perform slightly better than our penalized SQVAR estimator when $b$ is small.
It is not surprising in general that an unconstrained estimator achieves smaller RMSE than a constrained estimator.
However, as $b$ gets larger, this relationship reverses.
In particular, when the sample size is smaller, the robustness of our shape-restricted method to weak identification becomes more apparent.
Considering the fact that the standard QR estimator suffers from the quantile crossing problem especially when the sample size is small, as reported in Table \ref{tab:MC1nx}, we may conclude that the SQVAR method should be a promising alternative to the standard QR method.

\begin{table}[ht]
\centering
\caption{Average frequency of quantile crossing of QVAR models}
\label{tab:MC1nx}
\begin{tabular}{rccc}
\toprule
$b$ & $T = 200$ & $T = 600$ & $T = 1200$\\
\midrule
1 & 23.79 & 12.54 & 6.17\\
2 & 24.22 & 12.73 & 6.04\\
3 & 24.38 & 12.74 & 6.05\\
4 & 24.38 & 12.76 & 6.05\\
5 & 24.40 & 12.74 & 6.06\\
6 & 24.43 & 12.74 & 6.07\\
\bottomrule
\end{tabular}

\footnotesize
Note: The numbers reported are the average of $\frac{1}{T}\sum_{t = 1}^T \sum_{k = 1}^{98}\bm{1}\{\hat{Q}_{y_{1t}}(k/100 \mid \mathcal{F}_{t-1}) > \hat{Q}_{y_{1t}}((k + 1)/100 \mid \mathcal{F}_{t-1}) \}$.
\end{table}

\subsection{Simulation Study 2: Performance of the SQVAR method}

In this subsection, we assess the accuracy of the penalized SQVAR estimator and the BIC lag selection.
As in the previous subsection, the experimental setup and detailed results are relegated to Appendix \ref{app:MC}; here we report only the main findings.

Overall, we find that the estimator performs satisfactorily well in terms of RMSE.
Estimation errors decrease as the sample size $T$ increases.
The number of inner knots for the spline basis appears to have minimal impact on the accuracy.
Increasing the number of quantile grid points improves the estimation accuracy slightly.

For the performance of the BIC lag selection, we find that the frequency of correctly identifying the set of active lags increases with the sample size $T$, which is consistent with Theorem \ref{thm:BIC}.
In particular, the frequency that the selected active set contains the true $\mathcal S_1$ reaches 100\% when $T$ is large.
For the number of inner knots, a more parsimonious estimator tends to perform better possibly due to its smaller variance. 
For the choice of penalty parameter $\lambda$, although it is highly dependent on the DGP, the results suggest that exploring values in the neighborhood of $\lambda = c_\lambda \ln T / \sqrt{T}$ with $c_\lambda \in [0.5, 1]$ may be a reasonable default.

\subsection{Empirical study: An application to the U.S. ETFs}\label{sec:empir}

We apply the proposed model and method to daily returns of six exchange-traded funds (ETFs) traded in the U.S.
The six ETFs are the iShares 20+ Year Treasury Bond ETF (TLT), Energy Select Sector SPDR Fund (XLE), SPDR S\&P 500 ETF Trust (SPY), iShares Russell 2000 ETF (IWM), iShares MSCI EAFE ETF (EFA), and iShares U.S. Real Estate ETF (IYR).
These ETFs cover major segments of financial markets.
The data are obtained from Yahoo Finance using adjusted closing prices.

For each ETF, we first compute daily log returns in percentage points.
We then adjust these returns for day-specific effects by subtracting the mean for the corresponding weekday.
The data set covers the period from July 31, 2002 to December 30, 2024 and contains 5,643 daily observations.
The time series plots of the six outcome variables are given in Figure \ref{fig:data} in Appendix \ref{app:empir}.

\subsubsection{Estimation results}

We apply the proposed SQVAR estimator with the coordinate system given in \eqref{eq:max-min}.
The estimation procedure follows the same steps as those in the previous subsections.
For the choice of tuning parameters, we set the number of inner knots to one, $L=30$, and $p=6$.
The penalty parameter $\lambda$ is selected by minimizing $\mathrm{BIC}(\lambda)$ via a grid search.

\begin{figure}[ht]
\centering
\begin{subfigure}[b]{0.48\textwidth}
\centering
\includegraphics[width = 8.5cm]{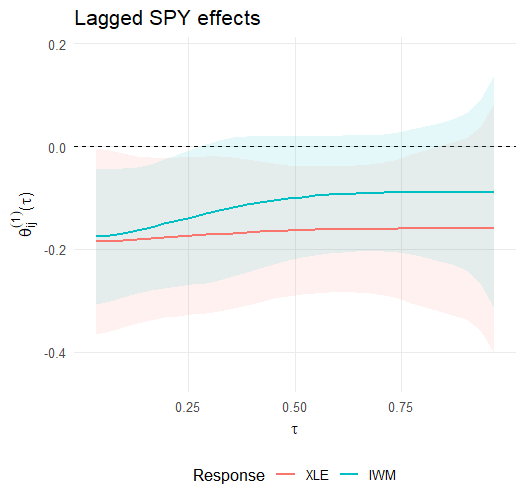}
\caption{Lagged SPY effects on XLE and IWM}
\label{fig:coef1}
\end{subfigure}
\begin{subfigure}[b]{0.48\textwidth}
\centering
\includegraphics[width = 8.5cm]{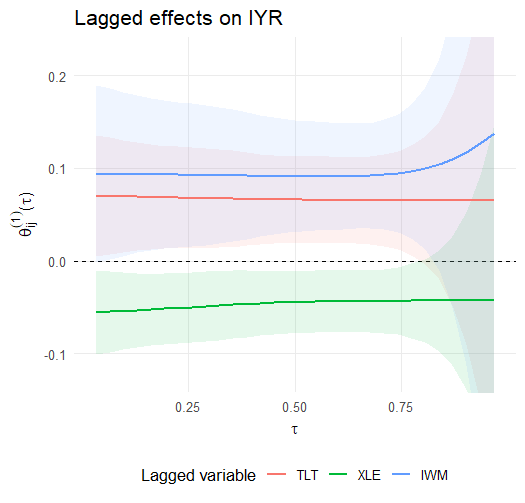}
\caption{Lagged effects on IYR}
\label{fig:coef2}
\end{subfigure}
\caption{Estimated first-order coefficient functions}
\label{fig:coef}
\end{figure}

For illustrative purposes, Figure \ref{fig:coef} reports two sets of estimated first-order coefficient functions.
The shaded areas represent pointwise 95\% confidence intervals.
In Figure \ref{fig:coef}, panel (a) shows the estimated effects of lagged SPY returns on XLE and IWM returns.
Both coefficient functions are negative over all quantile levels, indicating that a positive lagged SPY return, which proxies for broad-market movements, is associated with lower next-day returns of these ETFs.
Specifically, while the effect of SPY on XLE is relatively stable across quantiles, the effect on IWM, which represents small-cap U.S. equities, is more nonlinear and more negative in the lower quantiles.
This might suggest that the short-run effect of broad-market movements is stronger for the lower tail of the small-cap return distribution.

Panel (b) reports the estimated effects of lagged TLT, XLE, and IWM returns on IYR returns.
The lagged TLT and IWM effects are positive over most quantiles, whereas the lagged XLE effect is negative.
This contrast suggests that real estate ETF returns have different short-run dependence patterns with bond, energy, and small-cap equity returns.
The IWM effect also increases in the upper quantiles, indicating stronger positive effects on IYR in favorable return states.

\subsubsection{Scenario-based analysis}

To demonstrate our scenario-based impulse response analysis, we construct two scenarios corresponding to the following historical events:
\begin{enumerate}
\item The 2008 financial crisis (September 15, 2008)
\item The onset of the COVID-19 pandemic (March 11, 2020)
\end{enumerate}
To quantify the impact of these events, we define a baseline scenario that reflects stable market conditions.
Considering the market to be stable when the CBOE Volatility Index (VIX) is low (\url{https://fred.stlouisfed.org/series/VIXCLS}), we set December 13, 2023 as the starting date of the baseline period.
Below, we report the scenario-based SPY responses $y_{\text{SPY}, t+h}^*(\bm \tau_{ih})$ and $y_{\text{SPY}, t+h}^*(\bar{\bm \tau}_{ih})$, for $h = -5, \ldots, -1, 0 \text{ (event date) } , 1, \ldots, 22$, such that the difference between them corresponds to the average of $\text{IRF}_i(h,\bm \tau_{ih}, \bar{\bm \tau}_{ih})$.
For negative $h$, we simply report the estimated conditional quantile on that date.
The choice of $h$ = 22 corresponds roughly to the number of trading days in a month. 

\begin{figure}[ht]
\begin{center}
\includegraphics[width = 14cm]{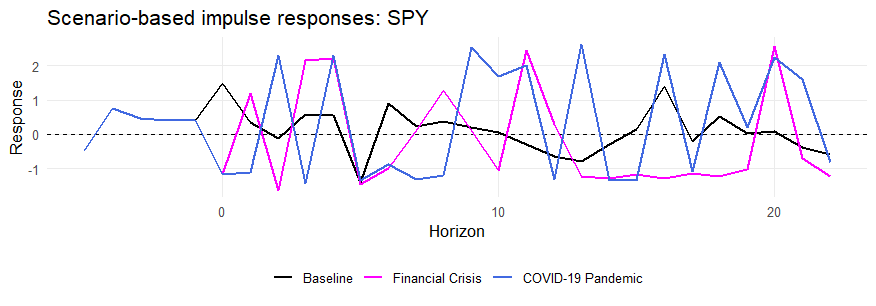}
\caption{Scenario-based impulse responses: SPY}
\label{fig:sb}
\end{center}
\end{figure}

Figure \ref{fig:sb} presents the scenario-based SPY responses under the 2008 financial crisis and the COVID-19 pandemic, each evaluated relative to the baseline scenario.
The corresponding impulse responses for the other ETFs are reported in Figure \ref{fig:sb_other} in Appendix \ref{app:empir} to save space.
Compared with the baseline path, both stress-event paths show much larger movements.
The COVID-19 scenario demonstrates repeated sharp swings between positive and negative responses throughout the forecast horizon.
The financial crisis scenario also exhibits large movements, but after the initial fluctuations, the response stays near the lower side of its range for several days.
These patterns illustrate that the scenario-based impulse response analysis can capture distinct forms of market stress dynamics.

\section{Conclusion}\label{sec:conclude}

In this paper, we introduced a simplex quantile VAR (SQVAR) framework that enforces non-crossing quantile curves in QVAR models.
For estimation, we developed a SCAD-penalized monotone series method.
We established asymptotic properties of the estimator, including the rate of convergence, asymptotic normality, and the consistency of model selection under some regularity conditions.
In addition, we derived a BIC-type criterion for choosing the SCAD penalty parameter and proved its theoretical validity.
Furthermore, we extended impulse response analysis to the proposed framework by developing two novel approaches.
An empirical application to the U.S. ETF data demonstrated that the proposed method captures heterogeneous and nonlinear dynamics across return distributions and provides useful tools for examining market stress episodes through scenario-based impulse response analysis.

There are several directions for future research.
An important open question is to consider the joint estimation of multiple time series, which could improve efficiency by explicitly accounting for cross-sectional dependence (e.g., \citealp{jun2009efficient, petrella2019joint}).
In addition, several inferential problems remain open, such as statistical inference for the impulse responses, uniform inference for the estimated coefficient functions and quantile functions, and error variance decomposition in the QVAR context.
Addressing these topics would further enhance the scope and applicability of the proposed framework.

\bigskip
\begin{center}
	{\large\bf ACKNOWLEDGMENTS}
\end{center}

The authors would like to thank the participants of the Taiwan Econometric Society Conference 2026 and the International Symposium on Econometric Theory and Applications (SETA) 2026 for their valuable comments and suggestions.
This paper has also benefited from discussions with Yacine A\"{i}t-Sahalia, Ker-Chau Li, and Takuya Ura.
Any remaining errors are solely our responsibility.
Ando's research was supported by the Australian Research Council Discovery Grant DP240101009. 
Hoshino's work was supported by the Japan Society for the Promotion of Science KAKENHI 23KK0226.

\clearpage

\appendix

\small
\begin{center}
\Large Appendix
\end{center}

Appendix \ref{app:prep} provides additional notations and preliminary mathematics used. 
Appendix \ref{app:max-min} verifies Example \ref{ex:max-min} and provides the proof of Lemma \ref{lem:MA}. 
Appendix \ref{app:proof-sqvar} and Appendix \ref{app:proof-bic} provide the proofs of Theorems \ref{thm:convrate}, \ref{thm:lag}, and \ref{thm:normality} and Theorem \ref{thm:BIC}, respectively.
Appendix \ref{app:MC} provides detailed information of the Monte Carlo experiments presented in Section \ref{sec:MC}. 
Appendix \ref{app:empir} provides supplementary figures for the empirical analysis in Section \ref{sec:empir}. 
Appendix \ref{app:screen} discusses a variable pre-screening procedure for high-dimensional time series. 

\section{Preliminary}\label{app:prep}

\paragraph{Additional notations.}

For two numbers $a$ and $b$, we write $a \lesssim b$ if and only if $a = O(b)$.
For a finite set $\mathcal B$, $|\mathcal B|$ denotes its  cardinality.
Split $\underbracket{C_t}_{(N + 1) \times 1}  \coloneqq (c_0, c_{1,t-1}, \ldots, c_{n,t-1}, \ldots, c_{1,t-p}, \ldots, c_{n,t-p})^\top$ into the active part $\underbracket{C_{1,it}}_{(|\mathcal{S}_{1i}| + 1) \times 1} \coloneqq (c_0, \{c_{l,t-j}\}_{(l,j) \in \mathcal{S}_{1i}})^\top$ and the inactive part $\underbracket{C_{2,it}}_{|\mathcal{S}_{0i}| \times 1} \coloneqq (\{c_{l,t-j}\}_{(l,j) \in \mathcal{S}_{0i}})^\top$.
Similarly, $\underbracket{\bm \gamma_i}_{H(N + 1) \times 1}$ is split into $\underbracket{\bm \beta_{1i}}_{H(|\mathcal{S}_{1i}| + 1) \times 1} \coloneqq (\gamma_{0i}, \{\gamma_{li}^{(j)}\}_{(l,j) \in \mathcal{S}_{1i}})^\top$ and $\underbracket{\bm \beta_{2i}}_{H|\mathcal{S}_{0i}| \times 1} \coloneqq (\{\gamma_{li}^{(j)}\}_{(l,j) \in \mathcal{S}_{0i}})^\top$.
With a slight abuse of notation, we often write $\bm \gamma_i = (\bm \beta_{1i}, \bm \beta_{2i})$.
Define
\begin{itemize}
    \item $m_{it}(\tau) \coloneqq C_t^\top \bm{\Phi}_i(\tau)$
    \item $\epsilon^*_{it}(\tau) \coloneqq y_{it} - m_{it}(\tau)$, $\quad \epsilon_{it}(\tau) \coloneqq y_{it} - \xi_t(\tau)^\top \bm{\gamma}_i$
    \item $v_{it}(\tau) \coloneqq \xi_t(\tau)^\top \bm{\gamma}_i - m_{it}(\tau)$
    \item $\xi_{1,it}(\tau) = C_{1,it} \otimes \bm b_H(\tau)$, $\quad \xi_{2,it}(\tau) = C_{2,it} \otimes \bm b_H(\tau)$
    \item $\bm{\Xi}_{1, i\tau} = (\xi_{1,i1}(\tau), \ldots, \xi_{1,iT}(\tau))^\top$, $\bm{\Xi}_{2, i\tau} = (\xi_{2,i1}(\tau), \ldots, \xi_{2,iT}(\tau))^\top$, $\bm{F}_{\tau} = \text{diag}\{f_{it}(m_{it}(\tau) \mid \mathcal{F}_{t-1})\}_{t = 1}^T$
    \item $\bar{\bm{\Xi}}_{1,i\tau} = \bm{\Xi}_{1, i\tau} -  \bm{\Xi}_{2, i\tau} \left(\sum_{\ell \in [L]} \bm{\Xi}_{2, i\tau_\ell}^\top \bm{F}_{\tau_\ell} \bm{\Xi}_{2, i\tau_\ell}\right)^{-1} \sum_{\ell \in [L]} \bm{\Xi}_{2, i\tau_\ell}^\top \bm{F}_{\tau_\ell} \bm{\Xi}_{1, i\tau_\ell}$
    \item $\bar \xi_{1,it}(\tau)$: $t$-th row of $\bar{\bm{\Xi}}_{1,i\tau}$
    \item $\bm{J}_{i, TL} \coloneqq \frac{1}{LT} \sum_{\ell \in [L]} \mathbb{E}\left( \bar{\bm{\Xi}}_{1,i\tau_\ell}^\top \bm{F}_{\tau_\ell} \bar{\bm{\Xi}}_{1,i\tau_\ell} \right)$
    \item $\bm V_{i, TL} \coloneqq \left( \frac{1}{L^2 T} \sum_{\ell, \ell' \in [L]} \sum_{t \in [T]} \mathbb{E}\left[ \bar \xi_{1,it}(\tau_\ell) \bar \xi_{1,it}(\tau_{\ell'})^\top  \right]\min\{\tau_\ell, \tau_{\ell'}\}(1 - \max\{\tau_\ell, \tau_{\ell'}\}) \right)$.
\end{itemize}
Note that $\sup_t ||v_{it}||_\infty \lesssim H^{-\pi}$ under Assumption \ref{as:basis}(ii).
With these notations, for a generic $\bm \gamma \in \mathcal G_H$, we can write
\begin{align}
    y_{it} - \xi_t(\tau)^\top \bm{\gamma}
    & = \epsilon_{it}^*(\tau) + C_t^\top \bm{\Phi}_i(\tau) - \xi_t(\tau)^\top \bm{\gamma} \\
    & = \epsilon_{it}^*(\tau) - \xi_{1,it}(\tau)^\top (\bm{\beta}_1 - \bm{\beta}_{1i}) - \xi_{2,it}(\tau)^\top (\bm{\beta}_2 - \bm{\beta}_{2i}) - v_{it}(\tau) \\
    & = \epsilon_{it}^*(\tau) - \bar\xi_{1,it}(\tau)^\top (\bm{\beta}_1 - \bm{\beta}_{1i}) - \xi_{2,it}(\tau)^\top \bm{\eta}(\bm{\beta}_1, \bm{\beta}_2)  - v_{it}(\tau),
\end{align}
where $\bm{\beta}_1$ and $\bm{\beta}_2$ are subvectors of $\bm \gamma$, defined analogously to $(\bm{\beta}_{1i}, \bm{\beta}_{2i})$, and
\begin{align}
    \bm{\eta}(\bm{\beta}_1, \bm{\beta}_2) = \bm{\beta}_2 - \bm{\beta}_{2i} + \left(\sum_{\ell \in [L]} \bm{\Xi}_{2, i\tau_\ell}^\top \bm{F}_{\tau_\ell} \bm{\Xi}_{2, i\tau_\ell}\right)^{-1} \sum_{\ell \in [L]} \bm{\Xi}_{2, i\tau_\ell}^\top \bm{F}_{\tau_\ell} \bm{\Xi}_{1, i\tau_\ell} (\bm{\beta}_1 - \bm{\beta}_{1i}).
\end{align}
Let $\mathbb{S}_0$ be the $H \times H(|\mathcal S_{1i}| + 1)$ dimensional matrix that selects the first $H$ elements of $\bm \beta_1$: e.g., $\mathbb{S}_0 \bm \beta_{1i} = \bm \gamma_{0i}$. 
Similarly, define $\mathbb{S}_{l}^{(j)}$ so that $\mathbb{S}_{l}^{(j)} \bm \beta_{1i} = \bm \gamma_{li}^{(j)}$ holds.
Let
\begin{align}
    [\sigma_{0i}(\tau)]^2
    & \coloneqq \bm b_H(\tau)^\top \mathbb{S}_0 \bm{J}_{i, TL}^{-1} \bm V_{i, TL} \bm{J}_{i, TL}^{-1} \mathbb{S}_0^\top \bm b_H(\tau) \\
    [\sigma_{li}^{(j)}(\tau)]^2
    & \coloneqq \bm b_H(\tau)^\top \mathbb{S}_{l}^{(j)}  \bm{J}_{i, TL}^{-1} \bm V_{i, TL} \bm{J}_{i, TL}^{-1} \mathbb{S}_{l}^{(j)\top} \bm b_H(\tau) \\
   [\nu_{0i}(\tau)]^2
   & \coloneqq \bm b_H(\tau)^\top \left( \kappa_0 \mathbb{S}_0 - \sum_{(l, j) \in \mathcal S_{1i}} \kappa_l \mathbb{S}_l^{(j)} \right) \bm{J}_{i, TL}^{-1} \bm V_{i, TL} \bm{J}_{i, TL}^{-1} \left( \kappa_0 \mathbb{S}_0 - \sum_{(l, j) \in \mathcal S_{1i}} \kappa_l \mathbb{S}_l^{(j)} \right)^\top \bm b_H(\tau)\\
   [\nu_{li}^{(j)}(\tau)]^2
   & \coloneqq \bm b_H(\tau)^\top \left(\mathbb{S}_l^{(j)} - \mathbb{S}_0\right) \bm{J}_{i, TL}^{-1} \bm V_{i, TL} \bm{J}_{i, TL}^{-1} \left(\mathbb{S}_l^{(j)} - \mathbb{S}_0\right)^\top \bm b_H(\tau),
\end{align}
where $\kappa_0 \coloneqq 1 + \sum_{(l, j) \in \mathcal S_{1i}} \frac{\text{lb}_l}{N \Delta_l}$, and $\kappa_l \coloneqq \frac{\text{lb}_l}{N \Delta_l}$.

\paragraph{Unconstrained estimator.}

To investigate the asymptotic properties of our monotone QR estimator, we define its unconstrained counterpart as follows.
\begin{align}
        \tilde{\bm{\gamma}}_i = \argmin_{\bm{\gamma} \in \mathcal{R}_H} Q_{TL}(\bm{\gamma})
\end{align}
where 
\begin{align}
        Q_{TL}(\bm{\gamma}) \coloneqq \frac{1}{LT} \sum_{\ell \in [L]} \sum_{t \in [T]} \rho_{\tau_\ell}\left( y_{it} - \xi_t(\tau_\ell)^\top \bm{\gamma}\right) + P_{\lambda}(\bm{\gamma})
\end{align}
and $\mathcal{R}_H = \prod_{j = 1}^{N + 1}\mathcal{R}_{j,H}$ is a compact parameter space.
Each $\mathcal{R}_{j,H}$ is constructed by dropping the sign and order restrictions from $\mathcal{G}_{j,H}$ (i.e., $\mathcal{G}_H \subseteq \mathcal{R}_H$).
Then, define $(\tilde \theta_{li}^{(j)}, \tilde \theta_{0i})$ in a similar manner to $(\hat \theta_{li}^{(j)}, \hat \theta_{0i})$.
Furthermore, define 
\begin{align} 
    \tilde{\bm{\alpha}}_i = \argmin_{\bm{\alpha} \in \mathcal{R}_H^\dagger} Q^\dagger_{TL}(\bm{\alpha}),
\end{align}
where 
\begin{align}
    Q^\dagger_{TL}(\bm{\alpha}) \coloneqq \frac{1}{LT} \sum_{\ell \in [L]} \sum_{t \in [T]} q_{it, \tau_\ell}(\bm{\alpha}) + P_{\lambda}(\bm{\gamma}_i + \bm{\alpha}),
\end{align}
$q_{it, \tau}(\bm{\alpha}) \coloneqq \rho_\tau( y_{it} - \xi_t(\tau)^\top \bm{\gamma}_i - \xi_t(\tau)^\top\bm{\alpha} ) - \rho_\tau( y_{it} - \xi_t(\tau)^\top \bm{\gamma}_i)$, and $\mathcal{R}_H^\dagger \coloneqq \{\bm{\alpha} : \bm{\gamma}_i + \bm{\alpha} \in \mathcal{R}_H\}$.
Observe that $\tilde{\bm{\gamma}}_i = \bm{\gamma}_i + \tilde{\bm{\alpha}}_i$ holds.

\paragraph{Knight's identity (\citealp{knight1998limiting})}
\begin{align}\label{eq:knight}
    \rho_\tau(x-y) - \rho_\tau(x) = - y \psi_\tau(x) + \int_0^y \left(\bm{1}\{x \le t\} - \bm{1}\{x \le 0\}\right) \text{d}t
\end{align}
where $\psi_\tau(x) \coloneqq \tau - \bm{1}\{x \le 0\}$.

\paragraph{Freedman's inequality (\citealp{freedman1975tail})}
Suppose that $\{X_t\}$ is a martingale difference sequence (MDS) with $\mathbb{E}[X_t \mid \mathcal{F}_{t-1}] = 0$, $|X_t| \le M$, and  $\sum_{t=1}^T \mathbb{E}[X_t^2 \mid \mathcal{F}_{t-1}] \le \sigma_T^2$.
Then, for any $\varepsilon > 0$, 
\begin{align}\label{eq:freedman}
        \Pr\left( \left| \sum_{t=1}^T X_t \right| \ge \varepsilon \right) \le 2 \exp\left( - \frac{\varepsilon^2}{2 \sigma_T^2 + \frac{2}{3} M \varepsilon} \right)
\end{align}
holds.

\section{Verification of Example \ref{ex:max-min} and Proof of Lemma \ref{lem:MA}}\label{app:max-min}

\begin{flushleft}
    \textbf{Verification of Example \ref{ex:max-min}}
\end{flushleft}

Recall that
\begin{align}
    \bm v_0 
    & = (1, \text{lb}_1, \ldots, \text{lb}_n, \ldots, \text{lb}_1, \ldots, \text{lb}_n) \\
    \bm v_l^{(j)}
    & = (1, \text{lb}_1, \ldots, \text{lb}_n, \ldots, \text{lb}_l + N \Delta_l, \ldots, \text{lb}_1, \ldots, \text{lb}_n).
\end{align}
We first verify that $W_t = c_{0t}  \bm v_0 + \sum_{(l, j) \in \mathcal S} c_{l, t - j} \bm v_l^{(j)}$ holds.
Since the first element of $\bm v_0$ and that of $\bm v_l^{(j)}$ are all one,
\begin{align}
    c_{0t} \cdot 1 + \sum_{(l, j) \in \mathcal S} c_{l, t - j} \cdot 1 = 1 \; (\text{first element of $W_t$}).
\end{align}
Similarly,
\begin{align}
    c_{0t} \cdot \text{lb}_l + \sum_{(i,j) \in [n] \times [p]} c_{i, t - j} \cdot \text{lb}_l + N \Delta_l c_{l, t - j}
    & = \text{lb}_l + N \Delta_l c_{l, t - j} \\
    & = \text{lb}_l + N \Delta_l \left( \frac{y_{l, t - j} - \text{lb}_l}{N \Delta_l} \right) = y_{l, t - j}.
\end{align}
Thus, these $c_{it}$'s are proper barycentric coordinates of $W_t$.

Next, we verify the forms of the coefficient functions $\bm \Phi_i$ and their monotonicity.
Without loss of generality, assume $n = 2$, $p = 2$, and $\mathcal{S}_{1i} = \{(1,1),(2,1)\}$.
Letting $\tilde y_{it} = y_{it} - \text{lb}_i$, we have
\begin{align}
    y_{it}
    & = \theta_{0i}(U_{it}) + \sum_{j = 1}^2\left[ y_{1,t-j} \theta^{(j)}_{1i}(U_{it}) + y_{2,t-j} \theta^{(j)}_{2i}(U_{it}) \right] \\
    & = \theta_{0i}(U_{it}) + \sum_{j = 1}^2\left[ \tilde y_{1,t-j} \theta^{(j)}_{1i}(U_{it}) + \tilde y_{2,t-j} \theta^{(j)}_{2i}(U_{it}) + \text{lb}_1 \theta^{(j)}_{1i}(U_{it}) + \text{lb}_2 \theta^{(j)}_{2i}(U_{it}) \right] \\
    & = \theta_{0i}(U_{it}) + \sum_{j = 1}^2\left[ c_{1,t-j} (N \Delta_1 \theta^{(j)}_{1i}(U_{it})) + c_{2,t-j} (N \Delta_2 \theta^{(j)}_{2i}(U_{it})) + \text{lb}_1 \theta^{(j)}_{1i}(U_{it}) + \text{lb}_2 \theta^{(j)}_{2i}(U_{it}) \right].
\end{align}
Moreover, letting $c_0 = 1 - \sum_{j=1}^2 \sum_{i=1}^2  c_{i,t-j}$ and $\phi_{0i}(U_{it}) \coloneqq \theta_{0i}(U_{it}) + \sum_{j=1}^2 \sum_{l=1}^2 \text{lb}_l \theta^{(j)}_{li}(U_{it})$,
\begin{align}
    y_{it}
    & = c_0 \phi_{0i}(U_{it}) + \sum_{j = 1}^2\left[ c_{1,t-j} (N \Delta_1 \theta^{(j)}_{1i}(U_{it})) + c_{2,t-j} (N \Delta_2 \theta^{(j)}_{2i}(U_{it}))  \right] + \sum_{j=1}^2 \sum_{i=1}^2  c_{i,t-j} \phi_{0i}(U_{it})\\
    & = c_0 \phi_{0i}(U_{it}) + \sum_{j = 1}^2\left[ c_{1,t-j} (N \Delta_1 \theta^{(j)}_{1i}(U_{it}) + \phi_{0i}(U_{it})) + c_{2,t-j} (N \Delta_2 \theta^{(j)}_{2i}(U_{it}) + \phi_{0i}(U_{it}))  \right].
\end{align}
By Assumption \ref{as:SMER}(ii), we have $\bm v_0 = (1, \text{lb}_1, \text{lb}_2, \text{lb}_1, \text{lb}_2) \in \text{SMER}_i$, ensuring that $u \mapsto \phi_{0i}(u)$ is monotonically increasing.
Moreover, since $\theta^{(2)}_{1i}(\cdot) = \theta^{(2)}_{2i}(\cdot) = 0$, we have $\phi_{0i}(U_{it}) \coloneqq \theta_{0i}(U_{it}) + \text{lb}_1 \theta^{(1)}_{1i}(U_{it}) + \text{lb}_2 \theta^{(1)}_{2i}(U_{it})$.
 
Next, define
\begin{align}
    \phi_{li}^{(j)}(U_{it}) \coloneqq N \Delta_l \theta^{(j)}_{li}(U_{it}) + \phi_{0i}(U_{it}).
\end{align}
For example when $j = 1$ and $l = 1$, 
\begin{align}
    \phi_{1i}^{(1)}(U_{it}) 
    & = N \Delta_1 \theta^{(1)}_{1i}(U_{it}) + \theta_{0i}(U_{it}) + \text{lb}_1 \theta^{(1)}_{1i}(U_{it}) + \text{lb}_2 \theta^{(1)}_{2i}(U_{it}) + \text{lb}_1 \theta^{(2)}_{1i}(U_{it}) + \text{lb}_2 \theta^{(2)}_{2i}(U_{it}) \\
    & = \theta_{0i}(U_{it}) + (\text{lb}_1 + N \Delta_1) \theta^{(1)}_{1i}(U_{it}) + \text{lb}_2 \theta^{(1)}_{2i}(U_{it}) + \text{lb}_1 \theta^{(2)}_{1i}(U_{it}) + \text{lb}_2 \theta^{(2)}_{2i}(U_{it}).
\end{align}
Again by Assumption \ref{as:SMER}(ii), $\bm v_1^{(1)} = (1, \text{lb}_1 + N \Delta_1, \text{lb}_2, \text{lb}_1, \text{lb}_2) \in \text{SMER}_i$, this implies the monotonicity of $\phi_{1i}^{(1)}$.

\qed

\begin{flushleft}
    \textbf{Proof of Lemma \ref{lem:MA}}
\end{flushleft}

It is sufficient to show that $\|\Gamma_{t,k}\|_2$ decays geometrically.
Decompose $\mathcal{A}(\bm{U}_t) = \mathcal{I} + \tilde{\mathcal{A}}_t$, where
\begin{align}
    \mathcal{I}
    \coloneqq \left(\begin{array}{cccc}
    \bm{0}_{n \times n} & \bm{0}_{n \times n} & \cdots & \bm{0}_{n \times n} \\
    I_n & \bm{0}_{n \times n} & \cdots & \bm{0}_{n \times n} \\
    \vdots & \ddots & \ddots & \vdots \\
    \bm{0}_{n \times n} & \cdots & I_n & \bm{0}_{n \times n}
    \end{array}\right), \quad 
    \tilde{\mathcal{A}}_t
    \coloneqq \left(\begin{array}{cccc}
    A_1(\bm{U}_t) & A_2(\bm{U}_t) & \cdots & A_p(\bm{U}_t) \\
    \bm{0}_{n \times n} & \bm{0}_{n \times n} & \cdots & \bm{0}_{n \times n} \\
    \vdots & \ddots & \ddots & \vdots \\
    \bm{0}_{n \times n} & \cdots & \bm{0}_{n \times n} & \bm{0}_{n \times n}
    \end{array}\right)
\end{align}
Recall that $\Gamma_{t,k} = \prod_{l = 1}^k \mathcal{A}(\bm{U}_{t - l + 1})$ for $k \ge 1$ and $\Gamma_{t,0} \coloneqq I_N$.
Noting that $\Gamma_{t,k} =  \mathcal{A}(\bm{U}_t) \prod_{l = 1}^{k-1} \mathcal{A}(\bm{U}_{t - l}) = \mathcal{A}(\bm{U}_t)\Gamma_{t-1,k-1}$ and $\mathcal{I}^p = \bm{0}$, for $k \ge p$, we have
\begin{align}
    \Gamma_{t,k} 
    & =  \mathcal{I} \Gamma_{t-1,k-1} +  \tilde{\mathcal{A}}_t \Gamma_{t-1,k-1} \\
    & =  \mathcal{I} \mathcal{A}(\bm{U}_{t-1})\Gamma_{t-2,k-2}  +  \tilde{\mathcal{A}}_t \Gamma_{t-1,k-1} \\
    & =  \mathcal{I}^2 \Gamma_{t-2,k-2}  + \mathcal{I} \tilde{\mathcal{A}}_{t-1} \Gamma_{t-2,k-2} + \tilde{\mathcal{A}}_t \Gamma_{t-1,k-1} \\
    & =  \mathcal{I}^3 \Gamma_{t-3,k-3}  + \mathcal{I}^2 \tilde{\mathcal{A}}_{t-2} \Gamma_{t-3,k-3} + \mathcal{I} \tilde{\mathcal{A}}_{t-1} \Gamma_{t-2,k-2} + \tilde{\mathcal{A}}_t \Gamma_{t-1,k-1} \\
    & \phantom{=} \vdots \\
    & = \mathcal{I}^k + \sum_{j = 1}^k \mathcal{I}^{j - 1} \tilde{\mathcal{A}}_{t-j+1} \Gamma_{t-j,k-j} \\
    & = \sum_{j \in [p]} \mathcal{I}^{j - 1} \tilde{\mathcal{A}}_{t-j+1} \Gamma_{t-j,k-j}.
\end{align}
Meanwhile, by assumption, $\left\|\tilde{ \mathcal{A}}_t \right\|_2 \le \varrho$ for all $t$.

Here, let $\text{i}_\ell \coloneqq \{\ell p, \ldots, (\ell + 1) p - 1\}$, and define $g_k \coloneqq \sup_t \left\| \Gamma_{t,k} \right\|_2$ and $\bar g_\ell \coloneqq \max_{k \in \text{i}_\ell} g_k$.
Then, for any $k \in \text{i}_{\ell + 1}$, we have 
\begin{align}
    \left\| \Gamma_{t,k} \right\|_2 
    & \le  \varrho \sum_{j \in [p]} \left\| \Gamma_{t - j, k - j} \right\|_2 \\
    & \le p \varrho \max_{k - p \le r \le k - 1} g_r \\
    & \le p \varrho \max_{r \in \text{i}_{\ell} \cup \text{i}_{\ell + 1}} g_r = p \varrho \max\{\bar g_\ell, \bar g_{\ell + 1}\},
\end{align}
where the third inequality follows from the fact that $k - p$ is in $\text{i}_{\ell}$ and $k - 1$ in either $\text{i}_{\ell + 1}$ or $\text{i}_{\ell}$.
Since the above inequality holds for any $k \in \text{i}_{\ell + 1}$, the right-hand side does not depend on $t$, and $p \varrho \in (0,1)$, we obtain 
\begin{align}\label{eq:contract0}
    \bar g_{\ell + 1} \le p \varrho \bar g_\ell 
\end{align}
Recall that $\Gamma_{t,0}$ is an identity matrix and thus $\bar g_0 \lesssim 1$.
Then, applying \eqref{eq:contract0} recursively, we obtain
\begin{align}\label{eq:contract}
    \left\| \Gamma_{t,k} \right\|_2 \lesssim (p \varrho)^{\lfloor k/p \rfloor} \to 0
\end{align}
as $k \to \infty$.
\qed

\section{Proofs of Theorems \ref{thm:convrate}, \ref{thm:lag}, and \ref{thm:normality}}\label{app:proof-sqvar}

\begin{lemma}\label{lem:NED}
    Suppose that Assumption \ref{as:coef}(i) holds.
    For all $i \in [n]$, $\{y_{it}\}$ is geometrically $L_q$-NED on $\{\bm{U}_t\}$ for $1 \le q < \infty$; that is, there exists $|\bar \varrho| < 1$ such that
\begin{align}
    \left\| y_{it} - \mathbb{E}[y_{it} \mid \mathcal{S}_{t-m}^{t+m}]\right\|_q \lesssim \bar \varrho^m,
\end{align}
where $\mathcal{S}_{t-m}^{t+m} \coloneqq \sigma\{\bm{U}_{t-m}, \bm{U}_{t-m+1}, \ldots, \bm{U}_{t+m} \}$.
\end{lemma}

\begin{proof}
Observe that
\begin{align}
    \bm{Y}_t - \mathbb{E}[ \bm{Y}_t \mid \mathcal{S}_{t-m}^{t+m}]
    & = \sum_{k = 0}^\infty \Gamma_{t,k}(\bm{U}_t, \bm{U}_{t-1}, \ldots, \bm{U}_{t-k+1}) \bm{E}(\bm{U}_{t-k})\\
    & \quad - \sum_{k = 0}^\infty \mathbb{E}[ \Gamma_{t,k}(\bm{U}_t, \bm{U}_{t-1}, \ldots, \bm{U}_{t-k+1}) \bm{E}(\bm{U}_{t-k}) \mid \mathcal{S}_{t-m}^{t+m}] \\
    & = \sum_{k = m + 1}^\infty \left( \Gamma_{t,k} \bm{E}(\bm{U}_{t-k}) - \mathbb{E}[ \Gamma_{t,k} \bm{E}(\bm{U}_{t-k}) \mid \mathcal{S}_{t-m}^{t+m}] \right) \\
    & = \sum_{k = m + 1}^\infty \Gamma_{t,k} \bm{E}(\bm{U}_{t-k}) - \sum_{k = m + 1}^\infty \mathbb{E}[ \Gamma_{t,k} \mid \mathcal{S}_{t-m}^{t+m}] \bm \mu,
\end{align}
where the last equality follows from Assumption \ref{as:U}.
We can write $y_{i,t} = \sum_{k = 0}^\infty \Gamma_{t,k}^{(i)} \bm{E}(\bm{U}_{t-k})$, where $\Gamma_{t,k}^{(i)}$ is the $i$-th row of $\Gamma_{t,k}$.
By Assumption \ref{as:MA}, there exists $\bar \varrho \in (0, 1)$ such that $|| \Gamma_{t,k}^{(i)} || \lesssim \bar \varrho^k$.
In addition, by Jensen's inequality, $|| \mathbb{E}[\Gamma_{t,k}^{(i)} \mid \mathcal{S}_{t-m}^{t+m}]|| \lesssim \bar \varrho^k$.
Hence, since $\bm{E}(\bm{U}_{t-k})$ and $\bm \mu$ are both bounded under Assumption \ref{as:coef}(i),
\begin{align}
    \left\| y_{it} - \mathbb{E}[y_{it} \mid \mathcal{S}_{t-m}^{t+m}]\right\|_q
    & \le \left\| \sum_{k = m+1}^\infty \Gamma_{t,k}^{(i)} \bm{E}(\bm{U}_{t-k}) \right\|_q + \left\| \sum_{k = m+1}^\infty \mathbb{E}[ \Gamma_{t,k} \mid \mathcal{S}_{t-m}^{t+m}] \bm \mu \right\|_q \\
    & \le \sum_{k = m+1}^\infty \left\{ \left( \mathbb{E} \left| \Gamma_{t,k}^{(i)} \bm{E}(\bm{U}_{t-k})  \right|^q \right)^{1/q} + \left( \mathbb{E} \left| \mathbb{E}[ \Gamma_{t,k} \mid \mathcal{S}_{t-m}^{t+m}] \bm \mu \right|^q \right)^{1/q} \right\} \\
    & \lesssim \sum_{k = m+1}^\infty \bar \varrho^k \lesssim  \bar \varrho^m.
\end{align}
This completes the proof.
\end{proof}

As a result of Lemma \ref{lem:NED}, we can also observe that $\{y_{i,t-j}\}$ is geometrically $L_q$-NED on $\{\bm{U}_t\}$ for any fixed $j$.
Moreover, by Corollary 4.3 of \cite{gallant1988unified}, $\{y_{i,t-j} y_{l,t-k}\}$ is also geometrically $L_q$-NED for any fixed $j$ and $k$.

\begin{lemma}\label{lem:Cov}
    Suppose that $\{x_t\}$ is uniformly bounded and geometrically $L_2$-NED on $\{\bm{U}_t\}$ with NED coefficient $\bar \varrho$.
    Then, $\left| \text{Cov}(x_t, x_s) \right| \lesssim \bar \varrho^{|t - s|}$ for $t \neq s$.
\end{lemma}

\begin{proof}
Decompose $x_t = x_{1t}^{(m)} + x_{2t}^{(m)}$, where
\begin{align}
    x_{1t}^{(m)} 
    & \coloneqq x_t - \mathbb{E}\left[x_t \mid \mathcal S_{t-m}^{t+m} \right]\\
    x_{2t}^{(m)}
    & \coloneqq \mathbb{E}\left[x_t \mid \mathcal S_{t-m}^{t+m} \right].
\end{align}
By assumption, we have $|| x_{1t}^{(m)}  ||_2 \lesssim \bar \varrho^m$.
Define $\Delta = \lfloor |t-s|/3 \rfloor$, so that $\mathcal S_{t-\Delta}^{t + \Delta} \cap \mathcal S_{s-\Delta}^{s + \Delta} = \varnothing$ and $\text{Cov} \left(x_{2t}^{(\Delta)},x_{2s}^{(\Delta)}\right)=0$ by Assumption \ref{as:U}.
Hence,
\begin{align}
\text{Cov}\left(x_t,x_s\right)
  &= \text{Cov}\left(x_{1t}^{(\Delta)}, x_{1s}^{(\Delta)}\right)
   + \text{Cov}\left(x_{1t}^{(\Delta)}, x_{2s}^{(\Delta)}\right)
   + \text{Cov}\left(x_{2t}^{(\Delta)}, x_{1s}^{(\Delta)}\right).
\end{align}
By Cauchy-Schwarz inequality, it is straightforward to see that each of the three terms on the right-hand side is of order $\bar \varrho^\Delta$, which implies the desired result.
\end{proof}


\begin{lemma}\label{lem:LLN}
Suppose that Assumptions \ref{as:coef}(i),  \ref{as:dist}(i), and \ref{as:basis}(i) hold.
Then,
\begin{itemize}
    \item[(i)] $\left\|\frac{1}{L T} \sum_{\ell \in [L]} \sum_{t \in [T]} \left\{ \xi_t(\tau_\ell) \xi_{t}(\tau_{\ell})^\top - \mathbb{E}\left[ \xi_t(\tau_\ell) \xi_{t}(\tau_{\ell})^\top  \right] \right\} \right\| = O_P(H/\sqrt{T})$
    \item[(ii)] $\left\|\frac{1}{L^2 T} \sum_{\ell, \ell' \in [L]} \sum_{t \in [T]}  \left\{ \xi_t(\tau_\ell) \xi_{t}(\tau_{\ell'})^\top  - \mathbb{E}\left[ \xi_t(\tau_\ell) \xi_{t}(\tau_{\ell'})^\top  \right] \right\} \min\{\tau_\ell, \tau_{\ell'}\}(1 - \max\{\tau_\ell, \tau_{\ell'}\}) \right\| = O_P(H/\sqrt{T})$
    \item[(iii)] $\left\|\frac{1}{L T} \sum_{\ell \in [L]} \sum_{t \in [T]}  \left\{ f_{it}(m_{it}(\tau) \mid \mathcal{F}_{t-1}) \xi_t(\tau_\ell) \xi_{t}(\tau_{\ell})^\top - \mathbb{E}\left[ f_{it}(m_{it}(\tau) \mid \mathcal{F}_{t-1}) \xi_t(\tau_\ell) \xi_{t}(\tau_{\ell})^\top  \right] \right\} \right\| = O_P(H/\sqrt{T})$
\end{itemize}
\end{lemma}

\begin{proof}
    We only prove (i).
    Results (ii) and (iii) can be proved analogously.
    Let $r_{t,ij,lk} \coloneqq y_{i,t-j}y_{l,t-k}$.
    Recall that each $c_{i,t-j}$ can be expressed as $c_{i,t-j} = a_{0j} + \sum_{(l,j') \in [n] \times [p]} a_{lj}^{(j')} y_{l, t-j'}$ for some constants $(a_{0j}, a_{1j}^{(1)}, \ldots, a_{nj}^{(p)})$.
    Then, a typical element of $\xi_t(\tau_\ell) \xi_t(\tau_\ell)^\top$ can be written as $b_{h_1}(\tau_\ell) b_{h_2}(\tau_\ell) c_{i,t-j} c_{l,t-k}$.
    Observe that
    \begin{align}
        \left| c_{i,t-j} c_{l,t-k} - \mathbb{E}[c_{i,t-j} c_{l,t-k}]\right| 
        & \lesssim \sum_{(l,j') \in [n] \times [p]} |a_{lj}^{(j')}| \cdot | y_{l, t-j'} - \mathbb{E}[y_{l, t-j'}]| \\
        & \quad + \sum_{(l,j') \in [n] \times [p]} |a_{lk}^{(j')}| \cdot | y_{l, t-j'} - \mathbb{E}[y_{l, t-j'}]| \\
        & \quad + \sum_{(l_1,j_1) \in [n] \times [p]} \sum_{(l_2,j_2) \in [n] \times [p]} |a_{l_1, j}^{(j_1)} a_{l_2, k}^{(j_2)}| \cdot |r_{t, l_1 j_1, l_2 j_2} - \mathbb{E}[r_{t, l_1 j_1, l_2 j_2}]|.
    \end{align}
     As discussed above, $\{y_{i,t-j}\}$ and $\{r_{t,ij,lk}\}$ are all $L_q$-NED on $\{\bm{U}_t\}$.
    For example, for $\{r_{t,ij,lk}\}$, by Lemma \ref{lem:Cov}, we have
    \begin{align}
        \mathbb{E}\left( \frac{1}{T}\sum_{t \in [T]} (r_{t,ij,lk} - \mathbb{E}[r_{t,ij,lk}]) \right)^2
        & = \frac{1}{T^2} \sum_{t \in [T]} \text{Var}\left( r_{t,ij,lk}\right) + \frac{1}{T^2} \sum_{t \in [T]} \sum_{s \neq t} \text{Cov}\left( r_{t,ij,lk}, r_{s,ij,lk}\right) \\
        & = \frac{1}{T^2} \sum_{t \in [T]} \text{Var}\left( r_{t,ij,lk}\right) + \frac{1}{T^2} \sum_{t \in [T]} \sum_{d = 1}^\infty \sum_{s: \: |t - s| = d} \text{Cov}\left( r_{t,ij,lk}, r_{s,ij,lk}\right)  \\
        & \lesssim \frac{1}{T} + \frac{1}{T^2} \sum_{t \in [T]} \sum_{d = 1}^\infty \bar \varrho^d  \lesssim \frac{1}{T}.
    \end{align}
    This leads to 
    \begin{align}
        \left|\frac{1}{T} \sum_{t \in [T]} (r_{t,ij,lk} - \mathbb{E}[r_{t,ij,lk}])  \right| = O_P(T^{-1/2})
    \end{align}
    by Markov's inequality.
    The same result applies to $\{y_{i,t-j}\}$.
    Hence,
    \begin{align}
        \left\|\frac{1}{T} \sum_{t \in [T]} \left\{ \xi_t(\tau_\ell) \xi_{t}(\tau_{\ell})^\top - \mathbb{E}\left[ \xi_t(\tau_\ell) \xi_{t}(\tau_{\ell})^\top  \right] \right\}\right\| = O_P(H/\sqrt{T})
    \end{align}
    holds. 
\end{proof}


\begin{lemma}[Identification]\label{lem:identification}
    Suppose that Assumptions \ref{as:dist}(i), (ii), \ref{as:basis}(ii), \ref{as:matrix}(i), and \ref{as:tune}(i) hold.
    Let $\bm{\alpha}$ be any element of $\mathcal{R}_H^\dagger$ such that $||\bm{\alpha}|| \ge c_\alpha$ for any fixed $c_\alpha > 0$.
    Then, for all sufficiently large $T$, $\mathbb{E}[Q^\dagger_{TL}(\bm{\alpha})] > \mathbb{E}[Q^\dagger_{TL}(\bm{0})]$.
\end{lemma}

\begin{proof}
First, note that $\mathbb{E}[Q^\dagger_{TL}(\bm{0})] = P_{\lambda}(\bm{\gamma}_i)$.
By \eqref{eq:knight},
\begin{align}
    \mathbb{E}[Q_{TL}^\dagger(\bm{\alpha})] - 
    \mathbb{E}[Q_{TL}^\dagger(\bm{0})] = \mathbb{E}[D_{1,TL}(\bm{\alpha})] + D_2 (\bm{\alpha}),
\end{align}
where
\begin{align}
    D_{1,TL}(\bm{\alpha})
    & \coloneqq \frac{1}{LT} \sum_{\ell \in [L]} \sum_{t \in [T]} \left[ \rho_{\tau_\ell}\left( \epsilon_{it}(\tau_\ell) - \xi_t(\tau_\ell)^\top \bm{\alpha}\right) - \rho_{\tau_\ell}\left( \epsilon_{it}(\tau_\ell) \right) \right] \\
    & = \frac{1}{LT} \sum_{\ell \in [L]} \sum_{t \in [T]} \underbracket{- \bm{\alpha}^\top \xi_t(\tau_\ell) \psi_{\tau_\ell} (\epsilon_{it}(\tau_\ell))}_{\eqqcolon \: d_{1,t,\tau_\ell}(\bm{\alpha})} \\
    & \quad + \frac{1}{LT} \sum_{\ell \in [L]} \sum_{t \in [T]}  \underbracket{\int_0^{\xi_t(\tau_\ell)^\top \bm{\alpha}} \left(\bm{1}\{\epsilon_{it}(\tau_\ell) \le x\} - \bm{1}\{\epsilon_{it}(\tau_\ell) \le 0\}\right) \text{d}x }_{\eqqcolon \: d_{2,t,\tau_\ell}(\bm{\alpha})}\\
    D_2 (\bm{\alpha})
    & \coloneqq P_{\lambda}(\bm{\gamma}_i + \bm{\alpha}) - P_{\lambda}(\bm{\gamma}_i).
\end{align}
Recall that for $(l, j) \in \mathcal S_{0i}$, we set $\tilde V_l^{(j)} \bm B_H(\cdot) \bm{\gamma}_i = 0$.
Thus, for $(l, j) \in \mathcal S_{0i}$, $s_\lambda ( || \tilde V_l^{(j)} \bm B_H(\cdot) \bm{\gamma}_i ||_2 ) = 0$.
Hence, 
\begin{align}
    D_2(\bm{\alpha})
    & = \sum_{(l, j) \in \mathcal S}   s_\lambda\left( \left\| \tilde V_l^{(j)} \bm B_H(\cdot) \bm{\gamma}_i + \tilde V_l^{(j)} \bm B_H(\cdot) \bm{\alpha} \right\|_2\right)  - \sum_{(l, j) \in \mathcal S}   s_\lambda\left( \left\| \tilde V_l^{(j)} \bm B_H(\cdot) \bm{\gamma}_i \right\|_2\right) \\
    & \ge - \sum_{(l, j) \in \mathcal S_{1i}}  s_\lambda\left( \left\| \tilde V_l^{(j)} \bm B_H(\cdot) \bm{\gamma}_i  \right\|_2\right)  \\
    & \ge - \frac{|\mathcal S_{1i}|(a + 1)\lambda^2}{2}.
\end{align}

Next, decompose $d_{1,t, \tau}(\bm{\alpha})$ further into the following two terms:
\begin{align}
    d_{1a,t,\tau}(\bm{\alpha})
    & \coloneqq - \bm{\alpha}^\top \xi_t(\tau) \psi_\tau(\epsilon^*_{it}(\tau)) \\
    d_{1b, t, \tau}(\bm{\alpha})
    & \coloneqq - \bm{\alpha}^\top \xi_t(\tau) (\psi_\tau(\epsilon_{it}(\tau)) - \psi_\tau(\epsilon^*_{it}(\tau))) \\
    & = \bm{\alpha}^\top \xi_t(\tau) (\bm{1}\{\epsilon_{it}(\tau) < 0\} - \bm{1}\{\epsilon^*_{it} (\tau) < 0\}).
\end{align}
Note that $\{\xi_t(\tau) \psi_{\tau}(\epsilon^*_{it}(\tau)) \}$ is MDS: 
\begin{align}
    \mathbb{E}[\xi_t(\tau)^\top \psi_{\tau_\ell}(\epsilon^*_{it}(\tau)) \mid \mathcal{F}_{t-1}] = \xi_t(\tau)^\top \mathbb{E}[\psi_\tau(\epsilon^*_{it}(\tau)) \mid \mathcal{F}_{t-1}] = 0.
\end{align}
By the law of iterated expectations, $\mathbb{E}[d_{1a,t,\tau}(\bm{\alpha})] = 0$.
Meanwhile, since
\begin{align}
    \mathbb{E}\left(\bm{1}\{\epsilon_{it}(\tau) < 0\} - \bm{1}\{\epsilon^*_{it} (\tau) < 0\} \mid \mathcal{F}_{t-1}\right)
    & = F_{it}(m_{it}(\tau) + v_{it}(\tau) \mid \mathcal{F}_{t-1}) - F_{it}(m_{it}(\tau) \mid \mathcal{F}_{t - 1}) \\
    & = O(H^{-\pi}),
\end{align}
 we have $\mathbb{E}[d_{1b,t,\tau}(\bm{\alpha})] = O(H^{-\pi})$ uniformly in $\bm \alpha \in \mathcal R_H^\dagger$ (recall that $\xi_t(\tau)^\top \bm{\alpha} = \xi_t(\tau)^\top \underbracket{(\bm \gamma_i + \bm{\alpha})}_{\in \mathcal R_H}  - \xi_t(\tau)^\top\underbracket{\bm \gamma_i}_{\in \mathcal R_H} = O(1)$ by assumption).

Moreover, by the mean value expansion,
\begin{align}
    \mathbb{E}\left(\bm{1}\{\epsilon_{it}(\tau) \le x\} - \bm{1}\{\epsilon_{it}(\tau) \le 0\} \mid \mathcal{F}_{t-1}\right)
    & = F_{it}(\xi_t(\tau)^\top \bm{\gamma}_i + x \mid \mathcal{F}_{t-1}) - F_{it}(\xi_t(\tau)^\top \bm{\gamma}_i \mid \mathcal{F}_{t-1}) \\
    & = f_{it}(\xi_t(\tau)^\top \bm{\gamma}_i + \lambda x \mid \mathcal{F}_{t-1}) x
\end{align}
for some $\lambda \in [0.1]$.
 Thus, we have
\begin{align}
    \frac{1}{LT} \sum_{\ell \in [L]} \sum_{t \in [T]} \mathbb{E}[d_{2,t, \tau_\ell} (\bm{\alpha}) \mid \mathcal{F}_{t-1}] 
    & = \frac{1}{L T} \sum_{\ell \in [L]} \sum_{t \in [T]} \int_0^{ \xi_t(\tau_\ell)^\top \bm{\alpha} } \mathbb{E}\left(\bm{1}\{\epsilon_{it}(\tau_\ell) \le x\} - \bm{1}\{\epsilon_{it}(\tau_\ell) \le 0\} \mid \mathcal{F}_{t-1}\right) \text{d}x \\
    & =\frac{1}{L T} \sum_{\ell \in [L]} \sum_{t \in [T]}  \int_0^{ \xi_t(\tau_\ell)^\top \bm{\alpha} } f_{it}(\xi_t(\tau)^\top \bm{\gamma}_i + \lambda x \mid \mathcal{F}_{t-1}) x \text{d}x \\
     & \ge \frac{\underbar{c}_f}{L T} \sum_{\ell \in [L]} \sum_{t \in [T]} \int_0^{ \xi_t(\tau_\ell)^\top \bm{\alpha} /C}  x \text{d}x \\
     & = \frac{\underbar{c}_f}{2 C^2 L T} \sum_{\ell \in [L]} \sum_{t \in [T]} \bm{\alpha}^\top \xi_t(\tau_\ell) \xi_t(\tau_\ell)^\top \bm{\alpha}
\end{align}
for a large constant $C$ by Assumption \ref{as:dist}(ii).
Thus, by the law of iterated expectations with Assumption \ref{as:matrix}(i), 
\begin{align}
    \frac{1}{LT} \sum_{\ell \in [L]} \sum_{t \in [T]} \mathbb{E}[d_{2,t, \tau_\ell} (\bm{\alpha}) ] \gtrsim \left\| \bm{\alpha} \right\|^2.
\end{align}
Finally, combining these results yields
\begin{align}
    \mathbb{E}[Q_{TL}^\dagger(\bm{\alpha})] - 
    \mathbb{E}[Q_{TL}^\dagger(\bm{0})] \gtrsim \left\| \bm{\alpha} \right\|^2 - H^{-\pi} - \lambda^2,
\end{align}
which implies the desired result.
\end{proof}


\begin{lemma}[Uniform convergence]\label{lem:uniform}
    Suppose that Assumptions \ref{as:coef}(i),  \ref{as:dist}(i), and \ref{as:basis}, \ref{as:matrix}(i), (ii), and \ref{as:tune}(ii) hold.
    Then, $\sup_{\bm{\alpha} \in \mathcal{R}_H^\dagger} \left| Q_{TL}^\dagger(\bm{\alpha}) - \mathbb{E}[ Q_{TL}^\dagger(\bm{\alpha})] \right| \overset{p}{\to} 0$ as $T \to \infty$.
\end{lemma}

\begin{proof}
    Observe that
    \begin{align}
    Q_{TL}^\dagger(\bm{\alpha}) - \mathbb{E}[ Q_{TL}^\dagger(\bm{\alpha})]
    & = \frac{1}{LT} \sum_{\ell \in [L]} \sum_{t \in [T]} \left( q_{it, \tau_\ell}(\bm{\alpha}) - \mathbb{E}[q_{it, \tau_\ell}(\bm{\alpha})] \right).
    \end{align} 
    Define
    \begin{align}
        A_{1,TL}(\bm \alpha)
        & \coloneqq \frac{1}{LT} \sum_{\ell \in [L]} \sum_{t \in [T]} \left( q_{it, \tau_\ell}(\bm{\alpha}) - \mathbb{E}[q_{it, \tau_\ell}(\bm{\alpha}) \mid \mathcal{F}_{t-1}] \right) \\
        A_{2,TL}(\bm \alpha)
        & \coloneqq \frac{1}{LT} \sum_{\ell \in [L]} \sum_{t \in [T]} \left( \mathbb{E}[q_{it, \tau_\ell}(\bm{\alpha}) \mid \mathcal{F}_{t-1}] - \mathbb{E}[q_{it, \tau_\ell}(\bm{\alpha}) ] \right).
    \end{align}
    For $A_{1,TL}(\bm \alpha)$, by \eqref{eq:knight}, we can decompose
    \begin{align}
        A_{1,TL}(\bm \alpha)
        & \coloneqq \frac{1}{LT} \sum_{\ell \in [L]} \sum_{t \in [T]} \left( d_{1a, t, \tau_\ell}(\bm{\alpha}) + d_{1b, t, \tau_\ell}(\bm{\alpha})  + h_{t, \tau_\ell}(\bm{\alpha})\right)  + O(H^{-\pi}),
    \end{align}
    where $h_{t, \tau}(\bm{\alpha}) \coloneqq d_{2, t, \tau}(\bm{\alpha}) - \mathbb{E}[d_{2, t, \tau}(\bm{\alpha}) \mid \mathcal{F}_{t-1}]$.

    First, observe that
    \begin{align}
    \sup_{\bm{\alpha} \in \mathcal{R}_H^\dagger} \left| \frac{1}{LT} \sum_{\ell \in [L]} \sum_{t \in [T]} d_{1a, t, \tau_\ell}(\bm{\alpha}) \right| \le \sup_{\bm{\alpha} \in \mathcal{R}_H^\dagger} \|\bm{\alpha}\| \cdot \left\| \frac{1}{LT} \sum_{\ell \in [L]} \sum_{t \in [T]} \xi_t(\tau_\ell) \psi_{\tau_\ell}(\epsilon^*_{it}(\tau_\ell)) \right\|
    \end{align}
    and that
     \begin{align}\label{eq:markov}
    & \mathbb{E} \left\| \frac{1}{LT} \sum_{\ell \in [L]} \sum_{t \in [T]} \xi_t(\tau_\ell) \psi_{\tau_\ell}(\epsilon^*_{it}(\tau_\ell)) \right\|^2 \\
    & = \frac{1}{L^2 T^2} \sum_{\ell, \ell' \in [L]} \sum_{t, t'\in [T]} \mathbb{E}\left[ \xi_t(\tau_\ell)^\top \psi_{\tau_\ell}(\epsilon^*_{it}(\tau_\ell))  \psi_{\tau_{\ell'}}(\epsilon^*_{it'}(\tau_{\ell'})) \xi_{t'}(\tau_{\ell'})  \right]   \\
    & =  \frac{1}{L^2 T^2} \sum_{\ell, \ell' \in [L]} \sum_{t \in [T]} \mathbb{E}\left[ \xi_t(\tau_\ell)^\top \xi_{t}(\tau_{\ell'}) \mathbb{E}\left[ \psi_{\tau_\ell}(\epsilon^*_{it}(\tau_\ell))  \psi_{\tau_{\ell'}}(\epsilon^*_{it}(\tau_{\ell'}))  \mid \mathcal{F}_{t-1} \right] \right] \\
    & \quad + \frac{1}{L^2 T^2} \sum_{\ell, \ell' \in [L]} \sum_{t, t' = 1: \: t > t'}^T \mathbb{E}\left[ \mathbb{E}\left[\xi_t(\tau_\ell)^\top \psi_{\tau_\ell}(\epsilon^*_{it}(\tau_\ell)) \mid \mathcal{F}_{t-1} \right] \psi_{\tau_{\ell'}}(\epsilon^*_{it'}(\tau_{\ell'})) \xi_{t'}(\tau_{\ell'})  \right]  \\
    & \quad + \frac{1}{L^2 T^2} \sum_{\ell, \ell' \in [L]} \sum_{t, t' = 1: \: t < t'}^T \mathbb{E}\left[ \xi_t(\tau_\ell)^\top \psi_{\tau_\ell}(\epsilon^*_{it}(\tau_\ell)) \mathbb{E}\left[ \psi_{\tau_{\ell'}}(\epsilon^*_{it'}(\tau_{\ell'})) \xi_{t'}(\tau_{\ell'}) \mid \mathcal{F}_{t'-1} \right] \right]  \\
    & = \frac{1}{L^2 T^2} \sum_{\ell, \ell' \in [L]} \sum_{t \in [T]} \mathbb{E}\left[ \xi_t(\tau_\ell)^\top \xi_{t}(\tau_{\ell'}) \right]\min\{\tau_\ell, \tau_{\ell'}\}(1 - \max\{\tau_\ell, \tau_{\ell'}\}).
    \end{align}
    Moreover, noting that $\xi_t(\tau) = C_t \otimes \bm b_H(\tau)$ and $C_t$ is bounded by Assumption \ref{as:SMER}(i), we have
    \begin{align}
    \mathbb{E} \left\| \frac{1}{LT} \sum_{\ell \in [L]} \sum_{t \in [T]} \xi_t(\tau_\ell) \psi_{\tau_\ell}(\epsilon^*_{it}(\tau_\ell)) \right\|^2 
    & \lesssim \frac{1}{L^2 T} \sum_{\ell, \ell' \in [L]} \bm b_H(\tau_\ell)^\top \bm b_H(\tau_{\ell'}) \min\{\tau_\ell, \tau_{\ell'}\}(1 - \max\{\tau_\ell, \tau_{\ell'}\}) \\
    & = \frac{1}{T}\left[\int_0^1 \int_0^1 \bm b_H(u)^\top \bm b_H(v) (\min\{u, v\} - uv)\text{d}u \text{d}v + o(1)\right] 
    \end{align}
    where the last equality follows from the construction of $\{\tau_\ell\}$ and $L \to \infty$.
    Using the identity
    \begin{align}
    \min\{u,v\}-uv = \int_0^1 \left(\bm 1\{s\le u\}-u\right)\left(\bm 1\{s\le v\}-v\right)\text{d}s,
    \end{align}
    we can write
    \begin{align}
        & \int_0^1 \int_0^1 \bm b_H(u)^\top \bm b_H(v) (\min\{u, v\} - uv)\text{d}u \text{d}v \\
        & = \int_0^1 \left(\int_0^1   \bm b_H(u)^\top \left(\bm 1\{s\le u\}-u\right)  \text{d}u \int_0^1 \left(\bm 1\{s\le v\}-v\right)\bm b_H(v)\text{d}v\right)\text{d}s,
    \end{align}
    Now, let 
    \begin{align}
    \kappa_H(s) & \coloneqq \int_0^1   \bm b_H(u)^\top \left(\bm 1\{s\le u\}-u\right)  \text{d}u \int_0^1 \left(\bm 1\{s\le v\}-v\right)\bm b_H(v)\text{d}v \\
    \kappa'_H(s) & \coloneqq \int_0^1   \bm b_H(u) \left(\bm 1\{s\le u\}-u\right)  \text{d}u \int_0^1 \left(\bm 1\{s\le v\}-v\right)\bm b_H(v)^\top \text{d}v, 
    \end{align}
    and note that $\kappa_H(s) = \bar{\text{eig}}(\kappa'_H(s))$.
    For any $\bm x \in \mathbb{R}^H$, Cauchy-Schwarz inequality gives
    \begin{align}
        \bm x^\top \kappa'_H(s) \bm x
        & = \left(\int_0^1 \bm x^\top \bm b_H(u)\left(\bm 1\{s\le u\}-u\right) \text{d}u\right)^2 \\
        & \le \int_0^1 (\bm x^\top \bm b_H(u))^2 \text{d}u \int_0^1 \left(\bm 1\{s\le u\}-u\right)^2\text{d}u \\
        & \lesssim \int_0^1 (\bm x^\top \bm b_H(u))^2 \text{d}u \\
        & \le \bar c_b \|\bm x\|^2
\end{align}
by Assumption \ref{as:basis}(iii).
This implies $\bar{\text{eig}}(\kappa'_H(s))\lesssim 1$ for any $s \in [0,1]$, and hence
\begin{align}
    \mathbb{E} \left\| \frac{1}{LT} \sum_{\ell \in [L]} \sum_{t \in [T]} \xi_t(\tau_\ell) \psi_{\tau_\ell}(\epsilon^*_{it}(\tau_\ell)) \right\|^2 \lesssim \frac{1}{T}.
\end{align}
Consequently, we have
\begin{align}\label{eq:markov1}
    \frac{1}{LT} \sum_{\ell \in [L]} \sum_{t \in [T]} d_{1a,t,\tau_\ell}(\bm{\alpha}) = \left\|\bm{\alpha}\right\| O_P(T^{-1/2})
\end{align}
uniformly in $\bm \alpha \in \mathcal{R}_H^\dagger$.

Next, by the monotonicity of the indicator function,
\begin{align}
    & \mathbb{E}\left[ \left\| \frac{1}{LT} \sum_{\ell \in [L]} \sum_{t \in [T]} d_{1b,t,\tau_\ell}(\bm{\alpha}) \right\| \right] \\
    & \le \frac{|| \bm{\alpha} ||}{L T} \sum_{\ell \in [L]} \sum_{t \in [T]} \mathbb{E}\left[  \left\| \xi_t(\tau_\ell) \right\| \cdot \left|  \bm{1}\{ y_{it} < \xi_t(\tau_\ell)^\top \bm{\gamma}_i \} - \bm{1}\{ y_{it} < C_t^\top \bm{\Phi}_i(\tau_\ell) \} \right| \right] \\
    & = \frac{|| \bm{\alpha} ||}{L T} \sum_{\ell \in [L]} \sum_{t \in [T]} \mathbb{E}\left[  \left\| \xi_t(\tau_\ell) \right\| \cdot \left|  \bm{1}\{ y_{it} < m_{it}(\tau_\ell) + v_{it}(\tau_\ell) \} - \bm{1}\{ y_{it} < m_{it}(\tau_\ell) \} \right| \right] \\
    &  \le \frac{|| \bm{\alpha} ||}{L T} \sum_{\ell \in [L]} \sum_{t \in [T]} \mathbb{E}\left[  \left\| \xi_t(\tau_\ell) \right\| \cdot \left(  \bm{1}\{ y_{it} < m_{it}(\tau_\ell) + || v_{it}||_\infty \} - \bm{1}\{ y_{it} < m_{it}(\tau_\ell) - || v_{it}||_\infty\} \right) \right] \\
    & \le \frac{|| \bm{\alpha} ||}{L T} \sum_{\ell \in [L]} \sum_{t \in [T]} \mathbb{E}\left[  \left\| \xi_t(\tau_\ell) \right\|  \mathbb{E}\left[ \bm{1}\{ y_{it} < m_{it}(\tau_\ell) + || v_{it}||_\infty \} - \bm{1}\{ y_{it} < m_{it}(\tau_\ell) - || v_{it}||_\infty\}  \mid \mathcal{F}_{t - 1} \right] \right]\\
    & = \frac{|| \bm{\alpha} ||}{L T} \sum_{\ell \in [L]} \sum_{t \in [T]} \mathbb{E}\Bigl[  \underbracket{\left\| \xi_t(\tau_\ell) \right\|}_{= \: O(H^{1/2})} \underbracket{\left( F_{it}(m_{it}(\tau_\ell) + || v_{it}||_\infty \mid \mathcal{F}_{t-1}) - F_{it}(m_{it}(\tau_\ell) - || v_{it}||_\infty \mid \mathcal{F}_{t-1} )\right)}_{= \: O(H^{-\pi})} \Bigr]\\
    & \lesssim || \bm{\alpha} ||H^{1/2 - \pi}
    \end{align}
    uniformly in $\bm \alpha \in \mathcal{R}_H^\dagger$, implying that
    \begin{align}\label{eq:markov2}
        \frac{1}{LT} \sum_{\ell \in [L]} \sum_{t \in [T]} d_{1b,t,\tau_\ell}(\bm{\alpha}) = \left\|\bm{\alpha}\right\| O_P(H^{1/2 - \pi})
\end{align}
by Markov's inequality.

\bigskip

Next, choose an arbitrary $e > 0$.
Since $\mathcal R_H^\dagger$ is a compact subset of $\mathbb{R}^{N+1}$, we can place lattice points in it with equal side length $e/\sqrt{H}$.
The total number of lattice points is bounded by $C(\sqrt{H}/e)^{N+1}$ for some large constant $C$.
Suppose $\bm \alpha' \in \mathcal R_H^\dagger$ is the nearest lattice point to a given $\bm \alpha$.
Then, we have
\begin{align}\label{eq:CS1}
\begin{split}
\left| \frac{1}{LT} \sum_{\ell \in [L]} \sum_{t \in [T]} ( h_{t, \tau}(\bm{\alpha}) - h_{t, \tau}(\bm{\alpha}')) \right|
& \le \frac{2}{LT} \sum_{\ell \in [L]} \sum_{t \in [T]} \left| \xi_t(\tau_\ell)^\top (\bm \alpha - \bm \alpha') \right|\\
& \le \frac{2}{\sqrt{LT}}  \sqrt{(\bm{\alpha} -\bm{\alpha}')^\top \sum_{\ell \in [L]} \sum_{t \in [T]} \xi_t(\tau_\ell) \xi_t(\tau_\ell)^\top (\bm{\alpha} - \bm{\alpha}')  }\\
& \lesssim \left\| \bm{\alpha} - \bm{\alpha}' \right\|  \le e,
\end{split}
\end{align}
with probability approaching one, where the second inequality is from Cauchy-Schwarz inequality, and the last inequality from Lemma \ref{lem:LLN}(i) and Assumption \ref{as:matrix}(i).

Now, by definition, $\{h_{1,t,\tau}(\bm \alpha)\}$ forms a MDS for each given $\bm \alpha \in \mathcal R_H^\dagger$.
Moreover, we can easily observe that $h_{1,t,\tau}(\bm \alpha)$ is bounded and that $\sum_{t \in [T]} \mathbb{E}[(h_{1,t,\tau}(\bm \alpha))^2 \mid \mathcal{F}_{t-1}] \lesssim T$.
Thus, for a given lattice point $\bm \alpha_k$, by Freedman's inequality \eqref{eq:freedman},
\begin{align}
    \Pr\left(\left| \sum_{t \in [T]}  h_{t, \tau}(\bm{\alpha}_k) \right|\ge T e\right) \le 2 \exp\left( - \frac{T e^2}{c_1 + c_2 e } \right).
\end{align}
Further, by Boole's inequality, noting that $N \asymp H$,
\begin{align}
    \Pr\left(\max_k \left| \sum_{t \in [T]}  h_{t, \tau}(\bm{\alpha}_k) \right|\ge T e\right) 
    & \le 2 C \left( \frac{\sqrt{H}}{e} \right)^{N + 1} \exp\left( - \frac{T e^2}{c_1 + c_2 e } \right) \to 0
\end{align}
under $(H \ln H)/T \to 0$.
Combining this with \eqref{eq:CS1}, we have
\begin{align}\label{eq:uniconv1}
    \sup_{\bm \alpha \in \mathcal R_H^\dagger} \left| \frac{1}{LT} \sum_{\ell \in [L]} \sum_{t \in [T]}  h_{t, \tau_\ell}(\bm{\alpha})\right| \overset{p}{\to} 0.
\end{align}
Consequently, from \eqref{eq:markov1}, \eqref{eq:markov2}, and \eqref{eq:uniconv1}, we obtain $\sup_{\bm \alpha \in \mathcal{R}_H^\dagger}\left|A_{1,TL}(\bm \alpha) \right| \overset{p}{\to} 0$.

\bigskip

We move on to $A_{2,TL}(\bm \alpha)$.
Similarly to the proof of Lemma \ref{lem:identification}, we have
\begin{align}
    A_{2,TL}(\bm \alpha)
    & = \frac{1}{LT} \sum_{\ell \in [L]} \sum_{t \in [T]} \left( \mathbb{E}[d_{2, t, \tau_\ell}(\bm{\alpha}) \mid \mathcal{F}_{t-1}] - \mathbb{E}[d_{2, t, \tau_\ell}(\bm{\alpha}) ] \right) + O(H^{-\pi}).
\end{align}
Recall 
\begin{align}
\mathbb{E}[d_{2, t, \tau}(\bm{\alpha}) \mid \mathcal{F}_{t-1}] 
& = \int_{0}^{\xi_t(\tau)^\top\bm{\alpha}}
\left[ F_{it}\left(m_{it}(\tau)+v_{it}(\tau)+x \mid \mathcal F_{t-1}\right) - F_{it} \left(m_{it}(\tau)+v_{it}(\tau) \mid \mathcal F_{t-1}\right)\right]\text{d}x.
\end{align}
For any $\bm \alpha \in \mathcal R_H^\dagger$, for $\bm \alpha_1, \bm \alpha_2 \in \mathcal R_H^\dagger$,
\begin{align}
    & \left|\mathbb{E}[d_{2, t, \tau}(\bm{\alpha}_1) \mid \mathcal{F}_{t-1}] - \mathbb{E}[d_{2, t, \tau}(\bm{\alpha}_2) \mid \mathcal{F}_{t-1}]\right| \\
    & = \left| \int_{\xi_t(\tau)^\top\bm{\alpha}_2}^{\xi_t(\tau)^\top\bm{\alpha}_1} \left[ F_{it}\left(m_{it}(\tau)+v_{it}(\tau)+x \mid \mathcal F_{t-1}\right) - F_{it} \left(m_{it}(\tau)+v_{it}(\tau) \mid \mathcal F_{t-1}\right)\right]\text{d}x \right|\\
    & = \sup_{x \in \mathbb{R}} \left| F_{it}\left(m_{it}(\tau)+v_{it}(\tau)+x \mid \mathcal F_{t-1}\right) - F_{it} \left(m_{it}(\tau)+v_{it}(\tau) \mid \mathcal F_{t-1}\right) \right| \cdot |\xi_t(\tau)^\top(\bm{\alpha}_1 -\bm{\alpha}_2) |\\
    & \lesssim |\xi_t(\tau)^\top(\bm{\alpha}_1 -\bm{\alpha}_2) |.
\end{align}
Hence, by the same argument as in \eqref{eq:CS1}, we can see that $(LT)^{-1} \sum_{\ell \in [L]} \sum_{t \in [T]} \mathbb{E}[d_{2,t,\tau_\ell}(\bm \alpha) \mid \mathcal{F}_{t-1}]$ is Lipschitz continuous.
By Lemma 1 of \cite{andrews1992generic}, this further implies that $(LT)^{-1} \sum_{\ell \in [L]} \sum_{t \in [T]}  \mathbb{E}[d_{2, t, \tau_\ell}(\bm{\alpha}) \mid \mathcal{F}_{t-1}]$ is stochastically equicontinuous in $\bm \alpha$.
Moreover, since NED property is preserved under Lipschitz transformation, $\mathbb{E}[d_{2, t, \tau}(\bm{\alpha}) \mid \mathcal{F}_{t-1}]$ is NED, and an NED law of large numbers holds at each $\bm \alpha$, as in Lemma \ref{lem:LLN}.
The point-wise convergence and stochastic equicontinuity lead to the uniform convergence (Theorem 1, \cite{andrews1992generic}), which concludes that
\begin{align}
    \sup_{\bm \alpha \in \mathcal{R}_H^\dagger}\left|\frac{1}{LT} \sum_{\ell \in [L]} \sum_{t \in [T]} \left( \mathbb{E}[d_{2, t, \tau_\ell}(\bm{\alpha}) \mid \mathcal{F}_{t-1}] - \mathbb{E}[d_{2, t, \tau_\ell}(\bm{\alpha}) ] \right)\right| \overset{p}{\to} 0.
\end{align}
Hence, $\sup_{\bm \alpha \in \mathcal{R}_H^\dagger}\left|A_{2,TL}(\bm \alpha) \right| \overset{p}{\to} 0$.
\end{proof} 


\begin{proposition}\label{prop:localmin_rate}
    Suppose that the assumptions in Lemma \ref{lem:identification} and \ref{lem:uniform}, and Assumptions \ref{as:coef}(ii) and \ref{as:basis}(iii) hold.
    Then, $\left\| \tilde{\bm{\gamma}}_i - \bm{\gamma}_i \right\| = O_P(1/\sqrt{T} + H^{1/2 - \pi})$. 
\end{proposition}

\begin{proof}
The identifiability of true $\bm{\alpha}$ (i.e., $\bm 0$) shown in Lemma \ref{lem:identification} is equivalent to the identifiability of $\bm{\gamma}_i$.
Then, the uniform convergence result in Lemma \ref{lem:uniform} gives the consistency of $\tilde{\bm{\gamma}}_i$ (see, e.g.,  the proof of Theorem 3.3 in \cite{su2016sieve}).
This ensures that for sufficiently large $T$, $\tilde \gamma_i$ exists in the neighborhood of $\bm{\gamma}_i$.

Given this fact, letting $r_{TH} \coloneqq 1/\sqrt{T} + H^{1/2 -  \pi}$, it suffices to show that for any given $\delta > 0$, there exists a large constant $K$ such that 
\begin{align}
    \Pr\left( \inf_{||\bm{k}|| = K} Q_{TL}(\bm{\gamma}_i + \bm{k} r_{TH}) > Q_{TL}(\bm{\gamma}_i)\right) \ge 1-\delta
\end{align}
for sufficiently large $T$, which implies that with probability at least $1-\delta$ there exists a local minimum in the ball $\{\bm{\gamma}_i + \bm{k} r_{TH}: ||\bm{k}|| \le K\}$.

Similar to Lemma \ref{lem:identification}, we decompose
\begin{align}
    Q_{TL}(\bm{\gamma}_i + \bm{k} r_{TH}) - Q_{TL}(\bm{\gamma}_i) = D_{1,TL}(\bm{k} r_{TH}) + D_2 (\bm{k} r_{TH}).
\end{align}
We first consider $D_2 (\bm{k} r_{TH})$.
Suppose $(l,j) \in \mathcal S_{1i}$ and write $\tilde V_l^{(j)} = (\tilde v_{0l}^{(j)}, \ldots \tilde v_{Nl}^{(j)})$.
Then,
\begin{align}
    & \left\| \tilde V_l^{(j)} \bm B_H(\cdot) \bm{\gamma}_i +  \tilde V_l^{(j)} \bm B_H(\cdot)\bm{k} r_{TH} \right\|_2 \\
    & = \left\| \sum_{a = 0}^N \tilde v_{al}^{(j)} \bm b_H(\cdot)^\top \gamma_{ai} + r_{TH}  \sum_{a = 0}^N \tilde v_{al}^{(j)} \bm b_H(\cdot)^\top k_a \right\|_2 \\
    & = \Biggl\| \underbracket{\sum_{a = 0}^N \tilde v_{al}^{(j)} \phi_{ai}(\cdot)}_{= \theta_{li}^{(j)}(\cdot)} + \sum_{a = 0}^N \tilde v_{al}^{(j)} \{ \bm b_H(\cdot)^\top \gamma_{ai} - \phi_{ai}(\cdot) \} + r_{TH}  \sum_{a = 0}^N \tilde v_{al}^{(j)} \bm b_H(\cdot)^\top k_a \Biggr\|_2 \\
    & \ge \left\| \theta_{li}^{(j)} \right\|_2 - \sum_{a = 0}^N \left| \tilde v_{al}^{(j)} \right| \left\|\{ \bm b_H(\cdot)^\top \gamma_{ai} - \phi_{ai}(\cdot) \}\right\|_2 - r_{TH} \sum_{a = 0}^N \left|  \tilde v_{al}^{(j)} \right| \left\| \bm b_H(\cdot)^\top k_a \right\|_2 \\
    & \gtrsim c - c_{\bm V} H^{-\pi} -  r_{TH} c_{\bm V}
\end{align}
by the reverse triangle inequality.
From this, for all sufficiently large $T$, the left-hand side term is bounded away from zero.
Clearly, the same result applies to $|| \tilde V_l^{(j)} \bm B_H(\cdot) \bm{\gamma}_i ||_2$ for $(l, j) \in \mathcal S_{1i}$.
Hence, since the SCAD penalty is non-negative and is flat for arguments larger than $a \lambda$, 
\begin{align}
    D_2(\bm{k} r_{TH})
    & =  \sum_{(l, j) \in \mathcal S}   s_\lambda\left( \left\| \tilde V_l^{(j)} \bm B_H(\cdot) \bm{\gamma}_i +  \tilde V_l^{(j)} \bm B_H(\cdot)\bm{k} r_{TH} \right\|_2\right) - \sum_{(l, j) \in \mathcal S}   s_\lambda\left( \left\|  \tilde V_l^{(j)} \bm B_H(\cdot) \bm{\gamma}_i \right\|_2\right) \\
    & \ge \sum_{(l, j) \in \mathcal S_{1i}}  s_\lambda\left( \left\| \tilde V_l^{(j)} \bm B_H(\cdot) \bm{\gamma}_i +  \tilde V_l^{(j)} \bm B_H(\cdot)\bm{k} r_{TH} \right\|_2\right)  - \sum_{(l, j) \in \mathcal S_{1i}}  s_\lambda\left( \left\|  \tilde V_l^{(j)} \bm B_H(\cdot) \bm{\gamma}_i \right\|_2\right)  \\
    & = 0
\end{align}
as $T \to \infty$ and $\lambda \to 0$.

Next, we move on to $D_{1,TL}(\bm{k} r_{TH})$.
By \eqref{eq:knight}, uniformly in $\bm k$, we can decompose
\begin{align}
    D_{1,TL}(\bm{k} r_{TH})
    & =  r_{TH} \underbracket{\frac{1}{LT} \sum_{\ell \in [L]} \sum_{t \in [T]} d_{1a,t,\tau_\ell}(\bm{k})}_{= ||\bm{k}|| O_P(T^{-1/2}) \; \text{by} \; \eqref{eq:markov1}} + r_{TH} \underbracket{\frac{1}{LT} \sum_{\ell \in [L]} \sum_{t \in [T]} d_{1b,t,\tau_\ell}(\bm{k})}_{=  ||\bm{k}|| O_P(H^{1/2 - \pi}) \; \text{by} \; \eqref{eq:markov2}} \\
    & \quad + \frac{1}{LT} \sum_{\ell \in [L]} \sum_{t \in [T]} d_{2,t,\tau_\ell}(r_{TH}\bm{k}).
\end{align}

By the mean value theorem,
\begin{align}
 F_{it}(m_{it}(\tau)+v_{it}(\tau) + x \mid \mathcal{F}_{t-1}) - F_{it}(m_{it}(\tau)+v_{it}(\tau) \mid \mathcal{F}_{t-1}) = f_{it}(m_{it}(\tau)+v_{it}(\tau) + \lambda x \mid \mathcal{F}_{t-1}) x
\end{align}
for some $\lambda \in [0,1]$.
Expanding the density around $m_{it}(\tau)$ gives
\begin{align}
    f_{it}(m_{it}(\tau)+v_{it}(\tau)+\lambda x \mid \mathcal{F}_{t-1}) = f_{it}(m_{it}(\tau) \mid \mathcal{F}_{t-1}) + O(|v_{it}(\tau)|+|x|)
\end{align}
by Assumption \ref{as:dist}(i), and therefore
\begin{align}
    & F_{it}(m_{it}(\tau)+v_{it}(\tau)+x \mid \mathcal{F}_{t-1}) - F_{it}(m_{it}(\tau)+v_{it}(\tau) \mid \mathcal{F}_{t-1}) \\
    & \quad = f_{it}(m_{it}(\tau) \mid \mathcal{F}_{t-1}) x + O(|v_{it}(\tau)| \cdot |x| + |x|^2).
\end{align}
Substituting this into $d_{2,t,\tau_\ell}(r_{TH}\bm{k})$ yields
\begin{align}
    \mathbb{E}\left[d_{2,t,\tau_\ell}(r_{TH}\bm{k}) \mid \mathcal{F}_{t-1}\right]
    & = \int_0^{r_{TH} \bm{k}^\top \xi_t(\tau_\ell) } \mathbb{E}\left(\bm{1}\{\epsilon_{it}(\tau_\ell) \le x\} - \bm{1}\{\epsilon_{it}(\tau_\ell) \le 0\} \mid \mathcal{F}_{t-1}\right) \text{d}x \\
    & = \int_{0}^{r_{TH}\bm{k}^\top\xi_t(\tau_\ell)}
    f_{it}(m_{it}(\tau_\ell) \mid \mathcal{F}_{t-1}) x \text{d} x + \int_{0}^{r_{TH}\bm{k}^\top\xi_t(\tau_\ell)} O(|v_{it}(\tau_\ell)|\cdot |x| + |x|^2) \text{d} x \\
    &= \frac{r_{TH}^2}{2} f_{it}(m_{it}(\tau_\ell) \mid \mathcal{F}_{t-1}) (\bm{k}^\top\xi_t(\tau_\ell))^2 \\
    & \quad + O\left(|v_{it}(\tau_\ell)| r_{TH}^2 (\bm{k}^\top\xi_t(\tau_\ell))^2 + r_{TH}^3 |\bm{k}^\top\xi_t(\tau_\ell)|^3\right).
\end{align}
Hence,
\begin{align}
    \frac{1}{LT}\sum_{\ell=1}^L\sum_{t=1}^T \mathbb{E}\left[d_{2,t,\tau_\ell}(r_{TH}\bm{k}) \mid \mathcal{F}_{t-1}\right]
    &= \frac{r_{TH}^2}{2}\bm{k}^\top \left( \frac{1}{LT}\sum_{\ell=1}^L\sum_{t=1}^T f_{it}(m_{it}(\tau_\ell) \mid \mathcal{F}_{t-1}) \xi_t(\tau_\ell)\xi_t(\tau_\ell)^\top \right)\bm{k} \\
    &\quad + O\left(\frac{r_{TH}^2}{LT}\sum_{\ell=1}^L\sum_{t=1}^T |v_{it}(\tau_\ell)| (\bm{k}^\top\xi_t(\tau_\ell))^2 \right) \\
    & \quad + O\left( \frac{r_{TH}^3}{LT}\sum_{\ell=1}^L\sum_{t=1}^T |\bm{k}^\top\xi_t(\tau_\ell)|^3 \right).
\end{align}
For the second term, since $\sup_t \|v_{it}\|_\infty \lesssim H^{-\pi}$ and $\bar{\mathrm{eig}}\left((LT)^{-1}\sum_{\ell \in [L]} \sum_{t \in [T]} \xi_t(\tau_\ell)\xi_t(\tau_\ell)^\top\right) = O_P(1)$ by Assumption \ref{as:matrix}(i) and Lemma \ref{lem:LLN}(i), it is of order $r_{TH}^2 \|\bm{k}\|^2 O(H^{-\pi})$.

For the third term, observe that
\begin{align}
    \frac{1}{LT}\sum_{\ell=1}^L\sum_{t=1}^T |\bm k^\top \xi_t(\tau_\ell)|^3 \le \left(\max_{t\in[T],\ell\in[L]}|\bm k^\top \xi_t(\tau_\ell)|\right) \bm k^\top\left(\frac{1}{LT}\sum_{\ell=1}^L\sum_{t=1}^T \xi_t(\tau_\ell)\xi_t(\tau_\ell)^\top\right)\bm k.
\end{align}
Moreover, by $\sup_t \|\xi_t\|_\infty = O(H^{1/2})$, $\max_{t,\ell}|\bm k^\top \xi_t(\tau_\ell)| = \|\bm k\| O(H^{1/2})$.
Hence, since $H^{1/2} r_{TH} \to 0$ by assumption, we have
\begin{align}
    \frac{r_{TH}^3}{LT}\sum_{\ell=1}^L\sum_{t=1}^T |\bm{k}^\top\xi_t(\tau_\ell)|^3 = r_{TH}^2 \cdot o\left( \bm{k}^\top\left( \frac{1}{L T} \sum_{\ell \in [L]} \sum_{t \in [T]} \xi_t(\tau_\ell) \xi_t(\tau_\ell)^\top \right)\bm{k} \right).
\end{align}
To summarize, uniformly in $\bm k$,
\small\begin{align}
    \frac{1}{LT} \sum_{\ell \in [L]} \sum_{t \in [T]} \mathbb{E}[d_{2,t,\tau_\ell}(r_{TH}\bm{k}) \mid \mathcal{F}_{t-1}]
    & = \frac{r_{TH}^2}{2} \bm{k}^\top\left( \frac{1}{L T} \sum_{\ell \in [L]} \sum_{t \in [T]} f_{it}(m_{it}(\tau_\ell) \mid \mathcal{F}_{t-1}) \xi_t(\tau_\ell) \xi_t(\tau_\ell)^\top \right)\bm{k} \\
    & \quad + r_{TH}^2 ||\bm{k}||^2 O\left( H^{- \pi} \right) + r_{TH}^2 \cdot o\left( \bm{k}^\top\left( \frac{1}{L T} \sum_{\ell \in [L]} \sum_{t \in [T]} \xi_t(\tau_\ell) \xi_t(\tau_\ell)^\top \right)\bm{k} \right).
\end{align}\normalsize
By Lemma \ref{lem:LLN}(iii) and Assumption \ref{as:matrix}(iii), the first term on the right-hand side is bounded below from $\underbar{c}_3 r^2_{TH} ||\bm{k}||^2/2$ with probability approaching one.
Similarly, Lemma \ref{lem:LLN}(i) with Assumption \ref{as:matrix}(i) implies that the third term is of order $r_{TH}^2 || \bm{k}||^2 o_P(1)$.

Meanwhile, writing $d_{2,t,\ell} = d_{2,t,\tau_\ell}(r_{TH}\bm{k})$ for notational simplicity,
\begin{align}
    & \mathbb{E}\left[ \left( \frac{1}{LT} \sum_{\ell \in [L]} \sum_{t \in [T]}\left( d_{2,t,\ell} - \mathbb{E}[d_{2,t,\ell} \mid \mathcal{F}_{t-1}]\right) \right)^2 \right] \\
    & = \frac{1}{L^2 T^2} \sum_{\ell, \ell' \in [L]} \sum_{t \in [T]} \mathbb{E}\left[ \left( d_{2,t,\ell} - \mathbb{E}[d_{2,t,\ell} \mid \mathcal{F}_{t-1}]\right) \left( d_{2,t,\ell'} - \mathbb{E}[ d_{2,t,\ell'} \mid \mathcal{F}_{t-1}]\right) \right] \\   
    & \quad + \frac{1}{L^2 T^2} \sum_{\ell, \ell' \in [L]} \sum_{t, t' = 1: \: t > t'}^T \mathbb{E}\left[ \left( d_{2,t,\ell} - \mathbb{E}[d_{2,t,\ell} \mid \mathcal{F}_{t-1}]\right) \left( d_{2,t',\ell'} - \mathbb{E}[ d_{2,t',\ell'} \mid \mathcal{F}_{t'-1}]\right) \right] \\
    & \quad + \frac{1}{L^2 T^2} \sum_{\ell, \ell' \in [L]} \sum_{t, t' = 1: \: t < t'}^T \mathbb{E}\left[ \left( d_{2,t,\ell} - \mathbb{E}[d_{2,t,\ell} \mid \mathcal{F}_{t-1}]\right) \left( d_{2,t',\ell'} - \mathbb{E}[ d_{2,t',\ell'} \mid \mathcal{F}_{t'-1}]\right) \right] \\
    & = \frac{1}{T}\mathbb{E}\left[ \frac{1}{L^2 T} \sum_{\ell, \ell' \in [L]} \sum_{t \in [T]}  \text{Cov}\left( d_{2,t,\ell},  d_{2,t,\ell'} \mid \mathcal{F}_{t-1}\right) \right].
\end{align}
Moreover, by applying Cauchy-Schwarz inequality repeatedly,
\begin{align} 
    \frac{1}{L^2 T} \sum_{\ell, \ell' \in [L]} \sum_{t \in [T]}  \left| \text{Cov}\left( d_{2,t,\ell},  d_{2,t,\ell'} \mid \mathcal{F}_{t-1}\right) \right| 
    & \le \frac{1}{L^2 T} \sum_{\ell, \ell' \in [L]} \sum_{t \in [T]}  \sqrt{\text{Var}( d_{2,t,\ell} \mid \mathcal{F}_{t-1})} \sqrt{\text{Var}( d_{2,t,\ell'} \mid \mathcal{F}_{t-1})} \\
    & = \frac{1}{L^2 T}  \sum_{t \in [T]} \left( \sum_{\ell \in [L]} \sqrt{\text{Var}( d_{2,t,\ell} \mid \mathcal{F}_{t-1})} \right)^2 \\
    & \le \frac{1}{L T}  \sum_{t \in [T]}  \sum_{\ell \in [L]} \text{Var}( d_{2,t,\ell} \mid \mathcal{F}_{t-1}) \\
    & \le r_{TH}^2 \bm{k}^\top \left( \frac{1}{L T}  \sum_{t \in [T]}  \sum_{\ell \in [L]}   \xi_t(\tau_\ell) \xi_t(\tau_\ell)^\top \right) \bm{k},
\end{align}
implying that $(LT)^{-1} \sum_{\ell \in [L]} \sum_{t \in [T]}( d_{2,t,\ell} - \mathbb{E}[d_{2,t,\ell} \mid \mathcal{F}_{t-1}]) = r_{TH} || \bm{k} || O_P(T^{-1/2})$.
Combining these results yields
\begin{align}\label{eq:d2}
\begin{split}
    \frac{1}{LT} \sum_{\ell \in [L]} \sum_{t \in [T]} d_{2,t,\tau_\ell}(r_{TH}\bm{k}) 
    & = \frac{r_{TH}^2}{2} \bm{k}^\top\left( \frac{1}{L T} \sum_{\ell \in [L]} \sum_{t \in [T]} f_{it}(m_{it}(\tau_\ell)\mid \mathcal{F}_{t-1}) \xi_t(\tau_\ell) \xi_t(\tau_\ell)^\top \right)\bm{k} \\
    & \quad + r_{TH}^2 \left\| \bm{k} \right\|^2 O\left( H^{- \pi} \right) + r_{TH}^2 \left\| \bm{k} \right\|^2 o_P(1) + r_{TH} \left\|\bm{k}\right\| O_P(T^{-1/2}),
\end{split}
\end{align}
and therefore, by taking a sufficiently large $K$,
\begin{align}
    Q_{TL}(\bm{\gamma}_i + \bm{k} r_{TH}) - Q_{TL}(\bm{\gamma}_i) 
    & = \frac{r_{TH}^2}{2} \bm{k}^\top\left( \frac{1}{L T} \sum_{\ell \in [L]} \sum_{t \in [T]} f_{it}(m_{it}(\tau_\ell)\mid \mathcal{F}_{t-1}) \xi_t(\tau_\ell) \xi_t(\tau_\ell)^\top \right)\bm{k} \\
    & \quad + r_{TH}^2 \left\| \bm{k} \right\|^2 O\left( H^{- \pi} \right) + r_{TH}^2 \left\| \bm{k} \right\|^2 o_P(1)   \\
    & \quad + r_{TH} \left\| \bm{k} \right\| O_P(T^{-1/2}) + r_{TH} \left\| \bm{k} \right\| O_P(H^{1/2 - \pi}) \\
    & = \frac{r_{TH}^2}{2} \bm{k}^\top\left( \frac{1}{L T} \sum_{\ell \in [L]} \sum_{t \in [T]} f_{it}(m_{it}(\tau_\ell)\mid \mathcal{F}_{t-1}) \xi_t(\tau_\ell) \xi_t(\tau_\ell)^\top \right)\bm{k} \\
    & \quad + r_{TH}^2 K^2 o_P(1) + r^2_{TH} K O_P(1) > 0
\end{align}
with probability approaching one.
\end{proof}


\begin{corollary}\label{cor:convrate}
Under the assumptions of Proposition \ref{prop:localmin_rate}, we have the following results:
\begin{align}
 \text{(i)} & \quad \left\| \tilde \phi_{0i} - \phi_{0i} \right\|_2 = O_P(1/\sqrt{T} + H^{1/2 -\pi}), 
 \quad \text{(ii)} \quad \left\| \tilde \phi_{li}^{(j)} - \phi^{(j)}_{li} \right\|_2 = O_P(1/\sqrt{T} + H^{1/2 -\pi}) \\
 \text{(iii)} & \quad \left\| \tilde \phi_{0i} - \phi_{0i} \right\|_\infty = O_P(\sqrt{H/T} + H^{1-\pi}),
 \quad \text{(iv)} \quad \left\| \tilde \phi_{li}^{(j)} - \phi^{(j)}_{li} \right\|_\infty = O_P(\sqrt{H/T} + H^{1-\pi}).
\end{align}
\end{corollary}

\begin{proof}
The results are immediate from Proposition \ref{prop:localmin_rate}.
\end{proof}


In the following, we postulate that the coordinate system \eqref{eq:max-min} is employed.
In this case, $\bm{\gamma}_i = (\bm{\beta}_{1i}, \bm{\beta}_{0i})$, where $\underbracket{\bm{\beta}_{1i}}_{H(|\mathcal S_{1i}| + 1) \times 1} = (\gamma_{0i}, \{\bm{\gamma}_{li}^{(j)}\}_{(l,j) \in \mathcal S_{1i}})^\top$, and $\underbracket{\bm{\beta}_{0i}}_{H |\mathcal S_{0i}| \times 1} = (\bm{\gamma}_{0i}^\top, \ldots, \bm{\gamma}_{0i}^\top)^\top$.

\begin{lemma}\label{lem:lagslct}
    Suppose that Assumptions \ref{as:coef} - \ref{as:tune} hold.
    Then, for any given $\bar{\bm \gamma}_i = (\bar{\bm{\beta}}_{1i}, \bar{\bm{\beta}}_{0i})$, with $\bar{\bm{\beta}}_{0i} = (\bar{\gamma}_{0i}^\top, \ldots, \bar{\gamma}_{0i}^\top)^\top$, such that $\left\| \bar{\bm{\gamma}}_i - \bm{\gamma}_i \right\| \lesssim r_{TH}$,
    \begin{align}
        \inf_{\bm{\beta}_2 \ne \bar{\bm{\beta}}_{0i}: \: \left\| \bm{\beta}_2 - \bar{\bm{\beta}}_{0i} \right\| \lesssim r_{TH} } Q_{TL}(\bar{\bm{\beta}}_{1i}, \bm{\beta}_2) = Q_{TL}(\bar{\bm{\beta}}_{1i}, \bar{\bm{\beta}}_{0i}),
    \end{align}
    with probability approaching one.
\end{lemma}

\begin{proof}
Observe that
\begin{align}
    & Q_{TL}(\bar{\bm{\beta}}_{1i}, \bar{\bm{\beta}}_{0i}) - Q_{TL}(\bm{\beta}_{1i},  \bm{\beta}_{0i}) \\ 
    & = \frac{1}{LT} \sum_{\ell \in [L]} \sum_{t \in [T]} q_{it,\tau_\ell}(\bar{\bm{\beta}}_{1i} - \bm{\beta}_{1i}, \bar{\bm{\beta}}_{0i} - \bm{\beta}_{0i}) + \sum_{(l, j) \in \mathcal S_{1i}}  s_\lambda\left( \left\| \bm{b}_H(\cdot)^\top\{\bar \gamma_{li}^{(j)} - \bar \gamma_{0i} \} \right\|_2\right) \\
    & \quad - \sum_{(l, j) \in \mathcal S_{1i}}  s_\lambda\left( \left\| \bm{b}_H(\cdot)^\top\{\gamma_{li}^{(j)} - \gamma_{0i} \} \right\|_2\right)
\end{align}
and that
\begin{align}
    & Q_{TL}(\bar{\bm{\beta}}_{1i}, \bm{\beta}_2) - Q_{TL}(\bm{\beta}_{1i},  \bm{\beta}_{0i}) \\ 
    & = \frac{1}{LT} \sum_{\ell \in [L]} \sum_{t \in [T]} q_{it,\tau_\ell}(\bar{\bm{\beta}}_{1i} - \bm{\beta}_{1i}, \bm{\beta}_2 - \bm{\beta}_{0i})  + \sum_{(l, j) \in \mathcal S_{1i}}  s_\lambda\left( \left\| \bm{b}_H(\cdot)^\top\{\bar \gamma_{li}^{(j)} - \bar \gamma_{0i} \} \right\|_2\right) \\
    & \quad + \sum_{(l, j) \in \mathcal S_{0i}}  s_\lambda\left( \left\| \bm{b}_H(\cdot)^\top\{ \gamma_l^{(j)} - \bar \gamma_{0i} \} \right\|_2\right) - \sum_{(l, j) \in \mathcal S_{1i}}  s_\lambda\left( \left\| \bm{b}_H(\cdot)^\top\{\gamma_{li}^{(j)} - \gamma_{0i} \} \right\|_2\right).
\end{align}
Hence,
\begin{align}
    & Q_{TL}(\bar{\bm{\beta}}_{1i}, \bm{\beta}_2) - Q_{TL}(\bar{\bm{\beta}}_{1i}, \bar{\bm{\beta}}_{0i})  \\
    & =  \frac{1}{LT} \sum_{\ell \in [L]} \sum_{t \in [T]} q_{it,\tau_\ell}(\bar{\bm{\beta}}_{1i} - \bm{\beta}_{1i}, \bm{\beta}_2 - \bm{\beta}_{0i}) - \frac{1}{LT} \sum_{\ell \in [L]} \sum_{t \in [T]} q_{it,\tau_\ell}(\bar{\bm{\beta}}_{1i} - \bm{\beta}_{1i}, \bar{\bm{\beta}}_{0i} - \bm{\beta}_{0i})\\
    & \quad + \sum_{(l, j) \in \mathcal S_{0i}}  s_\lambda\left( \left\| \bm{b}_H(\cdot)^\top\{ \gamma_l^{(j)} - \bar \gamma_{0i} \} \right\|_2\right).
\end{align}
Similarly as above, decompose $q_{it,\tau_\ell}(\bar{\bm{\beta}}_{1i} - \bm{\beta}_{1i}, \bm{\beta}_2 - \bm{\beta}_{0i}) = d_{1a,t,\tau_\ell}(\bar{\bm{\beta}}_{1i} - \bm{\beta}_{1i}, \bm{\beta}_2 - \bm{\beta}_{0i}) + d_{1b,t,\tau_\ell}(\bar{\bm{\beta}}_{1i} - \bm{\beta}_{1i}, \bm{\beta}_2 - \bm{\beta}_{0i}) + d_{2,t,\tau_\ell}(\bar{\bm{\beta}}_{1i} - \bm{\beta}_{1i}, \bm{\beta}_2 - \bm{\beta}_{0i})$.
As shown in \eqref{eq:markov1} and \eqref{eq:markov2}, we have
\begin{align}
    \frac{1}{LT} \sum_{\ell \in [L]} \sum_{t \in [T]} d_{1a,t,\tau_\ell}(\bar{\bm{\beta}}_{1i} - \bm{\beta}_{1i}, \bm{\beta}_2 - \bm{\beta}_{0i}) 
    & = O_P(r_{TH}T^{-1/2}) \\
    \frac{1}{LT} \sum_{\ell \in [L]} \sum_{t \in [T]} d_{1b,t,\tau_\ell}(\bar{\bm{\beta}}_{1i} - \bm{\beta}_{1i}, \bm{\beta}_2 - \bm{\beta}_{0i}) 
    & = O_P(r_{TH}H^{1/2 - \pi}).
\end{align}
Moreover, similar to \eqref{eq:d2},
\begin{align}
    \frac{1}{LT} \sum_{\ell \in [L]} \sum_{t \in [T]} d_{2,t,\tau_\ell}(\bar{\bm{\beta}}_{1i} - \bm{\beta}_{1i}, \bm{\beta}_2 - \bm{\beta}_{0i}) 
    & = O_P(r_{TH}^2) + O_P(r_{TH}T^{-1/2}). 
\end{align}
Hence,
\begin{align}
    \frac{1}{LT} \sum_{\ell \in [L]} \sum_{t \in [T]} q_{it,\tau_\ell}(\bar{\bm{\beta}}_{1i} - \bm{\beta}_{1i}, \bm{\beta}_2 - \bm{\beta}_{0i}) =  O_P(r_{TH}^2).
\end{align}
Applying the same argument, we obtain $(LT)^{-1} \sum_{\ell \in [L]} \sum_{t \in [T]} q_{it,\tau_\ell}(\bar{\bm{\beta}}_{1i} - \bm{\beta}_{1i}, \bar{\bm{\beta}}_{0i} - \bm{\beta}_{0i}) =  O_P(r_{TH}^2)$.

\bigskip

Meanwhile, by $|| \bm{\beta}_2 - \bar{\bm{\beta}}_{0i} || \lesssim r_{TH}$ and Assumption \ref{as:basis}(iii), there exists $c > 0$ such that
\begin{align}
    \sum_{(l, j) \in \mathcal S_{0i}}  s_\lambda\left( \left\| \bm{b}_H(\cdot)^\top\{ \gamma_l^{(j)} - \bar \gamma_{0i} \} \right\|_2\right) 
    \gtrsim s_\lambda\left( c \cdot r_{TH}\right) = c \cdot \lambda r_{TH} 
\end{align}
for sufficiently large $T$.
To summarize,
\begin{align}
    Q_{TL}(\bar{\bm{\beta}}_{1i}, \bm{\beta}_2) - Q_{TL}(\bar{\bm{\beta}}_{1i}, \bar{\bm{\beta}}_{0i}) \gtrsim O_P(r_{TH}^2) + c \cdot \lambda r_{TH},
\end{align}
which implies that $Q_{TL}(\bar{\bm{\beta}}_{1i}, \bm{\beta}_2) \ge Q_{TL}(\bar{\bm{\beta}}_{1i}, \bar{\bm{\beta}}_{0i})$ as $T$ increases under Assumption \ref{as:tune}(i).
This implies the desired result.
\end{proof}


\begin{corollary}\label{cor:lag}
    Under the assumptions of Lemma \ref{lem:lagslct}, for $(l, j) \in \mathcal S_{0i}$, $\Pr\left(\left\|\tilde \theta_{li}^{(j)}\right\|_2 = 0\right) \to 1$.
\end{corollary}

\begin{proof}
    Combined with Proposition \ref{prop:localmin_rate}, Lemma \ref{lem:lagslct} implies that $\Pr(\tilde \gamma_{li}^{(j)} = \tilde \gamma_{0i}) \to 1$ for all $(l, j) \in \mathcal S_{0i}$.
    Hence, $\tilde \theta_{li}^{(j)}(\tau) = (\tilde \phi_{li}^{(j)}(\tau) - \tilde \phi_{0i}(\tau)) / (N \Delta_l) = \bm{b}_H(\tau)^\top (\tilde \gamma_{li}^{(j)} - \tilde \gamma_{0i}) / ( N \Delta_l) = 0$ occurs with probability approaching one.
\end{proof}


Define $\tilde{\bm{\alpha}}_{1i}^* \coloneqq \bm{J}_{i, TL}^{-1} \sum_{\ell \in [L]} \sum_{t \in [T]} \bar \xi_{1,it}(\tau_\ell) \psi_{\tau_\ell}(\epsilon^*_{it}(\tau_\ell))/(LT)$, $\tilde{\bm{\beta}}_{1i}^* = \bm{\beta}_{1i} + \tilde{\bm{\alpha}}_{1i}^*$.
The next lemma shows that $\sqrt{T}(\tilde{\bm{\beta}}_{1i} - \bm{\beta}_{1i})$ and $\sqrt{T}(\tilde{\bm{\beta}}_{1i}^* - \bm{\beta}_{1i})$ are asymptotically equivalent, where $\tilde{\bm{\beta}}_{1i}$ is a subvector of $\tilde{\bm \gamma}_i$, defined similarly to $\bm{\beta}_{1i}$.
    
\begin{lemma}\label{lem:bahadur}
    Suppose the Assumptions \ref{as:coef} - \ref{as:tune} hold.
    In addition, if $(H^{3/2}\ln T)/\sqrt{T} \to 0$ is satisfied
    \begin{align}
        \sqrt{T}(\tilde{\bm{\beta}}_{1i} - \bm{\beta}_{1i}) = \bm{J}_{i, TL}^{-1} \sum_{\ell \in [L]} \sum_{t \in [T]} \bar \xi_{1,it}(\tau_\ell) \psi_{\tau_\ell}(\epsilon^*_{it}(\tau_\ell))/(L T^{1/2}) + o_P(1).
    \end{align}
\end{lemma}

\begin{proof}
Let $\tilde{\bm{\eta}}_i = \bm{\eta}(\tilde{\bm{\beta}}_{1i}, \tilde{\bm{\beta}}_{2i})$ and observe that
\begin{align}
    y_{it} - \xi_t(\tau)^\top \tilde{\bm \gamma}_i
    = \epsilon_{it}^*(\tau) - \bar\xi_{1,it}(\tau)^\top (\tilde{\bm{\beta}}_{1i} - \bm{\beta}_{1i}) - \xi_{2,it}(\tau_\ell)^\top \tilde{\bm{\eta}}_i - v_{it}(\tau_\ell).
\end{align}
Hence, 
\begin{align}
    Q_{TL}(\tilde{\bm \gamma}_i) = \tilde Q_{TL}(\tilde{ \bm \beta}_{1i}, \tilde{\bm{\eta}}_i) + P_\lambda(\tilde{\bm \gamma}_i),
\end{align}
where
\begin{align}
    \tilde Q_{TL}(\bm{\beta}_1, \bm{\eta}) \coloneqq \frac{1}{LT} \sum_{\ell \in [L]} \sum_{t \in [T]} \rho_{\tau_\ell}\left( \epsilon_{it}^*(\tau_\ell) - \bar\xi_{1,it}(\tau_\ell)^\top (\bm{\beta}_1 - \bm{\beta}_{1i}) - \xi_{2,it}(\tau_\ell)^\top \bm{\eta} - v_{it}(\tau_\ell) \right).
\end{align}
By definition of $\tilde{\bm{\beta}}_{1i}$ and its consistency and oracle property, $\tilde Q_{TL}(\bm{\beta}_1, \tilde{\bm{\eta}}_i)$ is minimized at $\tilde{\bm{\beta}}_{1i}$ with probability approaching one.
Thus, since $||\tilde{\bm{\eta}}_i|| = O_P(r_{TH}) = O_P(T^{-1/2})$, if we can show
\begin{align}
    \Pr\left( \inf_{ ||\bm{\beta}_1 - \tilde{\bm{\beta}}^*_{1i}|| = e/\sqrt{T}, \: ||\bm \eta || \lesssim 1/\sqrt{T}} \tilde Q_{TL}(\bm{\beta}_1, \bm{\eta}) \ge \tilde Q_{TL}(\tilde{\bm{\beta}}_{1i}^*, \bm{\eta}) \right)\to 1
\end{align}
for any $e > 0$, we can conclude that $||\tilde{\bm{\beta}}_{1i} - \tilde{\bm{\beta}}^*_{1i}|| \le e/\sqrt{T}$ for any $e > 0$ with probability approaching one.

For notational simplicity, write
\begin{align}
    g_{it, \tau}(\bm \beta_1, \bm \eta)
    \coloneqq \bar\xi_{1,it}(\tau)^\top \bm{\beta}_1 + \xi_{2,it}(\tau)^\top \bm{\eta} + v_{it}(\tau).
\end{align}
Then, by setting
\begin{align}
    x & = \epsilon_{it}^*(\tau) - \bar\xi_{1,it}(\tau_\ell)^\top (\tilde{\bm{\beta}}_{1i}^* - \bm{\beta}_{1i}) - \xi_{2,it}(\tau)^\top \bm{\eta} - v_{it}(\tau) \\
    y & = \bar\xi_{1,it}(\tau_\ell)^\top \left( \bm{\beta}_1 - \tilde{\bm{\beta}}_{1i}^*\right) 
\end{align}
and applying \eqref{eq:knight}, for any $\bm{\beta}_1$ and $\bm{\eta}$,
\begin{align}
    \tilde Q_{TL}(\bm{\beta}_1, \bm{\eta}) - \tilde Q_{TL}(\tilde{\bm{\beta}}_{1i}^*, \bm{\eta}) 
    & = - \frac{1}{LT} \sum_{\ell \in [L]} \sum_{t \in [T]} \bar\xi_{1,it}(\tau_\ell)^\top \left( \bm{\beta}_1 - \tilde{\bm{\beta}}_{1i}^*\right) \psi_{\tau_\ell}(\epsilon_{it}^*(\tau_\ell) - g_{it, \tau_\ell}(\tilde{\bm{\beta}}_{1i}^* - \bm{\beta}_{1i}, \bm \eta))  \\
    & \quad + \frac{1}{LT} \sum_{\ell \in [L]} \sum_{t \in [T]} \int_0^{\bar\xi_{1,it}(\tau_\ell)^\top (\bm{\beta}_1 - \tilde{\bm{\beta}}_{1i}^*)} \left(\bm{1}\{\epsilon_{it}^*(\tau_\ell) \le g_{it, \tau_\ell}(\tilde{\bm{\beta}}_{1i}^* - \bm{\beta}_{1i}, \bm \eta) + s\} \right. \\
    & \phantom{+ \frac{1}{LT} \sum_{\ell \in [L]} \sum_{t \in [T]} }\left. - \bm{1}\{\epsilon_{it}^*(\tau_\ell) \le g_{it, \tau_\ell}(\tilde{\bm{\beta}}_{1i}^* - \bm{\beta}_{1i}, \bm \eta)\} \right) \text{d}s \\
    & = \frac{1}{LT} \sum_{\ell \in [L]} \sum_{t \in [T]} \int_0^{\bar\xi_{1,it}(\tau_\ell)^\top (\bm{\beta}_1 - \tilde{\bm{\beta}}_{1i}^*)} \left(\bm{1}\{\epsilon_{it}^*(\tau_\ell) \le g_{it, \tau_\ell}(\tilde{\bm{\beta}}_{1i}^* - \bm{\beta}_{1i}, \bm \eta) + s\} - \tau_\ell \right) \text{d} s \\
    & = \frac{1}{LT} \sum_{\ell \in [L]} \sum_{t \in [T]} \underbracket{\int_{g_{it, \tau_\ell}(\tilde{\bm{\beta}}_{1i}^* - \bm{\beta}_{1i}, \bm \eta)}^{g_{it, \tau_\ell}(\bm{\beta}_1 - \bm{\beta}_{1i}, \bm \eta)} \left(\bm{1}\{\epsilon_{it}^*(\tau_\ell) \le s\} - \tau_\ell \right)\text{d}s}_{\coloneqq A_{it,\tau_\ell}(\bm{\beta}_1 - \bm{\beta}_{1i}, \tilde{\bm{\beta}}_{1i}^* - \bm{\beta}_{1i}, \bm{\eta})}.
\end{align}

Decompose $A_{it,\tau}(\bm{\beta}_1, \bm{\beta}_1', \bm{\eta}) = A_{1,it,\tau}(\bm{\beta}_1, \bm{\beta}_1', \bm{\eta}) + A_{2,it,\tau}(\bm{\beta}_1, \bm{\beta}_1', \bm{\eta}) + A_{3,it,\tau}(\bm{\beta}_1, \bm{\beta}_1', \bm{\eta})$, where
\begin{align}
    A_{1,it,\tau}(\bm{\beta}_1, \bm{\beta}_1', \bm{\eta}) 
    & \coloneqq \mathbb{E}[A_{it,\tau}(\bm{\beta}_1, \bm{\beta}_1', \bm{\eta}) \mid \mathcal{F}_{t-1}]\\
    A_{2,it,\tau}(\bm{\beta}_1, \bm{\beta}_1') 
    & \coloneqq - (\bm{\beta}_1 - \bm{\beta}_1')^\top \bar\xi_{1,it}(\tau) \psi_{\tau}(\epsilon_{it}^*(\tau))\\
    A_{3,it,\tau}(\bm{\beta}_1, \bm{\beta}_1', \bm{\eta}) 
    & \coloneqq A_{it,\tau}(\bm{\beta}_1, \bm{\beta}_1', \bm{\eta}) - \mathbb{E}[A_{it,\tau}(\bm{\beta}_1, \bm{\beta}_1', \bm{\eta}) \mid \mathcal{F}_{t-1}] + (\bm{\beta}_1 - \bm{\beta}_1')^\top \bar\xi_{1,it}(\tau) \psi_{\tau}(\epsilon_{it}^*(\tau)).
\end{align}
First, by Taylor expansion and Lipschitz continuity of $f_{it}$, noting the orthogonality between $\bar \xi_1$ and $\xi_2$,
\begin{align}\begin{split}\label{eq:g^3}
    & \frac{1}{LT} \sum_{\ell \in [L]} \sum_{t \in [T]} A_{1,it,\tau_\ell}(\bm{\beta}_1, \bm{\beta}_1', \bm{\eta}) \\
    & = \frac{1}{LT} \sum_{\ell \in [L]} \sum_{t \in [T]} \int_{g_{it, \tau_\ell}(\bm{\beta}_1', \bm \eta)}^{g_{it, \tau_\ell}(\bm{\beta}_1, \bm \eta)} \left(F_{it}(m_{it}(\tau_\ell) + s \mid \mathcal{F}_{t-1})  - F_{it}(m_{it}(\tau_\ell) \mid \mathcal{F}_{t-1}) \right)\text{d}s \\
    & = \frac{1}{LT} \sum_{\ell \in [L]} \sum_{t \in [T]} \int_{g_{it, \tau_\ell}(\bm{\beta}_1', \bm \eta)}^{g_{it, \tau_\ell}(\bm{\beta}_1, \bm \eta)}\left( f_{it}(m_{it}(\tau_\ell) \mid \mathcal{F}_{t-1})s + O(1) s^2 \right)\text{d}s \\
    & = \frac{1}{2} \bm{\beta}_1^\top \left( \frac{1}{LT} \sum_{\ell \in [L]} \bar{\bm{\Xi}}_{1,i\tau_\ell}^\top \bm{F}_{\tau_\ell} \bar{\bm{\Xi}}_{1,i\tau_\ell} \right) \bm{\beta}_1 - \frac{1}{2} \bm{\beta}_1^{\prime \top} \left( \frac{1}{LT} \sum_{\ell \in [L]} \bar{\bm{\Xi}}_{1,i\tau_\ell}^\top \bm{F}_{\tau_\ell} \bar{\bm{\Xi}}_{1,i\tau_\ell} \right) \bm{\beta}_1' \\
    & \quad + \frac{1}{2LT} \sum_{\ell \in [L]} \sum_{t \in [T]} f_{it}(m_{it}(\tau_\ell) \mid \mathcal{F}_{t-1}) \bar\xi_{1,it}(\tau_\ell)^\top (\bm{\beta}_1 - \bm{\beta}_1') v_{it}(\tau_\ell) \\
    & \quad + O\Bigl( \frac{1}{LT} \sum_{\ell \in [L]} \sum_{t \in [T]} \left[ |\bar\xi_{1,it}(\tau_\ell)^\top (\bm{\beta}_1 - \bm{\beta}_1')|^3 + |\bar\xi_{1,it}(\tau_\ell)^\top (\bm{\beta}_1 - \bm{\beta}_1')|^2 \cdot |g_{it, \tau_\ell}(\bm{\beta}_1', \bm \eta)| \right. \\
    & \qquad \qquad \left. + |\bar\xi_{1,it}(\tau_\ell)^\top (\bm{\beta}_1 - \bm{\beta}_1')| \cdot |g_{it, \tau_\ell}(\bm{\beta}_1', \bm \eta)|^2 \right] \Bigr).
\end{split}\end{align}
For any $|| \bm{\beta}_1 - \bm{\beta}_1' || \lesssim 1/\sqrt{T}$, $||\bm{\beta}_1'|| \lesssim 1/\sqrt{T}$, and $||\bm{\eta}|| \lesssim 1/\sqrt{T}$, by a similar argument to \eqref{eq:CS1}, we can find that
\begin{align}
    & \frac{1}{LT} \sum_{\ell \in [L]} \sum_{t \in [T]} \left|  f_{it}(m_{it}(\tau_\ell) \mid \mathcal{F}_{t-1}) \bar\xi_{1,it}(\tau_\ell)^\top (\bm{\beta}_1 - \bm{\beta}_1') v_{it}(\tau_\ell) \right| \lesssim T^{-1/2} H^{-\pi} \\
    & \frac{1}{LT} \sum_{\ell \in [L]} \sum_{t \in [T]} |\bar\xi_{1,it}(\tau_\ell)^\top (\bm{\beta}_1 - \bm{\beta}_1')|^3 \lesssim H^{1/2} T^{-3/2} \\
    & \frac{1}{LT} \sum_{\ell \in [L]} \sum_{t \in [T]} |\bar\xi_{1,it}(\tau_\ell)^\top (\bm{\beta}_1 - \bm{\beta}_1')|^2 \cdot |g_{it, \tau_\ell}(\bm{\beta}_1', \bm \eta)| \lesssim T^{-1}(\sqrt{H/T} + H^{-\pi}) \\
    & \frac{1}{LT} \sum_{\ell \in [L]} \sum_{t \in [T]} |\bar\xi_{1,it}(\tau_\ell)^\top (\bm{\beta}_1 - \bm{\beta}_1')| \cdot |g_{it, \tau_\ell}(\bm{\beta}_1', \bm \eta)|^2 \lesssim T^{-1/2} (H/T + H^{-2\pi})
\end{align}
Hence, in view of Lemma \ref{lem:LLN}(iii), we have
\begin{align}
    &\frac{1}{LT} \sum_{\ell \in [L]} \sum_{t \in [T]} A_{1,it,\tau_\ell}(\bm{\beta}_1 - \bm{\beta}_{1i}, \tilde{\bm{\beta}}_{1i}^* - \bm{\beta}_{1i}, \bm{\eta})\\
    &= \frac{1}{2} (\bm{\beta}_1 - \bm{\beta}_{1i})^\top \bm{J}_{i, TL} (\bm{\beta}_1 - \bm{\beta}_{1i}) - \frac{1}{2} \tilde{\bm{\alpha}}_{1i}^{*\top} \bm{J}_{i, TL} \tilde{\bm{\alpha}}_{1i}^* + o_P(T^{-1}).
\end{align}

Next, recalling that $\tilde{\bm{\alpha}}_{1i}^* \coloneqq \bm{J}_{i, TL}^{-1} \sum_{\ell \in [L]} \sum_{t \in [T]} \bar \xi_{1,it}(\tau_\ell) \psi_{\tau_\ell}(\epsilon^*_{it}(\tau_\ell))/(LT)$,
\begin{align}
    \frac{1}{LT} \sum_{\ell \in [L]} \sum_{t \in [T]} A_{2,it,\tau_\ell}(\bm{\beta}_1 - \bm{\beta}_{1i}, \tilde{\bm{\beta}}_{1i}^* - \bm{\beta}_{1i})
    & = - (\bm{\beta}_1 - \tilde{\bm{\beta}}_{1i}^*)^\top \frac{1}{LT} \sum_{\ell \in [L]} \sum_{t \in [T]} \bar \xi_{1,it}(\tau_\ell) \psi_{\tau_\ell}(\epsilon_{it}^*(\tau_\ell)) \\
    & = - (\bm{\beta}_1 - \tilde{\bm{\beta}}_{1i}^*)^\top \bm{J}_{i, TL} \tilde{\bm{\alpha}}_{1i}^*.
\end{align}
Meanwhile, by direct calculation,
\begin{align}
    \frac{1}{2} (\bm{\beta}_1 - \bm{\beta}_{1i} - \tilde{\bm{\alpha}}_{1i}^*)^\top \bm{J}_{i, TL} (\bm{\beta}_1 - \bm{\beta}_{1i} - \tilde{\bm{\alpha}}_{1i}^*) 
    & = \frac{1}{2} (\bm{\beta}_1 - \bm{\beta}_{1i})^\top \bm{J}_{i, TL} (\bm{\beta}_1 - \bm{\beta}_{1i}) + \frac{1}{2} \tilde{\bm{\alpha}}_{1i}^{*\top} \bm{J}_{i, TL} \tilde{\bm{\alpha}}_{1i}^* \\
    & \quad -  ((\bm{\beta}_1 - \tilde{\bm{\beta}}_{1i}^*) + (\tilde{\bm{\beta}}_{1i}^* -  \bm{\beta}_{1i}))^\top \bm{J}_{i, TL} \tilde{\bm{\alpha}}_{1i}^* \\
    & = \frac{1}{2} (\bm{\beta}_1 - \bm{\beta}_{1i})^\top \bm{J}_{i, TL} (\bm{\beta}_1 - \bm{\beta}_{1i}) - \frac{1}{2} \tilde{\bm{\alpha}}_{1i}^{*\top} \bm{J}_{i, TL} \tilde{\bm{\alpha}}_{1i}^* \\
    & \quad -  (\bm{\beta}_1 - \tilde{\bm{\beta}}_{1i}^*)^\top \bm{J}_{i, TL} \tilde{\bm{\alpha}}_{1i}^*.    
\end{align}
Hence,
\begin{align}
    & \frac{1}{LT} \sum_{\ell \in [L]} \sum_{t \in [T]} \left\{ A_{1,it,\tau_\ell}(\bm{\beta}_1 - \bm{\beta}_{1i}, \tilde{\bm{\beta}}_{1i}^* - \bm{\beta}_{1i}, \bm{\eta}) + A_{2,it,\tau_\ell}(\bm{\beta}_1 - \bm{\beta}_{1i}, \tilde{\bm{\beta}}_{1i}^* - \bm{\beta}_{1i}) \right\} \\
    & \quad =  \frac{1}{2} (\bm{\beta}_1 - \tilde{\bm{\beta}}_{1i}^*)^\top \bm{J}_{i, TL} (\bm{\beta}_1 - \tilde{\bm{\beta}}_{1i}^*)  + o_P(T^{-1}).
\end{align}

Now, write
\begin{align}
    A_{3,it, \tau}(\bm{\beta}_1, \bm{\beta}_1', \bm{\eta})
    & =  A_{it,\tau}(\bm{\beta}_1, \bm{\beta}_1', \bm{\eta}) - \mathbb{E}[A_{it,\tau}(\bm{\beta}_1, \bm{\beta}_1', \bm{\eta}) \mid \mathcal{F}_{t-1}] + (\bm{\beta}_1 - \bm{\beta}_1')^\top \bar\xi_{1,it}(\tau) \psi_{\tau}(\epsilon_{it}^*(\tau)) \\
    & = \int_{g_{it, \tau}(\bm{\beta}_1', \bm{\eta})}^{g_{it, \tau}(\bm{\beta}_1, \bm{\eta})} \left(\bm{1}\{\epsilon_{it}^*(\tau) \le s\} - F_{it}(m_{it}(\tau) + s \mid \mathcal{F}_{t-1}) \right)\text{d}s \\
    & \quad + \int_0^{\bar \xi_{1,it}(\tau)^\top (\bm{\beta}_1 - \bm{\beta}_1')} (\tau - \bm{1}\{\epsilon_{it}^*(\tau) \le 0\}) \text{d}s \\
    & = \underbracket{\int_{g_{it, \tau}(\bm{\beta}_1', \bm{\eta})}^{g_{it, \tau}(\bm{\beta}_1, \bm{\eta})} \left(\bm{1}\{\epsilon_{it}^*(\tau) \le s\} - \bm{1}\{\epsilon_{it}^*(\tau) \le 0\} \right)\text{d}s}_{\eqqcolon \: A_{31,it, \tau}(\bm{\beta}_1, \bm{\beta}_1', \bm{\eta})} \\
    & \quad + \underbracket{\int_{g_{it, \tau}(\bm{\beta}_1', \bm{\eta})}^{g_{it, \tau}(\bm{\beta}_1, \bm{\eta})} (F_{it}(m_{it}(\tau) \mid \mathcal{F}_{t-1}) - F_{it}(m_{it}(\tau) + s \mid \mathcal{F}_{t-1})) \text{d}s}_{\eqqcolon \: A_{32,it, \tau}(\bm{\beta}_1, \bm{\beta}_1', \bm{\eta})}.
\end{align}
Clearly $\{ L^{-1} \sum_{\ell \in [L]} A_{3,it, \tau_\ell}(\bm{\beta}_1, \bm{\beta}_1', \bm{\eta}) \}$ forms a MDS.
Thus, we aim to apply Freedman's inequality \eqref{eq:freedman} to this.
First, it is easy to see that $|A_{3,it, \tau_\ell}(\bm{\beta}_1, \bm{\beta}_1', \bm{\eta})| \lesssim \sqrt{H}||\bm{\beta}_1 - \bm{\beta}_1'||$ holds.

Meanwhile, for a general random variable $X$ and $a < b$, 
\begin{align}
    I(X) \coloneqq \int_a^b ( \bm{1}\{X \le s\} - \bm{1}\{X \le 0\} ) \text{d}s 
    & =
    \begin{cases}
        \overbracket{b - a}^{\text{val}_j}, & \overbracket{0 < X \le a, \phantom{xxxxx}}^{\text{int}_j}\\
        b - X, & \max\{0,a\} < X \le b,\\
        a - X, & a < X \le \min\{0,b\},\\
        a - b, & b < X \le 0 \\
        0, & \text{otherwise.}
    \end{cases} \\
    & = \sum_{j = 1}^4 \text{val}_j \bm{1}\{ \text{int}_j\}
\end{align}
holds.
Thus,
\begin{align}
    \mathbb{E}[I(X)^2 \mid \mathcal{F}_{t-1}]
    & = \sum_{j = 1}^4 \text{val}_j^2 \Pr(\text{int}_j \mid \mathcal{F}_{t-1}) \le \sum_{j=1}^4 (b - a)^2 \Pr(\text{int}_j \mid  \mathcal{F}_{t-1}).
\end{align}
When $a > b$, we can derive an analogous expression.
Now, in the current context, we can set $X = \epsilon_{it}^*(\tau)$, $a = g_{it, \tau}(\bm{\beta}_1', \bm{\eta})$ and $b = g_{it, \tau}(\bm{\beta}_1, \bm{\eta})$.
Then, $(b - a)^2 = (\bar\xi_{1,it}(\tau)^\top (\bm{\beta}_1 - \bm{\beta}_1'))^2$ and $\Pr(\text{int}_j \mid \mathcal{F}_{t-1})$ is of order either $O(g_{it, \tau}(\bm{\beta}_1', \bm{\eta}))$, $O(g_{it, \tau}(\bm{\beta}_1, \bm{\eta}))$, or $O(\bar\xi_{1,it}(\tau)^\top (\bm{\beta}_1 - \bm{\beta}_1'))$.
Therefore,
\begin{align}
    &\frac{1}{LT}\sum_{\ell \in [L]} \sum_{t \in [T]} \mathbb{E}\left[A_{31,it, \tau_\ell}(\bm{\beta}_1, \bm{\beta}_1', \bm{\eta})^2 \mid \mathcal{F}_{t-1}\right] \\
    & \lesssim \frac{1}{LT}\sum_{\ell \in [L]}\sum_{t \in [T]} \left( |\bar\xi_{1,it}(\tau_\ell)^\top (\bm{\beta}_1 - \bm{\beta}_1')|^3 + |\bar\xi_{1,it}(\tau_\ell)^\top (\bm{\beta}_1 - \bm{\beta}_1')|^2 |g_{it, \tau_\ell}(\bm{\beta}_1, \bm{\eta})| \right.\\
    & \quad \left. + |\bar\xi_{1,it}(\tau_\ell)^\top (\bm{\beta}_1 - \bm{\beta}_1')|^2 |g_{it, \tau_\ell}(\bm{\beta}_1', \bm{\eta})| \right)  \lesssim H^{1/2} T^{-3/2}
\end{align}
under $|| \bm{\beta}_1 - \bm{\beta}_1' || \lesssim 1/\sqrt{T}$, $||\bm{\beta}_1'|| \lesssim 1/\sqrt{T}$, and $||\bm{\eta}|| \lesssim 1/\sqrt{T}$.
Moreover, by Cauchy-Schwarz inequality, we can easily find that
\begin{align}
    \frac{1}{LT}\sum_{\ell \in [L]} \sum_{t \in [T]} \mathbb{E}\left[A_{32,it, \tau_\ell}(\bm{\beta}_1, \bm{\beta}_1', \bm{\eta})^2 \mid \mathcal{F}_{t-1}\right] 
    & \le \frac{c}{LT}\sum_{\ell \in [L]} \sum_{t \in [T]} |\bar\xi_{1,it}(\tau_\ell)^\top (\bm{\beta}_1 - \bm{\beta}_1')| \left| \int_{g_{it, \tau}(\bm{\beta}_1', \bm{\eta})}^{g_{it, \tau}(\bm{\beta}_1, \bm{\eta})} s^2 \text{d} s\right| \\
    & = O(\sqrt{H/T}) \cdot O(H T^{-3/2})
\end{align}
under $||\bm{\beta}_1 - \bm{\beta}_1'|| \lesssim 1/\sqrt{T}$, similar to \eqref{eq:g^3}.
Hence, $L^{-1} \sum_{t \in [T]} \sum_{\ell \in [L]} \mathbb{E}\left[A_{3,it, \tau_\ell}(\bm{\beta}_1, \bm{\beta}_1', \bm{\eta})^2 \mid \mathcal{F}_{t-1}\right] \lesssim \sqrt{H/T}$.

Now, let
\begin{align}
    S_t(\vartheta)
    &\coloneqq \frac{1}{L}\sum_{\ell \in [L]} A_{3,it,\tau_\ell}(\vartheta) \\
    \mathcal{T}
    & \coloneqq \left\{\vartheta = (\bm\beta_1, \bm\beta_1', \bm\eta) \in \mathbb{R}^{\text{dim}(\vartheta)}:
      \|\bm\beta_1-\bm\beta_1'\|\le C_1/\sqrt{T},\;
      \|\bm\beta_1\|\le C_2/\sqrt{T},\;
      \|\bm\eta\|\le C_3/\sqrt{T}
     \right\}.
\end{align}
Choose an arbitrary $u > 0$ and place lattice points in $\mathcal T$ with equal side length $u/(T\sqrt{H})$.
The number of lattice points in $\mathcal T$ is bounded by $C(T\sqrt{H})/u)^{\text{dim}(\vartheta)}$ for some large constant $C$.
For a given lattice point $\vartheta_k$, by Freedman's inequality \eqref{eq:freedman},
\begin{align}
    \Pr\left(\left| \sum_{t \in [T]} S_t(\vartheta_k) \right|\ge u\right)
    \le 2 \exp\left( - \frac{u^2 \sqrt{T/H} }{c_1 + c_2 u} \right)
\end{align}
for any $u > 0$.
Thus, by Boole's inequality,
\begin{align}
    \Pr\left(\max_{k}\left| \sum_{t \in [T]} S_t(\vartheta_k) \right|\ge u
    \right) \le
    2 C(T\sqrt{H})/u)^{\text{dim}(\vartheta)}
    \exp\left( - \frac{u^2 \sqrt{T/H} }{c_1 + c_2 u} \right) \to 0
\end{align}
under $(H^{3/2} \ln T)/\sqrt{T} \to 0$.

On the other hand, if $\vartheta_k$ is the nearest lattice point to a given $\vartheta$,
\begin{align}
    \left|\sum_{t \in [T]} (S_t(\vartheta) - S_t(\vartheta_k)) \right| 
    & \lesssim \frac{1}{L}\sum_{\ell \in [L]}\sum_{t \in [T]} \left\{ |\bar\xi_{1,it}(\tau_\ell)^\top (\bm{\beta}_1 - \bm{\beta}_{1,k})| +  |\bar\xi_{1,it}(\tau_\ell)^\top (\bm{\beta}_1' - \bm{\beta}_{1,k}')|  + \xi_{2,it}(\tau_\ell)^\top (\bm{\eta} - \bm \eta_k) |\right\} \\
    & \lesssim T || \vartheta - \vartheta_k|| \le u.
\end{align}
Since $u$ is arbitrary, combining the above arguments yields
\begin{align}
    \sup_{\vartheta\in\mathcal T}\left| \frac{1}{T} \sum_{t \in [T]} S_t(\vartheta) \right| 
    & = \sup_{\|\bm\beta_1-\bm\beta_1'\|\le C_1/\sqrt{T},\;
      \|\bm\beta_1\|\le C_2/\sqrt{T},\;
      \|\bm\eta\|\le C_3/\sqrt{T}}\left| \frac{1}{LT}\sum_{\ell \in [L]}\sum_{t \in [T]} A_{3,it,\tau_\ell}(\bm\beta_1, \bm\beta_1', \bm \eta) \right| \\
    & = o_P(T^{-1}).
\end{align}
Consequently, for any $|| \bm{\beta}_1 - \tilde{\bm{\beta}}_{1i}^* || \lesssim 1/\sqrt{T}$ and $||\bm{\eta}|| \lesssim 1/\sqrt{T}$,
\begin{align}
    \tilde Q_{TL}(\bm{\beta}_1, \tilde{\bm{\eta}}_i) - \tilde Q_{TL}(\tilde{\bm{\beta}}_{1i}^*, \tilde{\bm{\eta}}_i)
    & = \frac{1}{LT} \sum_{\ell \in [L]} \sum_{t \in [T]} A_{it, \tau_\ell}(\bm \beta_1 - \bm \beta_{1i}, \tilde{\bm{\beta}}_{1i}^* - \bm \beta_{1i}, \bm \eta) \\
    & = \frac{1}{2} (\bm{\beta}_1 -  \tilde{\bm{\beta}}_{1i}^*)^\top \bm{J}_{i, TL} (\bm{\beta}_1 - \tilde{\bm{\beta}}_{1i}^*)  + o_P(T^{-1}) \\
    & \ge 0
\end{align}
with probability approaching one.
This implies the desired result.

\end{proof}

\begin{proposition}\label{prop:normality}
    Suppose the assumptions of Lemma \ref{lem:bahadur} hold. 
    In addition, if $T^{1/2} H^{-\pi}/ || \bm b_H(\tau) || \to 0$ for a given $\tau \in (0,1)$, for $(l, j) \in \mathcal S_{1i}$, 
    \begin{align}
    \text{(i)} \quad
    & \frac{\sqrt{T}(\tilde \phi_{0i}(\tau) - \phi_{0i}(\tau))}{\sigma_{0i}(\tau)} \overset{d}{\to} N(0, 1) \\
    \text{(ii)} \quad
    & \frac{\sqrt{T}(\tilde \phi_{li}^{(j)}(\tau) - \phi_{li}^{(j)}(\tau))}{\sigma_{li}^{(j)}(\tau)} \overset{d}{\to} N(0, 1)
    \end{align}
\end{proposition}

\begin{proof}
Since the proofs are similar, we only show (i).
Observe
\begin{align}
    \frac{\sqrt{T}(\tilde \phi_{0i}(\tau) - \phi_{0i}(\tau))}{\sigma_{0i}(\tau)}
    & = \frac{\sqrt{T}\bm b_H(\tau)^\top (\tilde{\bm \gamma}_{0i} - \bm \gamma_{0i})}{\sigma_{0i}(\tau)} + \frac{\sqrt{T} (\bm b_H(\tau)^\top \bm \gamma_{0i} - \phi_{0i}(\tau))}{\sigma_{0i}(\tau)} \\
    & = \frac{\bm b_H(\tau)^\top \mathbb{S}_0 \bm{J}_{i, TL}^{-1} \sum_{\ell \in [L]} \sum_{t \in [T]} \bar \xi_{1,it}(\tau_\ell) \psi_{\tau_\ell}(\epsilon^*_{it}(\tau_\ell))/(L T^{1/2})}{\sigma_{0i}(\tau)} \\
    & \quad + \frac{o_P(||\bm b_H(\tau)||)}{\sigma_{0i}(\tau)} + \frac{O(T^{1/2} H^{-\pi})}{\sigma_{0i}(\tau)} 
\end{align}
By Assumptions \ref{as:matrix}(ii) and (iii), we have $[\sigma_{0i}(\tau)]^2 \gtrsim ||\bm b_H(\tau) ||^2$.
Thus, under $T^{1/2} H^{-\pi}/ || \bm b_H(\tau) || \to 0$, we have 
\begin{align}
    \frac{\sqrt{T}(\tilde \phi_{0i}(\tau) - \phi_{0i}(\tau))}{\sigma_{0i}(\tau)}
    & = \frac{\bm b_H(\tau)^\top \mathbb{S}_0 \bm{J}_{i, TL}^{-1} \sum_{\ell \in [L]} \sum_{t \in [T]} \bar \xi_{1,it}(\tau_\ell) \psi_{\tau_\ell}(\epsilon^*_{it}(\tau_\ell))/(L T^{1/2})}{\sigma_{0i}(\tau)} + o_P(1).
\end{align}
Let $a_{it} \coloneqq \bm b_H(\tau)^\top \mathbb{S}_0 \bm{J}_{i, TL}^{-1} \sum_{\ell \in [L]} \bar \xi_{1,it}(\tau_\ell) \psi_{\tau_\ell}(\epsilon^*_{it}(\tau_\ell))/(\sigma_{0i}(\tau) L T^{1/2})$, so that
\begin{align}
    \frac{\sqrt{T}(\tilde \phi_{0i}(\tau) - \phi_{0i}(\tau))}{\sigma_{0i}(\tau)}  = \sum_{t \in [T]} a_{it} + o_P(1).
\end{align}
Since $\{a_{it}\}$ is a MDS, it suffices to verify the following two conditions for the central limit theorem for MDS by \cite{scott1973central}:
\begin{align}
    (1) \;\; & \sum_{t \in [T]} \mathbb{E}[a_{it}^2 \mid \mathcal{F}_{t - 1}] \overset{p}{\to} 1 \\
    (2) \;\; & \sum_{t \in [T]} \mathbb{E}[a_{it}^2 \bm{1}\{|a_{it}| \ge e\} \mid \mathcal{F}_{t - 1}] \overset{p}{\to} 0 \;\; \text{for any $e> 0$}
\end{align}
From a similar calculation as in \eqref{eq:markov}, (1) can be easily verified.
To verify (2), it suffices to show that $\mathbb{E}[a_{it}^4 \mid \mathcal{F}_{t - 1}] = o(T^{-1})$:
\begin{align}
    \mathbb{E}[a_{it}^4 \mid \mathcal{F}_{t - 1}]
    & \lesssim \frac{\left( \sum_{\ell_1, \ell_2 = 1}^L \bm b_H(\tau)^\top \mathbb{S}_0 \bm{J}_{i, TL}^{-1}  \bar \xi_{1,it}(\tau_{\ell_1})  \bar \xi_{1,it}(\tau_{\ell_2})^\top \bm{J}_{i, TL}^{-1} \mathbb{S}_0^\top \bm b_H(\tau) \right)^2}{||\bm b_H(\tau) ||^4 L^4 T^2} \\
    & \lesssim \frac{H^2 \left( \bm b_H(\tau)^\top \mathbb{S}_0 \bm{J}_{i, TL}^{-1} \bm{J}_{i, TL}^{-1} \mathbb{S}_0^\top \bm b_H(\tau) \right)^2}{||\bm b_H(\tau) ||^4 T^2} \\  
    & \lesssim H^2/T^2 = o(T^{-1}).
\end{align}
This completes the proof.
\end{proof}


\begin{corollary}\label{cor:normality} 
    Under the assumptions of Proposition \ref{prop:normality},
    \begin{align}
    & \text{(i)} \quad
    \frac{\sqrt{T}(\tilde \theta_{0i}(\tau) - \theta_{0i}(\tau))}{\nu_{0i}(\tau)} \overset{d}{\to} N(0, 1) \\
    &\text{(ii)} \quad
    \frac{N \Delta_l\sqrt{T}(\tilde \theta_{li}^{(j)}(\tau) - \theta_{li}^{(j)}(\tau))}{\nu_{li}^{(j)}(\tau)} \overset{d}{\to} N(0, 1) \;\; \text{for $(l, j) \in \mathcal S_{1i}$} 
    \end{align}
\end{corollary}

\begin{proof}

First, recall the following definitions:
\begin{align}
    \tilde{\mathcal S}_{1i}
    & \coloneqq \left\{(l, j) \in \mathcal S: \: \left\| \tilde \phi_{li}^{(j)} - \tilde \phi_{0i}\right\|_2 \neq 0 \right\}\\
    \tilde \theta_{li}^{(j)}(\tau) 
    & \coloneqq \frac{\tilde \phi_{li}^{(j)}(\tau) - \tilde \phi_{0i}(\tau)}{N \Delta_l}  \\
    \tilde \theta_{0i}(\tau)
    & \coloneqq \tilde \phi_{0i}(\tau) - \sum_{(l, j) \in \tilde{\mathcal S}_{1i}} \text{lb}_l \tilde \theta_{li}^{(j)}(\tau) \\
    & = \tilde \phi_{0i}(\tau) - \sum_{(l, j) \in \tilde{\mathcal S}_{1i}} \frac{\text{lb}_l}{N \Delta_l} \tilde \phi_{li}^{(j)}(\tau) + \sum_{(l, j) \in \tilde{\mathcal S}_{1i}} \frac{\text{lb}_l}{N \Delta_l}\tilde \phi_{0i}(\tau) \\
    & = \tilde \kappa_0 \tilde \phi_{0i}(\tau) - \sum_{(l, j) \in \tilde{\mathcal S}_{1i}} \kappa_l \tilde \phi_{li}^{(j)}(\tau).
\end{align}

(i) Conditional on the event $\{\tilde{\mathcal S}_{1i} = \mathcal S_{1i}\}$,
\begin{align}
    &\sqrt{T}(\tilde \theta_{0i}(\tau) - \theta_{0i}(\tau))\\
    & = \kappa_0 \sqrt{T}(\tilde \phi_{0i}(\tau) - \phi_{0i}(\tau)) - \sum_{(l, j) \in \mathcal S_{1i}} \kappa_l \sqrt{T}(\tilde \phi_{li}^{(j)}(\tau) - \phi_{li}^{(j)}(\tau)) \\
    & = \bm b_H(\tau)^\top \left( \kappa_0 \mathbb{S}_0 - \sum_{(l, j) \in \mathcal S_{1i}} \kappa_l \mathbb{S}_l^{(j)} \right) \bm{J}_{i, TL}^{-1} \sum_{\ell \in [L]} \sum_{t \in [T]} \bar \xi_{1,it}(\tau_\ell) \psi_{\tau_\ell}(\epsilon^*_{it}(\tau_\ell))/(L T^{1/2}) + o_P(1) \\
    & \overset{d}{\to} N\left(0,\lim_{T \to \infty} [\nu_{0i}(\tau)]^2 \right).
\end{align}
In view of Corollaries \ref{cor:convrate} and \ref{cor:lag}, $\Pr(\{\tilde{\mathcal S}_{1i} = \mathcal S_{1i}\}) \to 1$, and thus the proof is completed.

\bigskip

(ii) Similarly to (i), we have
\begin{align}
    &\sqrt{T}(\tilde \theta_{li}^{(j)}(\tau) - \theta_{li}^{(j)}(\tau))\\
    & = \frac{\sqrt{T}(\tilde \phi_{li}^{(j)}(\tau) - \phi_{li}^{(j)}(\tau))}{N \Delta_l} - \frac{\sqrt{T}(\tilde \phi_{0i}(\tau) - \phi_{0i}(\tau))}{N \Delta_l} \\
    & = \frac{\bm b_H(\tau)^\top ( \mathbb{S}_l^{(j)} - \mathbb{S}_0 ) \bm{J}_{i, TL}^{-1} \sum_{\ell \in [L]} \sum_{t \in [T]} \bar \xi_{1,it}(\tau_\ell) \psi_{\tau_\ell}(\epsilon^*_{it}(\tau_\ell))/(L T^{1/2})}{N \Delta_l} + o_P(1) \\
    & \overset{d}{\to} N\left(0,\lim_{T \to \infty}\frac{[\nu_{li}^{(j)}(\tau)]^2}{(N \Delta_l)^2} \right).
\end{align}
\end{proof}


\begin{flushleft}
\textbf{Proof of Theorems \ref{thm:convrate}, \ref{thm:lag}, and \ref{thm:normality}}    
\end{flushleft}

By Proposition \ref{prop:localmin_rate}, and the interior point condition in Assumption \ref{as:basis}(ii), $\Pr\left(\tilde{\bm\gamma}_i\in\mathcal G_H\right) \to 1$.
On this event, the unconstrained minimizer is also a constrained minimizer because $\mathcal G_H\subset\mathcal R_H$ and thus they are asymptotically equivalent.
Theorems \ref{thm:convrate}, \ref{thm:lag}, and \ref{thm:normality} therefore follow respectively from Corollaries \ref{cor:convrate}, \ref{cor:lag}, and \ref{cor:normality}.
\qed


\section{Proof of Theorem \ref{thm:BIC}}\label{app:proof-bic}

\begin{align}
    \bm{\gamma}_i(\mathcal S_1)
    & = \argmin_{\bm{\gamma} \in \mathcal{G}_H} \frac{1}{LT} \sum_{\ell \in [L]} \sum_{t \in [T]} \mathbb{E}[\rho_{\tau_\ell}\left( y_{it} - \xi_t(\tau_\ell)^\top \bm{\gamma}\right)] \; \text{subject to} \; \gamma_l^{(j)} = \gamma_0 \; \text{for} \; (l,j) \not\in \mathcal S_1\\
    \hat{\bm{\gamma}}_i(\mathcal S_1)
    & = \argmin_{\bm{\gamma} \in \mathcal{G}_H} \frac{1}{LT} \sum_{\ell \in [L]} \sum_{t \in [T]} \rho_{\tau_\ell}\left( y_{it} - \xi_t(\tau_\ell)^\top \bm{\gamma}\right) \; \text{subject to} \; \gamma_l^{(j)} = \gamma_0 \; \text{for} \; (l,j) \not\in \mathcal S_1\\
    {\rm BIC} (\mathcal S_1)
    & \coloneqq \ln\left( \frac{1}{LT} \sum_{\ell \in [L]} \sum_{t \in [T]} \rho_{\tau_\ell}\left( y_{it} - \xi_t(\tau_\ell)^\top \hat{ \bm \gamma}_i(\mathcal S_1) \right) \right) + \frac{|\mathcal S_1|H \ln T}{2 T} \\
    {\rm BIC}_i
    & \coloneqq \ln\left( \frac{1}{LT} \sum_{\ell \in [L]} \sum_{t \in [T]} \rho_{\tau_\ell}\left( y_{it} - \xi_t(\tau_\ell)^\top \bm \gamma_i \right) \right) + \frac{|\mathcal S_{1i}| H \ln T}{2 T}
\end{align}
Note that $\bm{\gamma}_i(\mathcal S_{1i}) \neq \bm{\gamma}_i$ in general.
\begin{align}
    \xi_t(\tau)^\top \bm{\gamma} 
    & = c_{0t}b_H(\tau)^\top \gamma_0 + \sum_{(l,j) \in \mathcal S_1} c_{l,t-j} b_H(\tau)^\top \gamma_l^{(j)} + \sum_{(l,j) \not\in \mathcal S_1} c_{l,t-j} b_H(\tau)^\top \gamma_0 \\
    & \eqqcolon \xi_t(\tau, \mathcal S_1)^\top \bm \delta \\
    \xi_t(\tau)^\top \bm{\gamma}_i
    & = c_{0t}b_H(\tau)^\top \gamma_{0i} + \sum_{(l,j) \in \mathcal S_{1i}} c_{l,t-j} b_H(\tau)^\top \gamma_{li}^{(j)} + \sum_{(l,j) \not\in \mathcal S_{1i}} c_{l,t-j} b_H(\tau)^\top \gamma_{0i} \\
    & \eqqcolon \xi_t(\tau, \mathcal S_{1i})^\top \bm \beta_{1i}
\end{align}

\begin{align}
    \bm{\delta}_i(\mathcal S_1)
    & = \argmin_{\bm \delta \in \prod_{j=1}^{|\mathcal S_1| + 1}\mathcal{G}_{j,H}} \frac{1}{LT} \sum_{\ell \in [L]} \sum_{t \in [T]} \mathbb{E}[\rho_{\tau_\ell}\left( y_{it} - \xi_t(\tau_\ell, \mathcal S_1)^\top \bm \delta\right)] \\
    \hat{\bm \delta}_i(\mathcal S_1)
    & = \argmin_{\bm \delta \in \prod_{j=1}^{|\mathcal S_1| + 1}\mathcal{G}_{j,H}} \frac{1}{LT} \sum_{\ell \in [L]} \sum_{t \in [T]} \rho_{\tau_\ell}\left( y_{it} - \xi_t(\tau_\ell, \mathcal S_1)^\top \bm \delta\right) 
\end{align}
By construction,
\begin{align}
    \xi_t(\tau)^\top \bm{\gamma}_i(\mathcal S_1) = \xi_t(\tau, \mathcal S_1)^\top \bm{\delta}_i(\mathcal S_1) \; \text{and} \; \xi_t(\tau)^\top \hat{\bm \gamma}_i(\mathcal S_1) = \xi_t(\tau, \mathcal S_1)^\top \hat{\bm \delta}_i(\mathcal S_1)
\end{align}
hold.
Let $\epsilon_{it}(\tau, \mathcal S_1) \coloneqq y_{it} - \xi_t(\tau)^\top \bm{\gamma}_i(\mathcal S_1) = y_{it} - \xi_t(\tau, \mathcal S_1)^\top \bm{\delta}_i(\mathcal S_1)$.

\begin{lemma}\label{lem:underfit}
    Suppose that $\mathcal S_{1i}\nsubseteq\mathcal S_1$.
    Under Assumptions \ref{as:dist}(i), (ii), \ref{as:basis}(ii), \ref{as:matrix}(i), and \ref{as:BIC}(iii), we have
    \begin{align}
    \frac{1}{LT} \sum_{\ell \in [L]} \sum_{t \in [T]} \mathbb{E}[\rho_{\tau_\ell}\left( y_{it} - \xi_t(\tau_\ell)^\top \bm{\gamma}_i(\mathcal S_1) \right)] >  \frac{1}{LT} \sum_{\ell \in [L]} \sum_{t \in [T]} \mathbb{E}[\rho_{\tau_\ell}\left( y_{it} - \xi_t(\tau_\ell)^\top \bm{\gamma}_i \right)] 
    \end{align}
    for all sufficiently large $T$.
\end{lemma}

\begin{proof}
Observe that
\begin{align}
    & \frac{1}{LT} \sum_{\ell \in [L]} \sum_{t \in [T]} 
    \mathbb{E} \left[\rho_{\tau_\ell} \left( y_{it} - \xi_t(\tau_\ell)^\top \bm{\gamma}_i(\mathcal S_1) \right) 
    - \rho_{\tau_\ell} \left( y_{it} - \xi_t(\tau_\ell)^\top \bm{\gamma}_i \right)\right] \\
    & = \frac{1}{LT} \sum_{\ell \in [L]} \sum_{t \in [T]} 
    \mathbb{E} \left[\rho_{\tau_\ell}\left( \epsilon_{it}(\tau_\ell) - \xi_t(\tau_\ell)^\top \bigl(\bm{\gamma}_i(\mathcal S_1) - \bm{\gamma}_i \bigr) \right) - \rho_{\tau_\ell}\left( \epsilon_{it}(\tau_\ell) \right)\right] = \mathbb{E}\left[D_{1,TL}\left(\bm{\gamma}_i(\mathcal S_1) - \bm{\gamma}_i\right)\right],
\end{align}
where the definition of $D_{1,TL}(\cdot)$ is given in the proof of Lemma \ref{lem:identification}.  
There, it is shown that $\mathbb{E}[D_{1,TL}(\bm \alpha)] \gtrsim ||\bm \alpha ||^2 - H^{-\pi}$ for nonzero $\bm \alpha$.  
Since $||\bm{\gamma}_i(\mathcal S_1) - \bm{\gamma}_i|| \ge c > 0$ whenever $\mathcal S_{1i}\nsubseteq\mathcal S_1$ by Assumption \ref{as:BIC}(iii), the result follows.
\end{proof}

\begin{lemma}\label{lem:convrate2}
    Suppose that the assumptions of Theorem \ref{thm:BIC} hold.
    Then, for all $\mathcal S_{1i}\nsubseteq\mathcal S_1$, $||\hat{\bm \gamma}_i(\mathcal S_1) - \bm \gamma_i(\mathcal S_1) || = O_P(\sqrt{H/T})$.
\end{lemma}

\begin{proof}
Since unpenalized quantile regression is a convex minimization problem, it is sufficient to show that for any given $c > 0$, there exists a large constant $K$ such that 
\footnotesize\begin{align}
    \Pr\left( \inf_{||\bm{k}|| = K} \frac{1}{LT} \sum_{\ell, t} 
    \rho_{\tau_\ell} \left( y_{it} - \xi_t(\tau_\ell, \mathcal S_1)^\top (\bm{\delta}_i(\mathcal S_1) + \bm k \sqrt{H/T} )  \right) > \frac{1}{LT} \sum_{\ell, t} \rho_{\tau_\ell} \left( y_{it} - \xi_t(\tau_\ell, \mathcal S_1)^\top \bm{\delta}_i(\mathcal S_1) \right) \right) \ge 1-c
\end{align}\small
for sufficiently large $T$.
By \eqref{eq:knight},
\begin{align}
    & \frac{1}{LT} \sum_{\ell \in [L]} \sum_{t \in [T]}  \rho_{\tau_\ell} \left( y_{it} - \xi_t(\tau_\ell, \mathcal S_1)^\top (\bm{\delta}_i(\mathcal S_1) + \bm k  \sqrt{H/T} )  \right) - \frac{1}{LT} \sum_{\ell \in [L]} \sum_{t \in [T]} \rho_{\tau_\ell} \left( y_{it} - \xi_t(\tau_\ell, \mathcal S_1)^\top \bm{\delta}_i(\mathcal S_1) \right) \\
    & = - \bm k^\top \left(\frac{1}{LT} \sum_{\ell \in [L]} \sum_{t \in [T]} \xi_t(\tau_\ell, \mathcal S_1) \psi_{\tau_\ell}(\epsilon_{it}(\tau_\ell, \mathcal S_1)) \right) \sqrt{H/T} \\
    & \quad + \frac{1}{LT} \sum_{\ell \in [L]} \sum_{t \in [T]} \int_0^{\bm k^\top \xi_t(\tau_\ell, \mathcal S_1) \sqrt{H/T} }\left( \bm{1}\{\epsilon_{it}(\tau_\ell, \mathcal S_1) \le x \} - \bm{1}\{\epsilon_{it}(\tau_\ell, \mathcal S_1) \le 0 \}\right) \text{d}x.
\end{align}
By the first-order condition of minimization, $(LT)^{-1} \sum_{\ell \in [L]} \sum_{t \in [T]} \mathbb{E}[\xi_t(\tau_\ell, \mathcal S_1) \psi_{\tau_\ell}(\epsilon_{it}(\tau_\ell, \mathcal S_1))] = \bm 0$.
Then, the same argument as in Lemma \ref{lem:LLN}, the law of large numbers gives
\begin{align}
    \left\| \frac{1}{LT} \sum_{\ell \in [L]} \sum_{t \in [T]} \xi_t(\tau_\ell, \mathcal S_1) \psi_{\tau_\ell}(\epsilon_{it}(\tau_\ell, \mathcal S_1))\right\| = O_P(\sqrt{H/T}),
\end{align}
and thus the first term is of order $(H/T)||\bm k||O_P(1)$ uniformly in $\bm k$.

For the second term, letting $d_{t,\tau_\ell}(\bm{k}, \mathcal S_1) \coloneqq \int_0^{\bm k^\top \xi_t(\tau_\ell, \mathcal S_1) \sqrt{H/T} }\left( \bm{1}\{\epsilon_{it}(\tau_\ell, \mathcal S_1) \le x \} - \bm{1}\{\epsilon_{it}(\tau_\ell, \mathcal S_1) \le 0 \}\right) \text{d}x$, under Assumption \ref{as:BIC},
\begin{align}
    &\frac{1}{LT} \sum_{\ell \in [L]} \sum_{t \in [T]} \mathbb{E}[d_{t,\tau_\ell}(\bm{k}, \mathcal S_1) \mid \mathcal{F}_{t-1}]\\
    & = \frac{1}{L T} \sum_{\ell \in [L]} \sum_{t \in [T]} \int_0^{\bm k^\top \xi_t(\tau_\ell, \mathcal S_1) \sqrt{H/T}} \mathbb{E}\left(\bm{1}\{\epsilon_{it}(\tau_\ell, \mathcal S_1) \le x\} - \bm{1}\{\epsilon_{it}(\tau_\ell, \mathcal S_1) \le 0\} \mid \mathcal{F}_{t-1}\right) \text{d}x \\
    & = \frac{1}{L T} \sum_{\ell \in [L]} \sum_{t \in [T]} \int_0^{\bm k^\top \xi_t(\tau_\ell, \mathcal S_1) \sqrt{H/T}} \left( G_{\mathcal S_1, it}(x \mid \mathcal{F}_{t-1}) - G_{\mathcal S_1, it}(0 \mid \mathcal{F}_{t-1}) \right) \text{d}x \\
    & =\frac{1}{L T} \sum_{\ell \in [L]} \sum_{t \in [T]} g_{\mathcal S_1, it}( 0 \mid \mathcal{F}_{t-1})  \int_0^{\bm k^\top \xi_t(\tau_\ell, \mathcal S_1) \sqrt{H/T}} x \text{d}x  +  \frac{1}{L T} \sum_{\ell \in [L]} \sum_{t \in [T]} \int_0^{\bm k^\top \xi_t(\tau_\ell, \mathcal S_1) \sqrt{H/T}} o(x) \text{d}x \\
    & = \frac{H/T}{2} \bm{k}^\top\left( \frac{1}{L T} \sum_{\ell \in [L]} \sum_{t \in [T]} g_{\mathcal S_1, it}(0 \mid \mathcal{F}_{t-1}) \xi_t(\tau_\ell, \mathcal S_1) \xi_t(\tau_\ell, \mathcal S_1)^\top \right)\bm{k} \\
    & \quad + (H/T) \cdot o\left( \bm{k}^\top\left( \frac{1}{L T} \sum_{\ell \in [L]} \sum_{t \in [T]} \xi_t(\tau_\ell, \mathcal S_1) \xi_t(\tau_\ell, \mathcal S_1)^\top \right)\bm{k} \right) \\
    & \ge c (H/T) ||\bm k ||^2
\end{align}
for some constant $c >0$, with probability approaching one.
In addition, following the same argument as in the proof of Proposition \ref{prop:localmin_rate}, $(LT)^{-1} \sum_{\ell \in [L]} \sum_{t \in [T]}(d_{t,\tau_\ell}(\bm{k}, \mathcal S_1) - \mathbb{E}[d_{t,\tau_\ell}(\bm{k}, \mathcal S_1) \mid \mathcal{F}_{t-1}]) = \sqrt{H/T} || \bm k || O_P(T^{-1/2})$.
Combining these results yields the desired result.

\end{proof}

On the other hand, suppose that $\mathcal S_1 \supseteq \mathcal S_{1i}$ is true, and let $\bm{\delta}_{\mathcal S_1, i}$ denote a proper rearrangement of $(\gamma_{0i}, \{\gamma_{li}^{(j)}\}_{(l,j) \in \mathcal S_1}, \{\gamma_{0i}\}_{(l,j) \notin \mathcal S_1})$ such that $\xi_t(\tau, \mathcal S_1)^\top \bm{\delta}_{\mathcal S_1, i} = \xi_t(\tau)^\top \bm \gamma_i$.
Then, we can show that
\begin{align}
    \frac{1}{LT} \sum_{\ell, t} \rho_{\tau_\ell} ( y_{it} - \xi_t(\tau_\ell, \mathcal S_1)^\top (\bm{\delta}_{\mathcal S_1, i} + \bm k /\sqrt{T} )) > \frac{1}{LT} \sum_{\ell, t} \rho_{\tau_\ell} ( y_{it} - \xi_t(\tau_\ell, \mathcal S_1)^\top \bm{\delta}_{\mathcal S_1, i})
\end{align}
with probability arbitrarily close to one, by the same argument as in Proposition \ref{prop:localmin_rate}.
That is, $||\hat{\bm \gamma}_i(\mathcal S_1) - \bm \gamma_i || = O_P(1/\sqrt{T})$ holds in this case.

\begin{flushleft}
\textbf{Proof of Theorem \ref{thm:BIC}}    
\end{flushleft}

First, suppose that $\mathcal S_{1i}\nsubseteq\mathcal S_1$.  
By arguments similar to those used above, we have
\begin{align}
    & \frac{1}{LT} \sum_{\ell \in [L]} \sum_{t \in [T]}  \rho_{\tau_\ell} \left( y_{it} - \xi_t(\tau_\ell, \mathcal S_1)^\top \hat{\bm \delta}_i(\mathcal S_1) \right) - \frac{1}{LT} \sum_{\ell \in [L]} \sum_{t \in [T]} \rho_{\tau_\ell} \left( y_{it} - \xi_t(\tau_\ell, \mathcal S_1)^\top \bm{\delta}_i(\mathcal S_1) \right) \\
    & = \frac{1}{LT} \sum_{\ell \in [L]} \sum_{t \in [T]}  \rho_{\tau_\ell} \left( \epsilon_{it}(\tau_\ell, \mathcal S_1) - \xi_t(\tau_\ell, \mathcal S_1)^\top \{ \hat{\bm \delta}_i(\mathcal S_1) - \bm \delta_i(\mathcal S_1) \}\right) - \frac{1}{LT} \sum_{\ell \in [L]} \sum_{t \in [T]} \rho_{\tau_\ell} \left( \epsilon_{it}(\tau_\ell, \mathcal S_1)\right) \\
    & = - \{ \hat{\bm \delta}_i(\mathcal S_1) - \bm \delta_i(\mathcal S_1) \}^\top \left(\frac{1}{LT} \sum_{\ell \in [L]} \sum_{t \in [T]} \xi_t(\tau_\ell, \mathcal S_1) \psi_{\tau_\ell}(\epsilon_{it}(\tau_\ell, \mathcal S_1)) \right) \\
    & \quad + \frac{1}{LT} \sum_{\ell \in [L]} \sum_{t \in [T]} \int_0^{\xi_t(\tau_\ell, \mathcal S_1)^\top \{ \hat{\bm \delta}_i(\mathcal S_1) - \bm \delta_i(\mathcal S_1) \}}\left( \bm{1}\{\epsilon_{it}(\tau_\ell, \mathcal S_1) \le x \} - \bm{1}\{\epsilon_{it}(\tau_\ell, \mathcal S_1) \le 0 \}\right) \text{d}x \\
    & = O_P(H/T)
\end{align}
by Lemma \ref{lem:convrate2}.  
Hence,
\footnotesize\begin{align}
    {\rm BIC}(\mathcal S_1) - {\rm BIC}_i
    & = \ln\left(1 + \frac{\frac{1}{LT} \sum_{\ell, t} \rho_{\tau_\ell}\left( y_{it} - \xi_t(\tau_\ell)^\top \hat{ \bm \gamma}_i(\mathcal S_1) \right) - \frac{1}{LT} \sum_{\ell, t} \rho_{\tau_\ell}\left( y_{it} - \xi_t(\tau_\ell)^\top \bm \gamma_i \right)}{ \frac{1}{LT} \sum_{\ell, t} \rho_{\tau_\ell}\left( y_{it} - \xi_t(\tau_\ell)^\top \bm \gamma_i \right)} \right) \\
    & \quad + \frac{(|\mathcal S_1| - |\mathcal S_{1i}|) H \ln T}{2 T} \\
    & = \ln\left(1 + \frac{\frac{1}{LT} \sum_{\ell, t} \rho_{\tau_\ell}\left( y_{it} - \xi_t(\tau_\ell)^\top \bm \gamma_i(\mathcal S_1) \right) - \frac{1}{LT} \sum_{\ell, t} \rho_{\tau_\ell}\left( y_{it} - \xi_t(\tau_\ell)^\top \bm \gamma_i \right) + O_P(H/T)}{ \frac{1}{LT} \sum_{\ell, t} \rho_{\tau_\ell}\left( y_{it} - \xi_t(\tau_\ell)^\top \bm \gamma_i \right)} \right) \\
    & \quad + \frac{(|\mathcal S_1| - |\mathcal S_{1i}|) H \ln T}{2 T} \\
    & > 0
\end{align}\small
with probability approaching one, where the last inequality follows from the law of large numbers and Lemma \ref{lem:underfit}.  
Since ${\rm BIC}(\mathcal S_{1i}) \le {\rm BIC}_i$ by definition, it follows that ${\rm BIC}(\mathcal S_1) - {\rm BIC}(\mathcal S_{1i}) > 0$.  

Next, consider the case $\mathcal S_1 \supset \mathcal S_{1i}$.  
Noting that $(LT)^{-1} \sum_{\ell, t}  [\rho_{\tau_\ell}( y_{it} - \xi_t(\tau_\ell)^\top \hat{\bm \gamma}_i(\mathcal S_1)) - \rho_{\tau_\ell}( y_{it} - \xi_t(\tau_\ell)^\top \bm \gamma_i )] = O_P(T^{-1})$ for all $\mathcal S_1 \supseteq \mathcal S_{1i}$, we have 
\begin{align}
    {\rm BIC}(\mathcal S_1) - {\rm BIC}(\mathcal S_{1i})
    & = \ln\left(1 + \frac{\frac{1}{LT} \sum_{\ell, t} \rho_{\tau_\ell}\left( y_{it} - \xi_t(\tau_\ell)^\top \hat{ \bm \gamma}_i(\mathcal S_1) \right) - \frac{1}{LT} \sum_{\ell, t} \rho_{\tau_\ell}\left( y_{it} - \xi_t(\tau_\ell)^\top \hat{ \bm \gamma}_i(\mathcal S_{1i}) \right)}{ \frac{1}{LT} \sum_{\ell, t} \rho_{\tau_\ell}\left( y_{it} - \xi_t(\tau_\ell)^\top \hat{ \bm \gamma}_i(\mathcal S_{1i}) \right)} \right) \\
    & \quad + \frac{(|\mathcal S_1| - |\mathcal S_{1i}|) H \ln T}{2 T} \\
    & = \ln\left(1 + O_P(T^{-1}) \right) + \frac{(|\mathcal S_1| - |\mathcal S_{1i}|) H \ln T}{2 T} \\
    & > 0
\end{align}
with probability approaching one.  
The above discussion shows that for both underfitted and overfitted cases with $\mathcal S_1 \neq \mathcal S_{1i}$, we have ${\rm BIC}(\mathcal S_1) - {\rm BIC}(\mathcal S_{1i}) > 0$ with probability approaching one.  

Meanwhile, let $\lambda^*$ be chosen to satisfy Assumption \ref{as:tune}(i).
Then, by Theorems \ref{thm:convrate} and \ref{thm:lag},
\begin{align}
    \Pr\left(\hat{\mathcal S}_{1i}(\lambda^*) = \mathcal S_{1i}\right) \to 1.
\end{align}
On this event, the penalized estimator with penalty $\lambda^*$ coincides with the unpenalized estimator under the model $\mathcal S_{1i}$, and therefore ${\rm BIC}(\lambda^*) = {\rm BIC}(\mathcal S_{1i})$ with probability approaching one.
Noting that, for any $\lambda \ge 0$ such that $\hat{\mathcal S}_{1i}(\lambda) \ne \mathcal S_{1i}$, ${\rm BIC}(\lambda) \ge {\rm BIC}(\hat{\mathcal S}_{1i}(\lambda))$ holds by construction, we have
\begin{align}
    {\rm BIC}(\lambda) \ge {\rm BIC}(\hat{\mathcal S}_{1i}(\lambda))
    >
    {\rm BIC}(\mathcal S_{1i})
    =
    {\rm BIC}(\lambda^*)
\end{align}
with probability approaching one.
Thus, if $\lambda$ selects an incorrect active set, it is not a minimizer of ${\rm BIC}(\lambda)$; that is, the minimizer $\hat\lambda$ of \eqref{eq:BIClambda} satisfies
\begin{align}
    \Pr\left(\hat{\mathcal S}_{1i}(\hat\lambda)=\mathcal S_{1i}\right)\to 1.
\end{align}
\qed

\section{Details of the Monte Carlo experiments in Section \ref{sec:MC}}\label{app:MC}

\subsection{Comparison with the standard QR approach}

In this experiment, we consider the following trivariate QVAR model with lag order $p=2$:
\begin{align}
    y_{it} = \theta_0(U_{it}) + \sum_{l = 1}^3 \theta_l^{(1)}(U_{it}) y_{l,t-1} + \sum_{l = 1}^3 \theta_l^{(2)}(U_{it}) y_{l,t-2}, \qquad i \in \{1,2,3\},
\end{align}
where $(U_{1t},U_{2t},U_{3t})$ are independent over time and jointly distributed according to a Gaussian copula with correlation $0.3$ and $\text{Uniform}[0,1]$ marginals, and
\begin{align}
    & \theta_0(\tau) = 1 + F_\beta^{-1}(\tau), \quad \theta_1^{(1)}(\tau) = (0.1 \tau + 0.2 \sqrt{\tau})/b, \\
    & \theta_2^{(1)}(\tau) = (0.1 \tau + 0.2 F_\beta^{-1}(\tau))/b, \quad \theta_3^{(1)}(\tau) = (0.1 \tau + 0.2\tau^2)/b,
\end{align}
and $\theta_1^{(2)}(\cdot) = \theta_2^{(2)}(\cdot) = \theta_3^{(2)}(\cdot) = 0$, where $F_\beta$ denotes the CDF of beta distribution with parameters $\alpha = 2$ and $\beta = 2$, and $b \in \{1,2, \ldots, 6\}$.
The true lag order is one: $\mathcal S_{1i} = \{(1,1), (2,1), (3,1)\}$, for $i = 1, 2, 3$.
Based on this DGP, we generate $T$ observations for each time series after a short ``burn-in'' phase, where $T \in \{200, 600, 1200\}$.

For estimation, we transform the model into SQVAR using the barycentric coordinate in \eqref{eq:max-min}, and 
estimate the SQVAR coefficients via a cubic I-spline approximation with $\sharp k \in \{1,2\}$ inner knots (i.e., $H \in \{5,6\}$).
The number of quantile grid points is $L = 30$.
To perform model selection based on the proposed BIC criterion, solving $\min_\lambda \mathrm{BIC}(\lambda)$ directly can be computationally burdensome.
Therefore, in this analysis we set $\lambda = c_\lambda \ln T / \sqrt{T}$ with $c_\lambda \in \{0.5,1,1.5,2,2.5,3\}$, and select the active lags by minimizing BIC over these six $\lambda$ values.
The coefficient estimates reported below are computed under this (sub)optimal model.

To evaluate the estimation performance, we fix the evaluation points at $\tau_k = k/100$ for $k = 1, \ldots, 99$ and define the average pointwise RMSE and overall RMSE as
\begin{align}
\text{RMSE}(\tau)
& = \left[ \frac{1}{7 \times R}\sum_{r=1}^{R} \left\{ \left(\hat \theta_{r, 0}(\tau) - \theta_{0}(\tau)\right)^2 + \sum_{(i,j) \in [3] \times [2] }\left(\hat \theta_{r, i}^{(j)}(\tau) - \theta_i^{(j)}(\tau)\right)^2 \right\}\right]^{1/2}\\
\text{RMSE}_\text{all}
& = \left[ \frac{1}{7 \times 99 \times R}\sum_{k = 1}^{99} \sum_{r=1}^{R} \left\{ \left(\hat \theta_{r, 0}(\tau_k) - \theta_{0}(\tau_k)\right)^2 + \sum_{(i,j) \in [3] \times [2] }\left(\hat \theta_{r, i}^{(j)}(\tau_k) - \theta_i^{(j)}(\tau_k)\right)^2 \right\}\right]^{1/2},
\end{align}
where the subscript $r$ indicates that it is computed on the $r$-th replication, and $R = 500$ is the total number of replications.
Since the three time series are symmetric and share an identical DGP, we report results for series $i=1$ only.
The boxplots reported in Figure \ref{fig:box} in the main text are created based on $\text{RMSE}_\text{all}$ values with $\sharp k = 1$.

Table \ref{tab:MC1rmse} presents $\text{RMSE}(0.05)$, $\text{RMSE}(0.50)$, $\text{RMSE}(0.95)$, and $\text{RMSE}_\text{all}$ across all setups.
As reported in the main text, we observe that as $b$ increases, the benefit of using SQVAR over standard QR becomes apparent.
However, as expected, when the sample size is large, the benefit is relatively small.
Interestingly, the performance of the SQVAR estimator in the upper tail is clearly worse than in the lower tail.
This may be due to the DGP used and properties of the I-spline basis.

\begin{table}[!ht]
\centering\footnotesize
\caption{RMSEs for SQVAR and standard QR}
\label{tab:MC1rmse}
\begin{tabular}{rrrcccccccc}
\toprule
\multicolumn{3}{c}{ } & \multicolumn{2}{c}{$\tau = 0.05$} & \multicolumn{2}{c}{$\tau = 0.50$} & \multicolumn{2}{c}{$\tau = 0.95$} & \multicolumn{2}{c}{All} \\
\cmidrule(l{3pt}r{3pt}){4-5} \cmidrule(l{3pt}r{3pt}){6-7} \cmidrule(l{3pt}r{3pt}){8-9} \cmidrule(l{3pt}r{3pt}){10-11}
$T$ & $b$ & $\sharp k$ & SQVAR & QR & SQVAR & QR & SQVAR & QR & SQVAR & QR\\
\midrule
200 & 1 & 1 & 0.162 & 0.132 & 0.161 & 0.175 & 0.302 & 0.142 & 0.207 & 0.174\\
200 & 1 & 2 & 0.170 & 0.132 & 0.153 & 0.175 & 0.311 & 0.142 & 0.204 & 0.174\\
200 & 2 & 1 & 0.087 & 0.130 & 0.148 & 0.157 & 0.352 & 0.137 & 0.201 & 0.158\\
200 & 2 & 2 & 0.092 & 0.130 & 0.142 & 0.157 & 0.388 & 0.137 & 0.207 & 0.158\\
200 & 3 & 1 & 0.068 & 0.132 & 0.110 & 0.153 & 0.263 & 0.139 & 0.152 & 0.155\\
200 & 3 & 2 & 0.075 & 0.132 & 0.108 & 0.153 & 0.284 & 0.139 & 0.156 & 0.155\\
200 & 4 & 1 & 0.057 & 0.134 & 0.087 & 0.151 & 0.217 & 0.140 & 0.125 & 0.154\\
200 & 4 & 2 & 0.067 & 0.134 & 0.088 & 0.151 & 0.227 & 0.140 & 0.127 & 0.154\\
200 & 5 & 1 & 0.054 & 0.135 & 0.073 & 0.149 & 0.184 & 0.141 & 0.107 & 0.153\\
200 & 5 & 2 & 0.063 & 0.135 & 0.075 & 0.149 & 0.194 & 0.141 & 0.109 & 0.153\\
200 & 6 & 1 & 0.050 & 0.135 & 0.064 & 0.147 & 0.167 & 0.141 & 0.097 & 0.153\\
200 & 6 & 2 & 0.060 & 0.135 & 0.064 & 0.147 & 0.174 & 0.141 & 0.098 & 0.153\\
\addlinespace
600 & 1 & 1 & 0.100 & 0.070 & 0.093 & 0.107 & 0.150 & 0.075 & 0.112 & 0.101\\
600 & 1 & 2 & 0.111 & 0.070 & 0.093 & 0.107 & 0.164 & 0.075 & 0.119 & 0.101\\
600 & 2 & 1 & 0.055 & 0.070 & 0.112 & 0.095 & 0.270 & 0.076 & 0.150 & 0.092\\
600 & 2 & 2 & 0.055 & 0.070 & 0.116 & 0.095 & 0.299 & 0.076 & 0.162 & 0.092\\
600 & 3 & 1 & 0.038 & 0.072 & 0.094 & 0.092 & 0.222 & 0.077 & 0.123 & 0.090\\
600 & 3 & 2 & 0.034 & 0.072 & 0.095 & 0.092 & 0.239 & 0.077 & 0.127 & 0.090\\
600 & 4 & 1 & 0.029 & 0.073 & 0.073 & 0.090 & 0.178 & 0.078 & 0.097 & 0.090\\
600 & 4 & 2 & 0.028 & 0.073 & 0.073 & 0.090 & 0.186 & 0.078 & 0.099 & 0.090\\
600 & 5 & 1 & 0.027 & 0.073 & 0.059 & 0.089 & 0.147 & 0.079 & 0.081 & 0.089\\
600 & 5 & 2 & 0.026 & 0.073 & 0.060 & 0.089 & 0.154 & 0.079 & 0.082 & 0.089\\
600 & 6 & 1 & 0.023 & 0.074 & 0.049 & 0.088 & 0.125 & 0.080 & 0.068 & 0.089\\
600 & 6 & 2 & 0.023 & 0.074 & 0.049 & 0.088 & 0.132 & 0.080 & 0.069 & 0.089\\
\addlinespace
1200 & 1 & 1 & 0.076 & 0.049 & 0.068 & 0.074 & 0.105 & 0.051 & 0.082 & 0.070\\
1200 & 1 & 2 & 0.083 & 0.049 & 0.068 & 0.074 & 0.110 & 0.051 & 0.084 & 0.070\\
1200 & 2 & 1 & 0.045 & 0.050 & 0.086 & 0.067 & 0.194 & 0.051 & 0.112 & 0.064\\
1200 & 2 & 2 & 0.042 & 0.050 & 0.087 & 0.067 & 0.215 & 0.051 & 0.119 & 0.064\\
1200 & 3 & 1 & 0.030 & 0.051 & 0.091 & 0.065 & 0.206 & 0.053 & 0.115 & 0.063\\
1200 & 3 & 2 & 0.022 & 0.051 & 0.090 & 0.065 & 0.220 & 0.053 & 0.118 & 0.063\\
1200 & 4 & 1 & 0.025 & 0.052 & 0.070 & 0.064 & 0.162 & 0.054 & 0.089 & 0.062\\
1200 & 4 & 2 & 0.016 & 0.052 & 0.070 & 0.064 & 0.171 & 0.054 & 0.091 & 0.062\\
1200 & 5 & 1 & 0.023 & 0.052 & 0.057 & 0.063 & 0.130 & 0.054 & 0.072 & 0.062\\
1200 & 5 & 2 & 0.016 & 0.052 & 0.057 & 0.063 & 0.137 & 0.054 & 0.073 & 0.062\\
1200 & 6 & 1 & 0.023 & 0.053 & 0.048 & 0.062 & 0.110 & 0.054 & 0.061 & 0.062\\
1200 & 6 & 2 & 0.015 & 0.053 & 0.048 & 0.062 & 0.115 & 0.054 & 0.062 & 0.062\\
\bottomrule
\end{tabular}
\end{table}

\subsection{Performance of the SCAD-penalized SQVAR estimator}

We again consider a trivariate model with lag order $p = 2$.
Here, different from the previous subsection, we generate data through the SQVAR model by specifying the SQVAR coefficient functions as follows:
\begin{align}
    & \phi_0(\tau) = \Phi_{0.2}^{-1}(\tau), \quad \phi_1^{(1)}(\tau) = 3 \tau + 6 \sqrt{\tau}, \\
    & \phi_2^{(1)}(\tau) = 3 \tau + 6 \Phi_1(2 \tau - 1), \quad \phi_3^{(1)}(\tau) = 3 \tau + 6 \tau^2,
\end{align}
where $\Phi_a$ denotes the CDF of a normal distribution with mean zero and standard deviation $a$.
To recover the QVAR model from the SQVAR model, we compute $\text{lb}_i$ and $\text{ub}_i$ as the empirical minimum and maximum and update them at every $t$.
Then, at each $t$, we invert the simplex transformation to compute the corresponding QVAR coefficients and generate the next observation at $t+1$.
We iterate this procedure until the process reaches a stable state and further updates to the bounds and QVAR coefficients are negligible.
Although this data generation process is more complicated than the previous one, it yields a more realistic time series without imposing monotonicity directly on the QVAR coefficients.

In this experiment, we consider two sample sizes $T \in \{500, 2000\}$, two numbers of inner knots $\sharp k \in \{1, 2\}$, and two numbers of quantile grid points $L \in \{15, 30\}$.
The penalty parameter $\lambda$ is chosen as in the previous experiment.
For comparison, we also evaluate the penalized SQVAR estimator without the monotonicity constraint.
Table \ref{tab:MC2rmse} presents the pointwise and overall RMSE values for the two estimators under each experimental setup.
From this table, we observe that both estimators perform fairly well for all scenarios.
We do not find any significant difference in performance between the two estimators, which is consistent with our theory.

\begin{table}[!ht]
\centering\footnotesize
\caption{Estimation accuracy of QVAR coefficients}
\label{tab:MC2rmse}
\begin{tabular}{rrrcccccccc}
\toprule
\multicolumn{3}{c}{ } & \multicolumn{2}{c}{$\tau = 0.05$} & \multicolumn{2}{c}{$\tau = 0.50$} & \multicolumn{2}{c}{$\tau = 0.95$} & \multicolumn{2}{c}{All} \\
\cmidrule(l{3pt}r{3pt}){4-5} \cmidrule(l{3pt}r{3pt}){6-7} \cmidrule(l{3pt}r{3pt}){8-9} \cmidrule(l{3pt}r{3pt}){10-11}
$T$ & $\sharp k$ & $L$ & MN & UC & MN & UC & MN & UC & MN & UC\\
\midrule
500 & 1 & 15 & 0.067 & 0.061 & 0.048 & 0.050 & 0.083 & 0.111 & 0.059 & 0.066\\
500 & 1 & 30 & 0.064 & 0.056 & 0.048 & 0.049 & 0.079 & 0.097 & 0.058 & 0.061\\
500 & 2 & 15 & 0.059 & 0.054 & 0.048 & 0.052 & 0.102 & 0.119 & 0.063 & 0.068\\
500 & 2 & 30 & 0.057 & 0.050 & 0.048 & 0.050 & 0.091 & 0.102 & 0.060 & 0.063\\
500 & 3 & 15 & 0.056 & 0.050 & 0.049 & 0.053 & 0.113 & 0.116 & 0.066 & 0.069\\
500 & 3 & 30 & 0.055 & 0.048 & 0.048 & 0.051 & 0.099 & 0.098 & 0.062 & 0.063\\
\addlinespace
2000 & 1 & 15 & 0.057 & 0.047 & 0.025 & 0.023 & 0.045 & 0.032 & 0.036 & 0.030\\
2000 & 1 & 30 & 0.056 & 0.042 & 0.025 & 0.024 & 0.042 & 0.027 & 0.035 & 0.029\\
2000 & 2 & 15 & 0.051 & 0.039 & 0.026 & 0.023 & 0.053 & 0.033 & 0.037 & 0.029\\
2000 & 2 & 30 & 0.050 & 0.034 & 0.025 & 0.023 & 0.047 & 0.028 & 0.035 & 0.027\\
2000 & 3 & 15 & 0.049 & 0.036 & 0.026 & 0.024 & 0.063 & 0.034 & 0.039 & 0.030\\
2000 & 3 & 30 & 0.048 & 0.033 & 0.026 & 0.024 & 0.051 & 0.027 & 0.036 & 0.028\\
\bottomrule
\end{tabular}

\footnotesize

(MN: SQVAR estimator with monotonicity constraint; UC: unconstrained SQVAR estimator)
\end{table}

Table \ref{tab:MC2bic} summarizes the results of the BIC lag selection.
The frequency of correctly identifying the set of active lags increases with the sample size $T$.
In particular, the frequency that the selected active set contains the true $\mathcal S_1$ reaches 100\% when $T=2000$.
Regarding the number of inner knots and the number of quantile grid points, a more parsimonious (small $\sharp k$) and finer (large $L$) estimator tends to perform better.
The last six columns of Table \ref{tab:MC2bic} present the frequency with which each $c_\lambda$ is selected as optimal. 
The results suggest that choosing $c_\lambda$ around $0.5$ -- $1$ is a reasonable choice.

\begin{table}[ht]
\centering\footnotesize
\caption{Results of BIC lag selection}
\label{tab:MC2bic}
\begin{tabular}{rrrcccccccc}
\toprule
$T$ & $\sharp k$ & $L$ & $\Pr[\mathcal S_1 = \hat{\mathcal S}_1]$ & $\Pr[\mathcal S_1 \subseteq \hat{\mathcal S}_1]$ & $c_\lambda = 0.5$ & $c_\lambda = 1$ & $c_\lambda = 1.5$ & $c_\lambda = 2$ & $c_\lambda = 2.5$ & $c_\lambda = 3$\\
\midrule
500 & 1 & 15 & 0.596 & 0.992 & 0.350 & 0.244 & 0.160 & 0.110 & 0.088 & 0.048\\
500 & 1 & 30 & 0.680 & 0.992 & 0.296 & 0.236 & 0.176 & 0.158 & 0.092 & 0.042\\
500 & 2 & 15 & 0.462 & 0.996 & 0.320 & 0.234 & 0.182 & 0.136 & 0.086 & 0.042\\
500 & 2 & 30 & 0.556 & 0.994 & 0.330 & 0.258 & 0.178 & 0.118 & 0.086 & 0.030\\
500 & 3 & 15 & 0.372 & 0.992 & 0.332 & 0.250 & 0.166 & 0.104 & 0.096 & 0.052\\
500 & 3 & 30 & 0.462 & 0.998 & 0.348 & 0.208 & 0.186 & 0.126 & 0.086 & 0.046\\
\addlinespace
2000 & 1 & 15 & 0.822 & 1.000 & 0.286 & 0.252 & 0.184 & 0.108 & 0.088 & 0.082\\
2000 & 1 & 30 & 0.894 & 1.000 & 0.256 & 0.280 & 0.176 & 0.144 & 0.076 & 0.068\\
2000 & 2 & 15 & 0.600 & 1.000 & 0.346 & 0.270 & 0.172 & 0.106 & 0.066 & 0.040\\
2000 & 2 & 30 & 0.704 & 1.000 & 0.312 & 0.328 & 0.158 & 0.100 & 0.052 & 0.050\\
2000 & 3 & 15 & 0.524 & 1.000 & 0.318 & 0.270 & 0.202 & 0.094 & 0.064 & 0.052\\
2000 & 3 & 30 & 0.624 & 1.000 & 0.298 & 0.306 & 0.186 & 0.122 & 0.044 & 0.044\\
\bottomrule
\end{tabular}
\end{table}


\section{Supplementary Figures for the Empirical Application}\label{app:empir}

\begin{figure}[!ht]
\begin{center}
\includegraphics[width = 15cm]{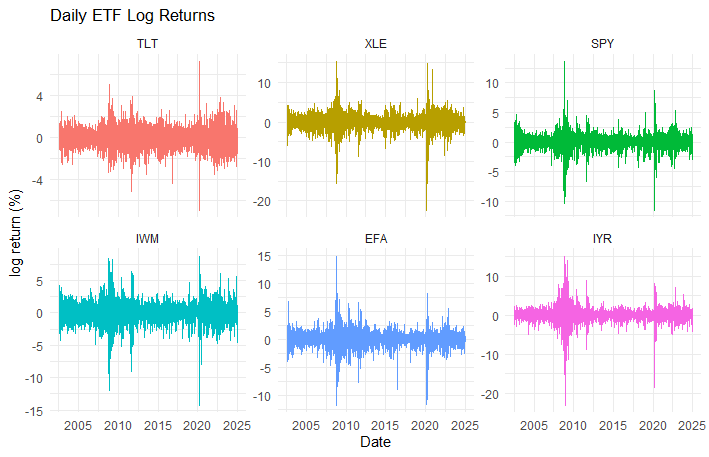}
\caption{Daily log returns of six U.S. ETFs from July 31, 2002 to December 30, 2024 ($T = 5643$)}
\label{fig:data}
\end{center}
\end{figure}

\begin{figure}[!ht]
\begin{center}
\includegraphics[width = 10cm]{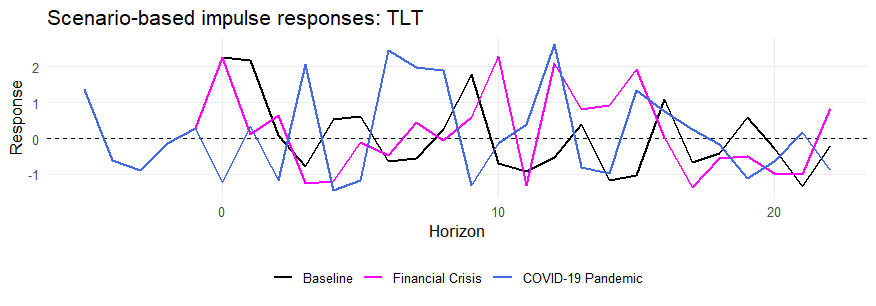}

\includegraphics[width = 10cm]{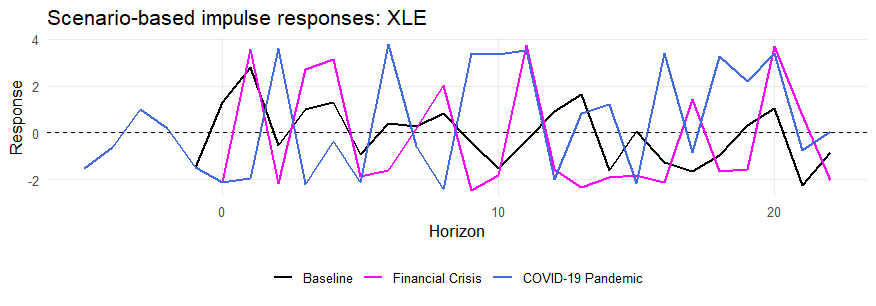}

\includegraphics[width = 10cm]{irf_spy.png}

\includegraphics[width = 10cm]{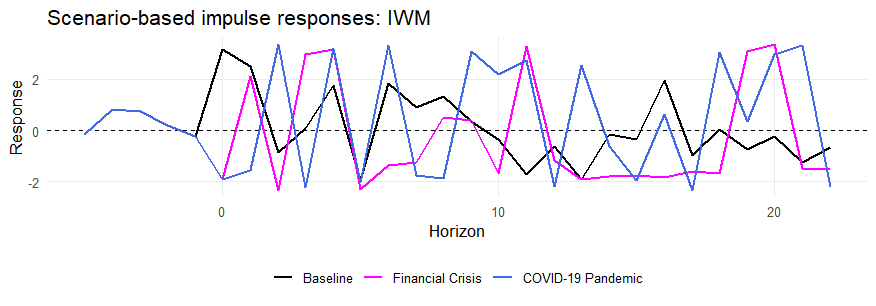}

\includegraphics[width = 10cm]{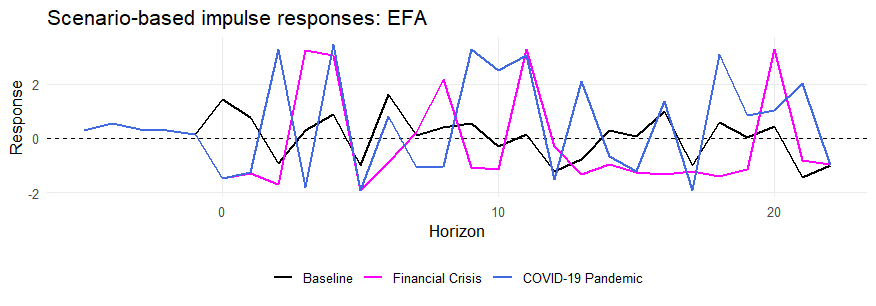}

\includegraphics[width = 10cm]{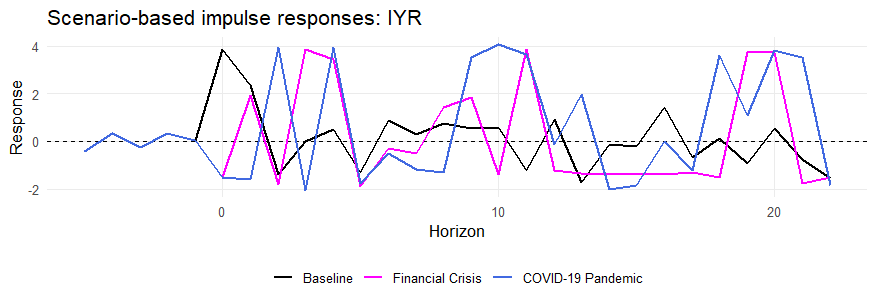}
\caption{Scenario-based impulse responses}
\label{fig:sb_other}
\end{center}
\end{figure}


\section{Pre-screening Irrelevant Variables in High-dimensional Time Series}\label{app:screen}

In some applications, it is natural to consider a high-dimensional QVAR model, where the number of cross-sectional units $n$ is large.
This section provides an additional result to address this challenging setting.
Note that the result presented below applies to a broad class of high-dimensional autoregressive models, including our QVAR as a special case.

We use a single index $m=(\ell,j)$ with $\ell\in [n]$ (series) and $j\in [p]$ (lag order), so that $X_{m,t} \coloneqq y_{\ell,t-j}$.
Let $\tau\in(0,1)$ be a fixed quantile level. 
For each time series $i \in [n]$ and predictor index $m=(\ell,j)$, we write the marginal conditional $\tau$-th quantile as $Q_{y_{it}}(\tau \mid X_{m,t}\big)$, and denote the unconditional $\tau$-th quantile of $y_{it}$ by $Q_{y_{it}}(\tau) \coloneqq \inf\{y:\Pr(y_{it}\le y)\ge\tau\}$. 
Based on the following observation 
\[
Y \text{ and } X_{m,t} \text{ are independent } \iff Q_{y_{it}}(\tau \mid X_{m,t}\big) - Q_{y_{it}}(\tau) = 0 \quad \text{for all } \tau \in (0, 1),
\] 
\cite{HeWangHong2013} developed a variable screening procedure by introducing a quantile-adaptive framework for high-dimensional heterogeneous data.
Similar to \cite{HeWangHong2013}, we define the active set for the $i$-th series by
\[
M_{i}=\{ m:\ Q_{y_{it}}(\tau \mid \mathcal F_{t-1})\ \text{functionally depends on}\ X_{m,t}\,
~
{\rm for~some}~\tau \in (0,1)\}. 
\]
To quantify the effect of $X_{m,t}$ on $y_{it}$ at a specific $\tau$ and similar to \cite{HeWangHong2013}, we consider the following measure:
\[
d_{i,m,\tau}(X_{m,t}) \coloneqq Q_{y_{it}}(\tau \mid X_{m,t}\big) - Q_{y_{it}}(\tau\big).
\] 
To estimate $Q_{y_{it}}(\tau \mid X_{m,t}\big)$ and $Q_{y_{it}}(\tau)$, we can use $\hat Q_{y_{it}}(\tau \mid X_{m,t}\big) = \hat b_{i,m,0}(\tau) + \hat b_{i,m,1}(\tau)X_{m,t}$ with $\{\hat b_{i,m,0}(\tau), \hat b_{i,m,1}(\tau)\}= \argmin_{b_{0},b_{1}} \sum_{t=1}^T \rho_\tau(y_{it} - b_{0}-X_{m,t}b_{1})$, and $\hat Q_{y_{it}}(\tau)=\hat F^{-1}_{y_i}(\tau)$, which is the $\tau$-th sample quantile function based on $\{y_{i1}, \dots, y_{iT}\}$.

Define
\[
\hat{d}_{i,m,\tau}(X_{m,t}) 
\coloneqq \hat Q_{y_{it}}(\tau \mid X_{m,t}\big)-\hat Q_{y_{it}}(\tau)
= \hat b_{i,m,0}(\tau) + \hat b_{i,m,1}(\tau)X_{m,t}-\hat F^{-1}_{y_i}(\tau).
\] 
We expect $\hat{d}_{i,m,\tau}(X_{m,t})$ to be close to zero if $X_{m,t}$ is independent of $y_{it}$.
The independence screening is based on the magnitude of the estimated marginal components
\[
\|\hat{d}_{i,m,\tau}\|_2^2 \coloneqq \frac{1}{T} \sum_{t=1}^T \hat{d}_{i,m,\tau}(X_{m,t})^2.
\] 
More specifically, we select the subset of variables 
$\|\hat{d}_{i,m,\tau}\|_2^2 \ge \nu_T $ for some $\tau \in (0,1)$ where $\nu_T$ is a predefined threshold. 
In practical implementation, we employ discrete points of $\tau \in (0,1)$ as $\mathcal T :=\{ \tau_1,\ldots,\tau_A\}$, and check the condition $\|\hat{d}_{i,m,\tau}\|_2^2 \ge \nu_T$ at each grid point: 
\[
\hat M_{i} = \left\{ m : \|\hat{d}_{i,m,\tau_a}\|_2^2 \ge \nu_T \quad {\rm for~some}~ \tau_a \in \mathcal T
\right\}. 
\] 
Note that unlike \cite{HeWangHong2013}, we are trying to detect the active set by considering the entire $\tau$ under dependent time series data. 
We impose the following assumptions.

\begin{assumption}[Conditions on $y_{it}$]
\label{assumption1}
There exists $q>0$ such that $||y_{it}||_{4+q} < \infty$ for all $i,t$. In addition, there exist constants $f_{\min}, f_{\max}>0$ such that 
    \[
    0 < f_{\min} \le f_{y_{it}|X_{m,t}}(Q_{y_{it}}(\tau \mid X_{m,t}) \mid X_{m,t}) \le f_{\max} < \infty,
    \]
    for all $i,t,m$ and $\tau \in (0,1)$.  
\end{assumption}

\begin{assumption}[Minimal marginal signal]
\label{assumption2}
For each $i$ and $m$, define the pseudo-true parameter $(b_{i,m,0}^*(\tau), b_{i,m,1}^*(\tau)) \coloneqq\argmin_{b_0,b_1} \mathbb{E}[\rho_\tau(y_{it} - b_0 - b_1 X_{m,t})]$ and $d_{i,m,\tau}^*(X_{m,t}) \coloneqq b_{i,m,0}^*(\tau) + b_{i,m,1}^*(\tau) X_{m,t} - Q_{y_{it}}(\tau)$. 
There exists constant $c_1>0$ and $\alpha>0$ such that for every active variable $m\in M_{i}$
\[
\|d_{i,m,\tau}^*(X_{m,t})\|_2^2 \;\ge\; c_1\, T^{-\alpha},
\]
where the expectation is under the stationary law.
\end{assumption}

\begin{assumption}[Weak dependence] 
\label{assumption3}
The process $\{Y_t\}$ is strictly stationary and each time series $y_{it}$ satisfies $\alpha$-mixing with coefficients $\alpha_i$ such that the summability condition $\sum_{m=1}^\infty \alpha_i(m)^{\delta/(2+\delta)} < \infty$ for some $\delta>0$ is satisfied for $i \in [n]$.  
\end{assumption}

\begin{theorem}
\label{thm:uniform_screening}
Suppose that Assumptions \ref{assumption1}--\ref{assumption3} 
hold.
Then there exist constants $C_1,C_2>0$ independent of $(T,n)$ such that, for sufficiently large $T$,
\[
\Pr\Big(
M_{i} \subseteq \hat{M}_{i} \, \text{for all} \; i \in [n]
\Big)
\ge 1 - C_1 n p \exp\!\Big(-C_2 T^{1 - 2\alpha} \Big).
\]
\end{theorem}

Theorem \ref{thm:uniform_screening} indicates that, with probability tending to one, the screened model contains all true active predictors for all quantile levels and all equations.
Therefore, when the dimension $n$ of QVAR is very high, we can apply the screening prior to performing the simplex transformation. 

\begin{remark}
The theorem implies that the probability bound will tend to one as $T \to \infty$ provided that $ n p \exp\!(-C_2 T^{1 - 2\alpha}) \to 0$.
An easy calculation reveals that, as long as $1 - 2\alpha > 0$, the bound holds uniformly for $np = \exp\!\big( o(T^{\,1 - 2\alpha}) \big)$, which allows $np$ to grow even at an exponential rate in a power of $T$.
Similar to \cite{HeWangHong2013}, the above screening procedure does not account for dependence among predictors. 
When two predictors are highly correlated, both are likely to be selected in $\hat M_i$ even though selecting only one may be sufficient. Although this may lead to over-selection in the presence of strong collinearity, the subsequent SCAD-penalized procedure with BIC (\ref{eq:BIC}) is capable of removing redundantly selected predictors. 
\end{remark}

Similar to \cite{HeWangHong2013}, we can study the number of selected variables after the screening. 
Define $Z_{m,it} \coloneqq\hat d_{i,m,\tau}(X_{m,t})^2 - \mathbb E[\hat d_{i,m,\tau}(X_{m,t})^2]$.  
For each candidate index $m$ define the screening score
\[
S_{i,m} = \mu_{i,m} + \frac{1}{T}\sum_{t=1}^T Z_{m,it},
\qquad \mu_{i,m} = \mathbb E[S_{i,m}].
\]

\begin{assumption}[Signal separation]
\label{assumption4}
For the null indices $m\notin M_i$ the population means satisfy, for some $\gamma_T$, $    \sup_{m\notin M_i} |\mu_{i,m}| = o(\gamma_T)$.
Moreover, the true active set $M_i$ has a fixed size $|M_i|=O(1)$, and for every $m\in M_i$ the population score satisfies $\mu_{i,m}\ge 2\gamma_T$ for all large $T$.
\end{assumption}

\begin{theorem}
\label{thm:uniform_screening2}
Suppose that Assumptions \ref{assumption1}--\ref{assumption4} 
hold.
Then there exists a deterministic finite constant $C$  such that
\[
\Pr\big(|\hat M_i|\le C\big) \;\to\; 1 \qquad (T\to\infty),
\]
where $\hat M_i=\{m:\; S_{i,m}>\gamma_T\}$.
\end{theorem}

The above theorem suggests that if a signal is well separated, then the model obtained after screening is $O(1)$. 
We can relax this result by allowing the size of active signals $|M_i|$ gradually increases with $T$. 
However, this is out of the scope of this paper. 

\subsection{Proof of Theorems \ref{thm:uniform_screening} and \ref{thm:uniform_screening2}}\label{app:proof-sis}

For each $i$ and $m$, define the pseudo-true parameter
\[
    (b_{i,m,0}^*(\tau), b_{i,m,1}^*(\tau)) \coloneqq \argmin_{b_0,b_1} \mathbb E[\rho_\tau(y_{it} - b_0 - b_1 X_{m,t})]. 
   \]
and
\[
d_{i,m,\tau}^*(X_{m,t}) \coloneqq b_{i,m,0}^*(\tau) + b_{i,m,1}^*(\tau) X_{m,t} - Q_{y_{it}}(\tau)
\]
and
\[
\hat d_{i,m,\tau}(X_{m,t}) \coloneqq\hat b_{i,m,0}(\tau) + \hat b_{i,m,1}(\tau) X_{m,t} - \hat Q_{y_i}(\tau).
\]
For each $i$ and $m$, we denote $\|d_{i,m,\tau}\|_2^2 \coloneqq \mathbb E\big[d_{i,m,\tau}^2(X_{m,t})\big]$, where $d_{i,m,\tau}(X_{m,t}) = Q_{y_{it}}(\tau \mid X_{m,t}) - Q_{y_{it}}(\tau)$.  

\begin{lemma}
\label{lem:Zt}
Let
\[
Z_{it} \coloneqq\hat d_{i,m,\tau}^2(X_{m,t})
- \mathbb{E}\!\left[\hat d_{i,m,\tau}^2(X_{m,t})\right],
\]
so that $\{Z_t\}_{t\ge1}$ is a stationary, centered sequence.  
Then there exist constants $C_1,C_2>0$ such that
\begin{equation}
\Pr\!\left(\Big|\frac{1}{T}\sum_{t=1}^T Z_{it}\Big| > u\right)
\le C_1 \exp(-C_2 T u^2), \quad \forall\, u>0,
\end{equation}
\end{lemma}

\begin{proof}
From boundedness of $\hat d_{i,m,\tau}$ together with the moment condition on $y_{it}$, there exists $\delta>0$ such that
$\mathbb{E}|Z_{it}|^{2+\delta} < \infty$.
Together with \cite{Davydov1968}'s inequality, we have 
\[
|\mathrm{Cov}(Z_{i1},Z_{i,1+h})| 
\le C \alpha(h)^{\delta/(2+\delta)}.
\]
Noting that the mixing coefficients satisfy the summability condition $ 
\sum_{h=1}^\infty \alpha(h)^{\delta/(2+\delta)} < \infty$, the variance of $\sum_{t=1}^T Z_{it}$ can be written as
\begin{align}
\mathrm{Var}\!\left(\sum_{t=1}^T Z_{it}\right)
& 
= T\,\mathrm{Var}(Z_{i1})
+ 2 \sum_{h=1}^{T-1} (T-h)\,\mathrm{Cov}(Z_{i1},Z_{i,1+h}). \\
&
\le C_1 T + 2C_2 \sum_{h=1}^{T-1} (T-h)\,\alpha(h)^{\delta/(2+\delta)}.
\\
&
\le C_1 T + C_3 T \sum_{h=1}^\infty \alpha(h)^{\delta/(2+\delta)}.
\end{align}
This implies $\mathrm{Var}\!\left(\frac{1}{T}\sum_{t=1}^T Z_{it}\right) = O(T^{-1})$.
By Chebyshev's inequality, we obtain $T^{-1}\sum_{t=1}^T Z_{it} = O_P(T^{-1/2})$. 

We now further show that $T^{-1}\sum_{t=1}^T Z_{it}$ has an exponential tail.
Define the filtration
\[
\mathcal{G}_{i,s} \coloneqq\sigma(\ldots, Z_{i,s-1}, Z_{is}), \quad s\in\mathbb{Z}.
\]
and the projective increments
\[
\Delta_{it}^{(k)} \coloneqq \mathbb E[Z_{it} \mid \mathcal{G}_{i,t-k}] - \mathbb E[Z_{it} \mid \mathcal{G}_{i,t-k-1}].
\]
for $k\ge0$. 
We decompose 
$\sum_{t=1}^{T} Z_{it} = M_{iT} + R_{iT}$, where 
\[
M_{iT} \coloneqq\sum_{t=1}^{T} \Delta_t^{(0)}, \qquad
R_{iT} \coloneqq\sum_{t=1}^{T} \sum_{k=1}^{\infty} \Delta_t^{(k)},
\]

We first study $R_T$. 
By stationarity and the triangle inequality,
\[
\mathbb E|R_{iT}| \le \sum_{t=1}^{T} \sum_{k=1}^{\infty} \mathbb E|\Delta_{it}^{(k)}|
\le \sum_{t=1}^{T} \sum_{k=1}^{\infty} \|\Delta_{it}^{(k)}\|_2
= T \sum_{k\ge1} \|\Delta_{i0}^{(k)}\|_2
\le C T,
\]
where we used the property of the mixing coefficients in our assumption. 
Thus, by Markov inequality, for any $u>0$,
\[
\Pr(|R_{iT}| > T u/2) \le \frac{2 C}{u}.
\]

Note that $\tau \in(0,1)$ and the parameter space is compact; thus we have $|Z_{it}|<M$ for some finite $M>0$. 
Thus, the martingale differences $\Delta_{it}^{(0)}$ satisfy $|\Delta_{it}^{(0)}| \le 2 M$. 
Applying Freedman's inequality \eqref{eq:freedman}, we obtain
\[
\Pr(M_{iT} > T u/2) \le \exp(- C_2 T u^2),
\]
for some $C_2>0$ depending only on $M$ and $\Var(Z_{i0})$.

Combining the above results, we have 
\[
\Pr\Big(\Big|\sum_{t=1}^{T} Z_{it}\Big| > T u\Big) \le \Pr(|M_{iT}| > T u/2) + \Pr(|R_{iT}| > T u/2) \le C_1 \exp(-C_2 T u^2),
\]
where we used the fact that the probability bound of $\Pr(|R_{iT}| > T u/2)$ is independent of $T$. 
This implies the claim. 
\end{proof}


\begin{flushleft}
\textbf{Proof of Theorem \ref{thm:uniform_screening}}    
\end{flushleft}

We first decompose the screening measure $d_{i,m,\tau}(X_{m,t})$ as
\[
\|\hat d_{i,m,\tau}\|_2^2 - \|d_{i,m,\tau}^*\|_2^2 
= (\|\hat d_{i,m,\tau}\|_2^2 - \mathbb E[\hat d_{i,m,\tau}^2]) + (\mathbb E[\hat d_{i,m,\tau}^2] - \|d_{i,m,\tau}^*\|_2^2) \coloneqq I_1 + I_2.
\]
We evaluate $I_1$ and $I_2$ separately.

Under standard results for linear quantile regression under $\alpha$-mixing, we have $\hat b_{i,m}(\tau) - b_{i,m}^*(\tau)=O_P(T^{-1/2})$ and $ \hat Q_{y_i}(\tau)- Q_{y_i}(\tau)=O_P(T^{-1/2})$ uniformly over $i,m,\tau \in \mathcal T$. 
Therefore, we have 
\[
\hat d_{i,m,\tau}(X_{m,t}) - d_{i,m,\tau}^*(X_{m,t}) = O_P(T^{-1/2}),
\]
which implies
\[
I_2 = \mathbb E[\hat d_{i,m,\tau}^2] - \|d_{i,m,\tau}^*\|_2^2 = O_P(T^{-1/2}).
\]

We next evaluate $I_1$. 
Define $Z_t \coloneqq\hat d_{i,m,\tau}^2(X_{m,t}) - \mathbb E[\hat d_{i,m,\tau}^2(X_{m,t})]$.  
From Lemma \ref{lem:Zt}, for some constants $C_1,C_2>0$ and all $u>0$, we have 
\[
\Pr\Big(\Big|\frac1T \sum_{t=1}^T Z_t\Big| > u\Big) \le C_1 \exp(-C_2 T u^2).
\]

Given $n$ series, $p$ lags, and $A$ grid points in $\mathcal T$, we have the union bound of the following probability,
\[
\Pr\Big(\max_{i,m,\tau_a} |\|\hat d_{i,m,\tau_a}\|^2_2 - \|d_{i,m,\tau_a}^*\|^2_2| > u\Big) \le n p A C_1 \exp(-C_2 T u^2).
\]
By the assumptions of Theorem \ref{thm:uniform_screening2}, for all $m \in M_i$, we have $\|d_{i,m,\tau}^*\|_2^2 \ge c_1 T^{-\alpha}$.
We now choose threshold $\nu_T = \frac12 c_1 T^{-\alpha}$.
Then, for $T$ large enough and $\alpha < 1/2$, 
\[
\|\hat d_{i,m,\tau}\|_2^2 \ge \nu_T \quad \text{for all active } m \in M_i.
\]
which implies $M_i \subseteq \hat M_i$ for all $i \in [n]$.
Finally, we take $u = c T^{-\alpha}$. 
Then
\[
\Pr\Big(M_i \subseteq \hat M_i \text{ for all} \; i \in [n]\Big) \ge 1 - C_1 n p A \exp(-C_2 T^{1-2\alpha}),
\]
which is the claim of theorem.

\qed 

\begin{flushleft}
\textbf{Proof of Theorem \ref{thm:uniform_screening2}}    
\end{flushleft}

Denote the selected set of variables after the screening by $\hat M_i = \{m:\; S_{i,m}>\gamma_T\}$. 
We decompose
\[
\hat M_i \;=\; M_i \cup F_{iT},
\qquad F_{iT} \coloneqq\{ m\notin M_i| 1( S_{i,m}>\gamma_T)\},
\]
and thus $|F_{iT}|$ is the number of false positives.

Fix any $m\notin M_i$. 
By definition
$ S_{i,m} = \mu_{i,m} + \frac{1}{T}\sum_{t=1}^T Z_{m,t}$.
Using the assumption of Theorem \ref{thm:uniform_screening2}, for large $T$ we have $|\mu_{i,m}|\le \gamma_T/2$.  
Therefore, for large $T$,
\begin{align*}
\Pr\big(S_{i,m}>\gamma_T\big)
&= \Pr\Big(\mu_{i,m} + \frac{1}{T}\sum_{t=1}^T Z_{m,it} > \gamma_T\Big) \\
&\le \Pr\Big(\frac{1}{T}\sum_{t=1}^T Z_{m,it} > \gamma_T - \mu_{i,m}\Big)
\le \Pr\Big(\Big|\frac{1}{T}\sum_{t=1}^T Z_{m,it}\Big|> \gamma_T/2\Big).
\end{align*}
From the result of Theorem \ref{thm:uniform_screening}, for all sufficiently large $T$ we have
\[
\Pr\big(S_{i,m}>\gamma_T\big) \le C_1 \exp\!\big(-C_2 T (\gamma_T/2)^2\big)
= C_1 \exp\!\Big(-\frac{C_2}{4} T \gamma_T^2\Big).
\]
By choosing $\gamma_T$ as $\gamma_T^2 \;=\; \frac{4\log p + C'}{C_2 T}$, for all large $T$ and all $m \notin M_i$,
\[
\Pr\big(S_{i,m}>\gamma_T\big) \le C_1 \exp\!\Big(-\log p - \frac{C}{4}\Big)
= C_1 e^{-C/4} p^{-1}.
\]
Summing over the $p-s_{i0}$ false indices, we have 
\[
\mathbb E[|F_{iT}|] \;=\; \sum_{m\notin M_i} \Pr(S_{i,m}>\gamma_T)
\le (p-s_{i0})\, C_1 e^{-C'/4} p^{-1} \le C. 
\]
Thus the expected number of false positives is bounded. 
Fix any $\varepsilon\in(0,1)$. By Markov's inequality,
\[
\Pr\big(|F_{iT}| > C/\varepsilon\big) \le \varepsilon.
\]
Because $\varepsilon>0$ is arbitrary, for every $\varepsilon>0$, there exists $C$ and $T_0$ such that for all $T\ge T_0$, $\Pr\big(|\hat M_i|\le C\big) \ge 1-\varepsilon$. 
This completes the proof of Theorem \ref{thm:uniform_screening2}. 

\qed 

\clearpage
\begin{small}
\bibliography{references.bib}
\end{small}

\end{document}